%% file: jnl-2020-tap-WS-1.tex
\documentclass[journal]{IEEEtran}
\usepackage{pgfplots}
\usepackage{subfig}
\input{myDefinitions.tex}

\usepackage{mathtools}
\usepackage{filecontents}
\usepackage{pgfplots}
 \usepackage{tikz} 

\usepackage[shortlabels]{enumitem}
\ifCLASSINFOpdf
\else
\fi
\newcommand{\revise}[1]{{#1}}

\begin{document}
%
\title{Wigner-Smith Time Delay Matrix for Electromagnetics: Theory and Phenomenology}
%
%
%

\author{Utkarsh~R.~Patel
        and~Eric~Michielssen \\
        Published in the IEEE Trans. on Antennas \& Propag., vol. 69, no. 2, Feb. 2021\\
        DOI: 10.1109/TAP.2020.3008650
\thanks{U. R. Patel and E. Michielssen are with the Department
of Electrical Engineering and Computer Science, University of Michigan, Ann Arbor,
MI, 48109 USA e-mail: urpatel@umich.edu and emichiel@umich.edu.}}

\markboth{}%
{}
%



\maketitle

\begin{abstract}
Wigner-Smith (WS) time delay concepts have been used extensively in quantum mechanics to characterize delays experienced by particles interacting with a potential well.  
This paper formally extends WS time delay theory to Maxwell’s equations and explores its potential applications in electromagnetics. 
The WS time delay matrix relates a lossless and reciprocal system’s scattering matrix to its frequency derivative and allows for the construction of modes that experience well-defined group delays when interacting with the system. 
The matrix’ entries for guiding, scattering, and radiating systems are energy-like overlap integrals of the electric and/or magnetic fields that arise upon excitation of the system via its ports. 
The WS time delay matrix has numerous applications in electromagnetics, including the characterization of group delays in multiport systems, the description of electromagnetic fields in terms of elementary scattering processes, and the characterization of frequency sensitivities of fields and multiport antenna impedance matrices.      
\end{abstract}


%
\IEEEpeerreviewmaketitle

\section{Introduction}

In 1960, Felix Smith published a seminal paper “Lifetime Matrix in Collision Theory,” a description of procedures to characterize the time delays experienced by particles during quantum mechanical interactions~\cite{Smith_1960}.  Starting from the Schr\"{o}dinger equation, Smith showed that the matrix
\begin{align}
    \matr{Q} = j \matr{S}^\dag \frac{\partial \matr{S}}{\partial \omega}
    \label{eq:WS_2}
\end{align}
where $\matr{S}$  is a potential well’s scattering matrix and  $\omega$ denotes angular frequency, fully characterizes the particles' average time of residence in the system.  Over the past 60 years, the “Wigner-Smith (WS) time delay matrix” $\matr{Q}$  (as it has come to be known) has found many applications in quantum mechanics, including the study of particle tunneling through potential barriers~\cite{Buttiker_1982}--\cite{Wardlaw_1988},  the characterization of  photoionization and photoemission time delays~\cite{Gallmann_2017}--\cite{Hockett_2016}, and the analysis of decaying  quantum systems~\cite{Dittes_2000}.  For an excellent review of the field, see~\cite{Texier_2016}.

References to Smith’s paper in the electromagnetics literature have been few and far in between, however. An exception is~\cite{Winful_2003}, where group delays of fields interacting with a two-port waveguide were characterized in terms of their WS dwell times.  In optics and photonics, WS time delay concepts have been used to describe wave propagation in multimode fibers \revise{and waveguides} ~\cite{Carpenter_2015}--\revise{\cite{fan2005principal}}, to optimize light storage in highly scattering environments~\cite{Durand_2019}, to shape the flow of light in disordered media~\cite{Brandstotter_2019}, and to characterize optical fields passing through complex cavities~\cite{Gerardin_2016}--\cite{Bohm_2018}. 
Another notable line of work involves the statistical characterization of 
chaotic fields in large enclosures and reverberation chambers by exploiting connections between WS time delays and random matrix theory~\cite{Texier_2013}--\nocite{Lewenkopf_1992,Cunden_2015}\cite{Orjubin_2007}.

This paper outlines a WS time delay theory for electromagnetics.  Its  contributions are threefold.  
\begin{itemize}[leftmargin=*]
\item First, it reviews WS theory from a system’s perspective, using equation~\eqref{eq:WS_2} to elucidate $\matr{Q}$’s central role in characterizing group delays in lossless and reciprocal electromagnetic systems. 
It also describes the so-called WS
modes that arise upon diagonalization of $\matr{Q}$ and experience well-defined group delays when interacting with a system.
\item Second, it introduces closed-form expressions for the entries of the electromagnetic WS time delay matrix for guiding, scattering, and radiating systems.  Indeed,  $\matr{Q}$’s defining equation notwithstanding, its computation may proceed without knowledge of $\partial \matr{S}/\partial \omega$. For guiding systems (e.g. closed multiport waveguide networks) excited by Transverse Electromagnetic (TEM) waves,  the elements of the WS time delay matrix~\eqref{eq:WS_2} can be expressed in terms of volume integrals of energy(-like) densities involving the electric and magnetic fields that arise upon excitation of the system’s ports. For guiding systems with non-TEM excitations, scattering systems (e.g. perfect electrically conducting surfaces excited by impinging waves) or radiating systems (e.g. antennas and arrays thereof), additional correction terms and renormalization procedures are called for to obtain~\eqref{eq:WS_2}. 
\item Third, it elucidates some important characteristics of WS modes and demonstrates the potential use of~\eqref{eq:WS_2} in the broadband characterization of antenna systems.  Specifically, it shows that WS modes naturally untangle resonant, corner/edge, and ballistic scattering phenomena as they are characterized by different dwell times within a system.  It also demonstrates that knowledge of $\matr{Q}$  and $\matr{S}$  allows for the computation of $\partial \matr{S}/\partial \omega$,  which in turn can be used to assess the frequency dependence of impedance matrices of multiport systems. The theory and methods presented in this paper therefore can be viewed as multiport extensions of procedures for characterizing the bandwidth, quality factor, and stored energy of single-port antennas, see \cite{Schab_2018}--\nocite{Chalas_2016, Best_2005, VDB_2010, Capek_2015}\cite{Gustafsson_2015}.
\end{itemize}
 The three topics above are detailed in Secs.~II--IV below.  Conclusions and avenues for future research are provided in Sec.~V.

Throughout this paper, a time dependence $e^{j\omega t}$ with $\omega = 2 \pi f$ is assumed. 
Additionally, $\,^\dag$, $\,^T$, $\,^*$, and $'$ represent adjoint, transpose, complex conjugate, and angular frequency derivative $\left(d/d\omega\right)$ operations, respectively.

\input{theory.tex}

\input{results.tex}

\section{Conclusions and Avenues For Future Research}
\label{sec:conclusions}

This paper presented a WS theory for electromagnetic fields.  Following a review of basic WS concepts, closed-form expressions for the entries of the WS time delay matrix involving energy-like overlap integrals of port-excited fields were presented. 
Furthermore, the nature of WS modes in guiding and scattering systems was elucidated, and the use of the WS time delay matrix for characterizing the frequency sensitivity of antenna impedance matrices was illustrated.  
 
Applications of the WS time delay concepts abound in electromagnetics.  The authors foresee many more uses of WS methods, including
\begin{itemize}
    \item the design of multiport and distributed system that exhibit precise delays, including filter banks, antenna arrays, and (meta)materials; 
    \item the systematic phenomenological classification of fields interacting with guiding, scattering, and radiating systems, e.g. in the identification of scattering centers for radar cross section analysis;
    \item the broadband characterization of multiport antenna systems;
    \item the construction of fast frequency-sweep computational methods for characterizing broadband electromagnetic phenomena.
\end{itemize}
Work on several of the above topics is in progress and will be reported in future papers. \revise{WS relationships for lossy systems are being studied as well.}


%

\appendices

\section{Waveguide Modes}
\label{app:waveguide_ports}

Fields inside waveguides can be expanded in terms of TE, TM, and TEM modes. 

Consider the Helmholtz equation
\begin{align}
    \nabla_t^2 \Phi_p(u,v) + k_{c,p}^2 \Phi_p(u,v) &= 0\,.
    \label{eq:App_Waveguide_Phi}
\end{align}
The TM mode profile is expressed as~\cite{Marcuvitz}
\begin{align}
    \vect{\cal X}_{p,\mathrm{TM}}(u,v) = -\nabla_t \Phi_p(u,v) \,,
\end{align}
where $\Phi_p(u,v)$ is the solution to \eqref{eq:App_Waveguide_Phi} subject to boundary condition
\begin{align}
    \Phi_p(u,v) = 0 
 \end{align}
on the cross-section's boundary $d{\cal S}$.
The TE mode profile is given by
\begin{align}
    \vect{\cal X}_{p,\mathrm{TE}}(u,v) = 
-\unit{w} \times \nabla_t \Phi_p(u,v) 
\end{align}
where $\Phi_p(u,v)$ is the solution to~\eqref{eq:App_Waveguide_Phi} subject to the boundary condition
\begin{align}
    \frac{\partial \Phi_p (u,v)}{\partial n} = 0\,
\end{align}
on $d{\cal S}$, where $(\partial/\partial n)$ denotes derivative w.r.t. the outward normal direction.
Finally, the TEM mode profile is given by
\begin{align}
    \vect{\cal X}_{p,\mathrm{TEM}}(u,v) = -\nabla_t \Phi_p(u,v)  
\end{align}
subject to the boundary condition
\begin{align}
    \frac{\partial \Phi_p(u,v)}{\partial \tau} = 0
\end{align}
where $(\partial/\partial n)$ denotes derivative w.r.t. the tangential direction.
In~\eqref{eq:App_Waveguide_Phi}, $k_{c,p}$ is the cutoff wave number and $\beta_p = \sqrt{k^2 - k_{c,p}^2}$ is the propagation constant.
The wave impedance is
\begin{align}
    Z_p = \begin{cases}
    \frac{k Z}{\beta_p} & \text{if $p$ is a TE mode}\\
    \frac{\beta_p Z}{k} & \text{if $p$ is a TM mode} \\
    Z & \text{if $p$ is a TEM mode}
    \end{cases}\,.
\end{align}
\section{Modes for Scattering Systems}
\label{app:radiation_ports}

Free-space electric fields in spherical coordinate systems can be expanded in terms of vector spherical harmonics (VSH). 
Electric fields derived from VSH are either TE or TM (to $\unit{r}$) in nature.
\emph{Incoming} TE and TM fields can be expressed as~\cite{Kristensson, Hansen}
\begin{align}
\vect{\cal I}_{lm,\mathrm{TE}}(r,\theta,\phi)  &= j^{l+1} k Z^{1/2} h_l^{(1)}(k r) \vect{\cal X}_{1 lm} (\theta,\phi) \label{eq:VSH_TE}
 \\
\vect{\cal I}_{lm,\mathrm{TM}}(r,\theta,\phi)  &= \frac{j^l Z^{1/2}}{ r} \frac{\partial  k r h_l^{(1)}(k r)}{\partial (kr)}     \vect{\cal X}_{2 lm} (\theta,\phi)  \label{eq:VSH_TM} \\
&\quad + \sqrt{l(l+1)} \frac{j^l Z^{1/2}}{r}  h_l^{(1)}(k r) \vect{\cal X}_{3lm}(\theta,\phi)\,, \nonumber
\end{align}
where $l = \{ 1, \hdots, \infty \}$ and $m = \{-l,\hdots, l\}$ are mode indices and $h_l^{(1)}(z)$ is the $l$-th order spherical Hankel function of the first kind.  
The vector spherical harmonic $\vect{\cal X}_{ilm}(\theta,\phi)$ is
\begin{subequations}
\begin{align}
\vect{\cal X}_{1lm} (\theta, \phi) &= \frac{1}{\sqrt{l(l+1)}} \nabla \times \left( \vect{r} Y_{lm}(\theta,\phi)\right)  \\ 
\vect{\cal X}_{2lm}(\theta, \phi) &= \frac{1}{\sqrt{l(l+1)}} r \nabla Y_{lm}(\theta,\phi)\\
\vect{\cal X}_{3lm}(\theta,\phi) &= \unit{r} Y_{lm}(\unit{r})\,.
\end{align}
\end{subequations}
where the scalar spherical harmonic $Y_{lm}(\theta,\phi)$ is~\cite{Kristensson, Hansen}
\begin{align}
Y_{lm}(\theta,\phi) = (-1)^m \sqrt{\frac{2l+1}{4 \pi} \frac{(l-m)!}{(l+m)!}} P_l^m(\cos \theta) e^{j m \phi} .
\label{eq:Ylm}
\end{align}
Here, $P_{l}^{m}(x)$ is the associated Legendre polynomial of degree $l$ and order $m$~\cite{Abr64}.

As $r \rightarrow \infty$, incoming VSH electric fields \eqref{eq:VSH_TE}--\eqref{eq:VSH_TM} can be approximated as
\begin{subequations}
\begin{align}
\lim_{r\rightarrow \infty}\vect{\cal I}_{lm,\mathrm{TE}}(r,\theta,\phi)  &\cong  Z^{1/2}\frac{e^{j k r}}{r} \vect{\cal X}_{1 lm} (\theta,\phi) \label{eq:I_TE_rinf} \\
\lim_{r\rightarrow \infty} \vect{\cal I}_{lm,\mathrm{TM}}(r,\theta,\phi)  &\cong Z^{1/2} \frac{e^{j k r}}{r} \vect{\cal X}_{2 lm} (\theta,\phi) \label{eq:I_TM_rinf}
\end{align}
\end{subequations}
where the large argument approximation of the spherical Hankel functions was used~\cite{Abr64}.
Note that $\vect{\cal X}_{1lm}$ and $\vect{\cal X}_{2lm}$ are transverse to $\unit{r}$.
Equation~\eqref{eq:scatter_Einc} directly follows from~\eqref{eq:I_TM_rinf}.
The outgoing vector spherical waves in \eqref{eq:scatter_Eout} are obtained by complex conjugating $\vect{\cal I}_{lm,\mathrm{TE}}(r,\theta,\phi)$ and $\vect{\cal I}_{lm,\mathrm{TM}}(r,\theta,\phi)$.



\ifCLASSOPTIONcaptionsoff
  \newpage
\fi



\bibliographystyle{IEEEtran}
\bibliography{IEEEabrv,biblio}
%

%









\end{document}

%% file: myDefinitions.tex
\usepackage{tikz}
\usetikzlibrary{calc}
\usetikzlibrary{arrows}
\usetikzlibrary{spy}
\usepackage{relsize}
\usepackage{wrapfig}
\usepackage{eurosym}
\usepackage[siunitx]{circuitikz}
\usepackage{amsmath}
\usepackage{amssymb}
\usepackage{amsfonts}
\usepackage{esint}
\usepackage{mathrsfs}
\usepackage{amsthm}
\usepackage{amscd}
\usepackage{listings}
\usepackage{booktabs,longtable}
\usepackage{graphicx}
\usepackage{multirow}
\usepackage{textcomp}

\newcommand{\vect}[1]{\boldsymbol{#1}}

\newcommand{\matr}[1]{\mathbf{#1}}
\newcommand{\abs}[1]{\left \lvert #1 \right\rvert }

\newcommand{\bfrac}[2]{\left( \frac{#1}{#2} \right)}

\newcommand{\vr}[0]{\vect{r}}

 \newcommand{\unit}[1]{\vect{\hat{#1}}}

\newcommand{\junk}[1] {}

\def\XXint#1#2#3{{\setbox0=\hbox{$#1{#2#3}{\int}$}
\vcenter{\hbox{$#2#3$}}\kern-.5\wd0}}

\newcommand*\widebar[1]{%
  \hbox{%
    \vbox{%
      \hrule height 0.5pt 
      \kern0.3ex
      \hbox{%
        \kern-0.05em
        \ensuremath{#1}%
        \kern-0.05em
      }%
    }%
  }%
}

\usepackage[siunitx]{circuitikz}
\usepackage{etoolbox}
\makeatletter

\pgfcircdeclarebipole{}
    {\ctikzvalof{bipoles/tline/height}}{tline}
    {\ctikzvalof{bipoles/tline/height}}
    {\ctikzvalof{bipoles/tline/width}}
{   
    \pgfpointdiff{\pgfpointanchor{\ctikzvalof{bipole/name}start}{center}}
                 {\pgfpointanchor{\ctikzvalof{bipole/name}end}{center}}
    \pgfmathparse{veclen(\the\pgf@x,\the\pgf@y)}
    \pgftransformxshift{\pgfmathresult/2-30pt}
    \pgf@circ@res@left=\dimexpr-\pgfmathresult pt+40pt\relax
    \pgf@circ@res@step=.2\pgf@circ@res@right 
    \pgfsetlinewidth{\pgfkeysvalueof{/tikz/circuitikz/bipoles/thickness}\pgfstartlinewidth}
    \pgfpathellipse{\pgfpoint{\pgf@circ@res@right-\pgf@circ@res@step}{0}}
                   {\pgfpoint{\pgf@circ@res@step}{0}}
                   {\pgfpoint{0}{-\pgf@circ@res@up}}
    \pgfpathmoveto{\pgfpoint{\pgf@circ@res@right-\pgf@circ@res@step}{\pgf@circ@res@up}}
    \pgfpathlineto{\pgfpoint{\pgf@circ@res@left+\pgf@circ@res@step}{\pgf@circ@res@up}}
    \pgfpatharc{-90}{90}{-\pgf@circ@res@step and -\pgf@circ@res@up}
    \pgfpathlineto{\pgfpoint{\pgf@circ@res@right-\pgf@circ@res@step}{\pgf@circ@res@down}}
    \pgfsetfillcolor{white}
    \pgfusepath{draw,fill}
    \pgfpathmoveto{\pgfpoint{\pgf@circ@res@right-2*\pgf@circ@res@step}{0pt}}
    \pgfpathlineto{\pgfpoint{\pgf@circ@res@right}{0pt}}
    \pgfusepath{draw}
}

\newcommand{\myparallel}{{\mkern1mu\vphantom{\perp}\vrule depth 0pt\mkern2mu\vrule depth 0pt\mkern3mu}}

\allowdisplaybreaks

%% file: theory.tex
\newcommand{\Hquad}{\hspace{0.05em}} 

\section{WS Theory: Systems Perspective}

\input{graphics/TM_Line}

\subsection{Group Delay in a One-Port System}

Consider the linear, time-invariant, lossless, and reciprocal one-port system shown in Fig.~\ref{fig:TM_Line}, with $M=1$. Assume that the line supports the time-harmonic \emph{incoming} signal
\begin{align}
    {\cal E}^i(w,\omega) = e^{j \beta(\omega) w}
\end{align}
where $w$ is the distance away from the port and $\beta(\omega)$ represents the line's propagation constant.
The system generates the outgoing signal
\begin{align}
    {\cal E}^o(w,\omega) = S(\omega) e^{-j \beta(\omega) w} = S(\omega) \left({\cal E}^i(w,\omega)\right)^*
\end{align}
where $S(\omega)$ is the scattering coefficient. Because the system is lossless, $\abs{S(\omega)} = 1$, i.e. $S(\omega) = e^{-j\gamma(\omega)}$. $\gamma(\omega)$'s frequency dependence is key to describing the system's response to transient excitations. Indeed, assume that the line supports the incoming narrowband pulse  $\widetilde{{\cal E}^i}(w,t)$ with center frequency $\omega_o$, bandwidth $2\Delta \omega$, and real envelope \revise{$\widetilde{A}(t) = \int_{-\Delta\omega}^{\Delta \omega} A(\omega) e^{j\omega t} d\omega $}
given by
\begin{align}
    \widetilde{{\cal E}^{i}}(w,t) &= \mathrm{Re} \left[ \int_{\omega_o - \Delta \omega}^{\omega_o + \Delta \omega} A(\omega - \omega_o) e^{j \left(\omega t + \beta(\omega) w\right)} d\omega\right] \nonumber \\
    &\cong \widetilde{A}\left(t + \beta'(\omega_o) w\right) \cos \left(\omega_o t + \beta(\omega_o) w \right)
\end{align}
where use was made of $\beta(\omega) \cong \beta(\omega_o) + \beta'(\omega_o) (\omega - \omega_o)$. 
The system generates the outgoing pulse
\begin{align}
    \widetilde{{\cal E}}\Hquad^{o}(w,t) &= \mathrm{Re} \left[ \int_{\omega_o - \Delta \omega}^{\omega_o + \Delta \omega} S(\omega) A(\omega-\omega_o) e^{j \left(\omega t + \beta(\omega) w\right)} d\omega \right] \nonumber \\
    &\cong \widetilde{A}\left(t - \beta'(\omega_o) w - \gamma'(\omega_o)\right)\nonumber \\
    &\quad \quad \cos\left(\omega_o t - \beta(\omega_o) w - \gamma(\omega_o) \right) \label{eq:Eout_pulse}
\end{align}
\revise{where use was made of $\gamma(\omega) \cong \gamma(\omega_o) + \gamma'(\omega_o) (\omega - \omega_o)$}. 
\revise{Pulse distortions not captured by \eqref{eq:Eout_pulse} will occur if the above approximations for $\beta(\omega)$ and $\gamma(\omega)$ are violated.}
The outgoing pulse's group velocity and group delay are $\partial \omega / \partial \beta(\omega)$ and
\begin{align}
    Q &= \gamma'(\omega) \nonumber \\
    &= j S^*(\omega) S'(\omega)
    \label{eq:Oneport_Narrowband_qp}
\end{align}
evaluated for $\omega = \omega_o$, respectively~\cite{Pozar_2005}.

\subsection{Group Delay in a Multi-Port System}

The above scenario is easily generalized to the linear, time-invariant, lossless, and reciprocal $M$-port system in Fig.~\ref{fig:TM_Line}, where all lines are assumed identical. 
Assume that line $p$ supports the time-harmonic \emph{incoming} signal
\begin{align}
    {\cal E}_p^i(w,\omega) = e^{j\beta(\omega) w} \,.
\end{align}
On line $1 \le m \le M$ the system generates the outgoing signal
\begin{equation}
    {\cal E}_p^o(w,m,\omega) = \matr{S}_{mp} e^{-j\beta (\omega) w} = \matr{S}_{mp} \left({\cal E}_m^i(w, \omega)\right)^* \,.
    \end{equation}
The $M\times M$ scattering matrix $\matr{S}$ is \revise{unitary} and symmetric, i.e.
\begin{subequations}
\begin{gather}
    \matr{S}^\dag \matr{S} = \matr{I}_M \label{eq:SSI}\\
    \matr{S} = \matr{S}^T \label{eq:SSt}
\end{gather}
\end{subequations}
where $\matr{I}_M$ is the $M \times M$ identity matrix.

Next, assume that port $p$ is excited by the incoming narrowband pulse
\begin{align}
    \widetilde{{\cal E}_p^i}(w,t) \equiv \widetilde{{\cal E}^i}(w,t)\,.
\end{align}
Using $\matr{S}_{mp}(\omega) = \abs{\matr{S}_{mp}(\omega)} e^{-j\gamma_{mp}(\omega)}$, the outgoing pulse on line $m$ is
\begin{align}
 \widetilde{{\cal E}}\Hquad_p^o\Hquad(w,m,t) &= \mathrm{Re} \left[ \int_{\omega_o - \Delta \omega}^{\omega_o +\Delta \omega} A(\omega - \omega_o) \matr{S}_{mp}(\omega) e^{j\left(\omega t - \beta(\omega) w\right)} d\omega \right] \nonumber \\
 &\cong \abs{\matr{S}_{mp}(\omega_o)} \widetilde{A}\left(t - \beta'(\omega_o) w - \gamma_{mp}'(\omega_o) \right) \nonumber \\
 &\quad \quad \quad  \cos \left( \omega_o t - \beta(\omega_o) w - \gamma_{mp}(\omega_o) \right) \nonumber \\
 & \quad + \abs{\matr{S}_{mp}(\omega_o)}' \frac{\partial \widetilde{A}\left(t - \beta'(\omega_o) w \right)}{\partial t} \nonumber \\
 & \quad \quad \quad \sin \left( \omega_o t - \beta(\omega_o) w - \gamma_{mp}(\omega_o)\right) \,. \label{eq:Multiport_Narrowband_Eo}
\end{align}
Given the narrowband and smooth nature of $\widetilde{A}(t)$, the second term in~\eqref{eq:Multiport_Narrowband_Eo} can be neglected. It follows that the outgoing signal's group delay on line $m$ due to an incoming signal on line $p$ is 
\begin{align}
    q_{mp} &= \gamma_{mp}'(\omega) \nonumber \\
    &= j \frac{1}{\matr{S}_{mp}(\omega)} \matr{S}_{mp}'(\omega)
    \label{eq:Multiport_Narrowband_qp}
\end{align}
evaluated for $\omega = \omega_o$.
To generalize~\eqref{eq:Oneport_Narrowband_qp} to multiport systems, Smith put forward the time delay matrix defined in~\eqref{eq:WS_2}.
To interpret $\matr{Q}$, he introduced the weighted average time delay experienced by $\widetilde{{\cal E}}\Hquad_p^i(w,t)$ as it makes its way from input port $p$ to all output ports:
\begin{align}
    \left < q_{mp} \right > \equiv \sum_{m=1}^{M} \abs{\matr{S}_{mp}(\omega)}^2 q_{mp}\,.
    \label{eq:Multiport_Narrowband_qp2}
\end{align}
The weighting coefficient $\abs{\matr{S}_{mp}(\omega)}^2$ is the fraction of the power carried by $\widetilde{{\cal E}}\Hquad_p^i(w,t)$ transferred to port $m$. 
Inserting~\eqref{eq:Multiport_Narrowband_qp} into~\eqref{eq:Multiport_Narrowband_qp2} yields
\begin{align}
    \left < q_{mp} \right > &\equiv j \sum_{m=1}^{M} \abs{\matr{S}_{mp}(\omega)}^2 \frac{1}{\matr{S}_{mp}(\omega)} \matr{S}_{mp}'(\omega) \nonumber \\
    &= \matr{Q}_{pp}(\omega)
    \label{eq:Multiport_Narrowband_qp3} 
\end{align}
where use was made of~\eqref{eq:SSI} and~\eqref{eq:SSt}. Equation~\eqref{eq:Multiport_Narrowband_qp3} implies that the diagonal elements of $\matr{Q}$ are the average time delays experienced by wave packets entering the system. $\matr{Q}$'s off-diagonal elements have no direct physical interpretation but are important in the transformations described below.

\subsection{Simultaneous Diagonalization of $\matr{Q}$ and $\matr{S}$, and WS Modes}
\label{sec:SimultaneousDiagonalization}

Important insights into the temporal delays imposed by the system can be obtained by diagonalizing $\matr{Q}$ and $\matr{S}$. 
It immediately follows from~\eqref{eq:WS_2} that the WS time delay matrix is Hermitian:
\begin{align}
    \matr{Q}^\dag &= \left(j \matr{S}^\dag \matr{S}' \right)^\dag \nonumber \\
    &= -j \left(\matr{S}^\dag\right)' \matr{S} \nonumber \\
    &= j \matr{S}^\dag \matr{S}' \nonumber \\
    &= \matr{Q}\,.
\end{align}
Here, use was made of $\left(\matr{S}^\dag\right)' \matr{S} = - \matr{S}^\dag \matr{S}'$, which follows from the frequency derivative of~\eqref{eq:SSI}. $\matr{Q}$ therefore can be diagonalized as
\begin{align}
\matr{Q} = \matr{W} \widebar{\matr{Q}}  \matr{W}^\dag
\label{eq:Q_diagonal}
\end{align}
where $\matr{W}$ is a unitary matrix whose columns are  $\matr{Q}$'s eigenvectors,
henceforth termed ``WS modes'', and $\widebar{\matr{Q}}$ is a diagonal matrix holding $\matr{Q}$'s real eigenvalues.
Without loss of generality, it is assumed that all eigenvalues are distinct\footnote{\revise{identical  eigenvalues  may occur when the system under consideration exhibits symmetries.  In practice, these symmetries can be broken by perturbing the structure (or shifting its position w.r.t. the reference plane/origin).  In the context of numerical simulations, it even can be broken via asymmetric discretization of the structure.}} and ordered \revise{$\widebar{\matr{Q}}_{11} < \widebar{\matr{Q}}_{22} < \hdots < \widebar{\matr{Q}}_{MM}$}. 
It is easily shown that $\matr{W}$ also factorizes $\matr{S}$ as
\begin{align}
    \matr{S} = \matr{W}^*  \widebar{\matr{S}} \matr{W}^\dag
    \label{eq:eq15}
\end{align}
where $\widebar{\matr{S}}$ is diagonal.
Indeed, substituting~\eqref{eq:eq15} and~\eqref{eq:Q_diagonal} into~\eqref{eq:WS_2} yields
\begin{align}
    \matr{S}' &= -j \left(\matr{W}^* \widebar{\matr{S}} \matr{W}^\dag \right) \left(\matr{W} \widebar{\matr{Q}} \matr{W}^\dag \right) \nonumber \\
    &= -j \matr{W}^* \widebar{\matr{S}}\, \widebar{\matr{Q}} \matr{W}^\dag \,.
    \label{eq:177}
\end{align}
Equation~\eqref{eq:SSt} implies $\left(\matr{S}' \right)^T = \matr{S}'$. 
It therefore follows from~\eqref{eq:177} that  $\widebar{\matr{S}}\, \widebar{\matr{Q}} = \widebar{\matr{Q}} \,\widebar{\matr{S}}$. 
Since $\widebar{\matr{S}}$ commutes with a diagonal matrix with distinct elements, it must be diagonal~\cite{Horn2012}.
The same result can be obtained using time reversal arguments. 
\revise{As the system is lossless,  $\abs{\widebar{\matr{S}}_{pp}} = 1$ and the phases of the columns of $\matr{W}$ can be chosen such that $\widebar{\matr{S}} = \matr{I}_M$. The 
simultaneous diagonalization of $\matr{Q}$ and $\matr{S}$ implies
that the WS modes are entirely decoupled and characterized by the well-defined time-delays in $\widebar{\matr{Q}}$.

The Courant-Fisher minimax theorem states that $\matr{W}$'s
 $q$-th column, $\matr{w}_q = \begin{bmatrix} \matr{W}_{1q} & \matr{W}_{2q} & \hdots & \matr{W}_{Mq} \end{bmatrix}^T$ minimizes the Rayleigh quotient~\cite{Horn2012}
 \begin{align}
     R(\matr{w}) = \frac{\matr{w}^T \matr{Q} \matr{w}}{\matr{w}^T \matr{w}}
 \end{align}
 in the space of vectors $\matr{w}$ orthogonal to $\mathrm{Span}\left( \matr{w}_1, \matr{w}_2, \hdots, \matr{w}_{q-1} \right)$, and that $\widebar{\matr{Q}}_{qq} = R\left(\matr{w}_q\right)$.
 Alternatively, $\partial R(\matr{w})/\partial \matr{w} = 0$ for each $\matr{w}_q$, and $R(\matr{w})$'s critical values are the WS time delays $\widebar{\matr{Q}}_{qq}$. 
 The WS modes therefore extremize  the amount of time that a pulse dwells in the system. Specifically, the first (last) incoming WS mode represents a combination of excitations that results in an outgoing pulse experiencing the smallest (largest) possible time delay while interacting with the system.

Next, let $\widehat{\matr{W}} \equiv \matr{W}(\omega_o)$ for a fixed frequency $\omega_o$, and define $\widehat{\matr{Q}}(\omega)$ and $\widehat{\matr{S}}(\omega)$ as
\begin{subequations}
\begin{align}
    \widehat{\matr{Q}}(\omega) &= \widehat{\matr{W}}^\dag \matr{Q}(\omega) \widehat{\matr{W}} \label{eq:Qhat_1}\\
    \widehat{\matr{S}}(\omega) &= \widehat{\matr{W}}^T \matr{S}(\omega) \widehat{\matr{W}}\,. \label{eq:Shat_1} 
\end{align}
\end{subequations}
Note that  $\widehat{\matr{Q}}(\omega) = \widebar{\matr{Q}}(\omega)$ and $\widehat{\matr{S}}(\omega) = \matr{I}_M$ when $\omega = \omega_o$.

Substituting \eqref{eq:Qhat_1}--\eqref{eq:Shat_1} into \eqref{eq:WS_2} yields
\begin{align}
    \widehat{\matr{Q}} = j \widehat{\matr{S}}^\dag \, \widehat{\matr{S}}\,'  \,.
    \label{eq:group_delay_MP}
\end{align}
Exciting the system with the time-harmonic incoming WS mode
\begin{align}
    {\cal E}_{\mathrm{WS},q}^i ( w, \omega_o) \equiv \sum_{p=1}^M \widehat{\matr{W}}_{pq} {\cal E}_{p}^i(w,\omega_o)    
\end{align}
results in the outgoing WS mode
\begin{align}
    {\cal E}_{\mathrm{WS},q}^o(w,m,\omega_o) =  \left({\cal E}_{\mathrm{WS},q}^i(w,\omega_o)\right)^* \,.
    \label{eq:motivation_WS_modes1}
\end{align}
A narrowband outgoing pulse built from WS mode $q$ therefore exhibits group delay $\widebar{\matr{Q}}_{qq}$ w.r.t. its incoming counterpart, uniformly across all lines.

 Finally, it is noted that if $\matr{S}(\omega_o)$ and $\matr{Q}(\omega_o)$ are known,~\eqref{eq:WS_2} and~\eqref{eq:group_delay_MP} yield
 \begin{align}
     \matr{S} \left(\omega_o + \delta \omega \right) &\cong \matr{S}(\omega_o) - j \, \delta\omega \, \matr{S}(\omega_o) \matr{Q}(\omega_o)  \nonumber \nonumber \\
     \widehat{\matr{S}}(\omega_o + \delta \omega) &\cong \matr{I}_M - j\, \delta\omega\,  \widebar{\matr{Q}}(\omega_o) \nonumber \\
     &\cong  e^{-j\,\delta\omega\, \widebar{\matr{Q}}(\omega_o)}\,.
 \end{align}
 These estimates of $\matr{S}(\omega_o + \delta\omega)$ and $\widehat{\matr{S}}(\omega_o + \delta \omega)$ in turn can be used to approximate the system's response at $\omega_o + \delta \omega$. For example, consider the incoming signal
 \begin{align}
    {\cal F}_q^i(w, \omega_o + \delta \omega) \equiv \sum_{p=1}^M \widehat{\matr{W}}_{pq} {\cal E}_p^i(w, \omega_o + \delta\omega)
 \end{align}
 obtained by evolving the frequency of the ${\cal E}_{p}^i(w,\omega_o)$ in the WS modes in~\eqref{eq:motivation_WS_modes1}, while keeping their combination constants fixed.
 The outgoing signal ${\cal F}_{q}^o(w,m,\omega_o + \delta \omega)$ generated in response to this excitation is
 \begin{align}
     {\cal F}_{q}^o&(w, m, \omega_o + \delta \omega) = \sum_{p=1}^{M} \widehat{\matr{W}}_{pq} {\cal E}_p^o(w, m, \omega_o+\delta \omega)  \nonumber \\
     &\cong  e^{-j\delta\omega \widebar{\matr{Q}}_{qq}(\omega_o)} \left( {\cal F}_{q}^i(w, \omega_o + \delta \omega)  \right)^* \,,
 \end{align}
showing that, to first order, WS modes do not couple when changing the frequency; they simply acquire an extra phase delay $\delta\omega\, \widebar{\matr{Q}}_{qq}(\omega_o)$.}

While the discussion so far assumed all lines were identical, almost all methods and conclusions presented above continue to hold true when this condition is violated.  
Even when the lines support waves traveling at different speeds, WS modes still describe wave packets that simultaneously exit the system, even though they disperse after that (example: non TEM waveguides).
\revise{In addition, while the above derivation and those that follow assume lossless systems, the results and phenomenology of WS modes qualitatively extend to slightly lossy systems as evidenced by the recent experimental observation of WS modes in real-world fiber optic and microwave systems~\cite{Carpenter_2015},\cite{DelHougne}--\cite{Horodynski2020}.}

\section{WS Theory: Electromagnetic Perspective}

The electromagnetic WS time delay matrix $\matr{Q}(\omega)$ can be evaluated from knowledge of the fields at frequency $\omega$ that exist throughout  the system for all possible port excitations.  
This section presents expressions for the entries of the WS time delay matrix $\matr{Q}(\omega)$ for guiding ($g$), scattering ($s$), and radiating ($r$) systems.  
For guiding systems fed by TEM waveguides, $\matr{Q}(\omega)$'s entries are expressed as energy-like overlap integrals of the system's electric and/or magnetic fields; correction factors involving the system's scattering matrix and waveguide impedances are used when the system is fed through non-TEM ports.  For scattering and radiating systems,  a renormalization procedure that extracts the system's far-fields is introduced.  
\subsection{Guiding Systems}
\label{sec:guiding_systems}
\input{graphics/closed_system.tex}
\subsubsection{Setup}
Consider a lossless microwave network with perfect electrically conducting (PEC) walls that is terminated by homogeneous waveguides of uniform cross section. 
Let $\Omega$ and $d\Omega$ denote the network’s volume and physical port surfaces, respectively. 
Next, consider the (global) curvilinear coordinate system $(u, v, w)$ shown in Fig.~\ref {fig:GuidingStructure_1}. On $d\Omega$, $w = 0$ and $(u, v, w)$ is  locally Cartesian.  Let $\unit{w}$ denote the outward pointing normal to $d\Omega$.  
The physical port surfaces are assumed far removed from waveguide discontinuities, so fields there can be expressed in terms of propagating modes.
Each physical port supports one or more propagating modes; 
let $M_g$ denote the total number of propagating modes in all physical ports.

Assume that the network is excited by 
an \emph{incoming} unit-power field with transverse electric and magnetic components 
\begin{subequations}
\begin{align}
    \vect{\cal E}_{p,\myparallel}^i(\vr,\omega) &=  n_p(\omega) e^{j \beta_p(\omega) w} \vect{\cal X}_p( u,v) \label{eq:guiding_Einc}\\
   \vect{\cal H}_{p,\myparallel}^i(\vr,\omega) &= \frac{ n_p(\omega)}{Z_p(\omega)} e^{j \beta_p(\omega) w} \left(-\unit{w} \times \vect{\cal X}_p(u,v)\right)
    \label{eq:guiding_Hinc}\,.
\end{align}
\end{subequations}
Here $\vr = (u,v,w)$ and $1 \le p \le M_g$ denotes the index of a TE, TM, or TEM propagating mode.
The mode's transverse profile $\vect{\cal X}_p(u,v)$ is supported on $d\Omega_p \subset d\Omega$.  
\revise{Note that \eqref{eq:guiding_Einc}--\eqref{eq:guiding_Hinc} only describe fields in the vicinity of $d\Omega_p$, and not everywhere inside $\Omega$ as the field there may not have an obvious modal decomposition.}
Note that $d\Omega_p = d\Omega_{p’}$ when mode $p$ and $p’$ share the same physical port. 
In~\eqref{eq:guiding_Einc}--\eqref{eq:guiding_Hinc}, $Z_p(\omega)$, $n_p(\omega) = \sqrt{Z_p(\omega)}$, and $\beta_p(\omega)$ are the $p$-th mode's impedance, power normalization factor, and propagation constant.
A procedure to obtain $\vect{\cal X}_p(u,v)$, $n_p(\omega)$, $Z_p(\omega)$, and $\beta_p(\omega)$ for arbitrarily shaped waveguides is outlined in Appendix~\ref{app:waveguide_ports}.
These modes are automatically orthogonal and normalized to satisfy 
\begin{align}
\int_{d\Omega} \vect{\cal X}_p(u,v) \cdot \vect{\cal X}_q^*(u,v) \, du\,dv = \delta_{pq}\,.
\label{eq:Closed_Orthogonality}
\end{align}
The $\vect{\cal X}_{q}(u,v)$ are purely real, i.e. $\vect{\cal X}_{q}^*(u,v) = \vect{\cal X}_q(u,v)$. 

Next, let $\vect{\cal E}_p(\vr,\omega)$ and $\vect{\cal H}_p(\vr,\omega)$ denote the field throughout $\Omega$ due to excitation \eqref{eq:guiding_Einc}--\eqref{eq:guiding_Hinc}, assuming all ports are matched.  Near $d\Omega$, these fields' transverse (to $\unit{w}$) components  can be expressed as
\begin{subequations}
\begin{align}
    \vect{\cal E}_{p,\myparallel}(\vr,\omega) &= \vect{\cal E}_{p,\myparallel}^i(\vr,\omega) + \vect{\cal E}_{p,\myparallel}^o(\vr,\omega) \label{eq:guiding_Etot}\\
    \vect{\cal H}_{p,\myparallel}(\vr,\omega) &= \vect{\cal H}_{p,\myparallel}^i(\vr,\omega) + \vect{\cal H}_{p,\myparallel}^o(\vr,\omega)\,, \label{eq:guiding_Htot}
\end{align}
\end{subequations}
where the \emph{outgoing} transverse fields are
\begin{subequations}
\begin{align}
    \vect{\cal E}_{p,\myparallel}^o(\vr,\omega) &= \sum_{m = 1}^{M_g} \matr{S}_{mp}(\omega) n_m(\omega) e^{-j \beta_m(\omega) w} \vect{\cal X}_m(u,v) \label{eq:guiding_Eout} \\
    \vect{\cal H}_{p,\myparallel}^o(\vr,\omega) &= - \sum_{m = 1}^{M_g} \matr{S}_{mp}(\omega) \frac{n_m(\omega)}{Z_m(\omega)} e^{-j \beta_m(\omega) w} \nonumber \\
    &\quad \quad \quad \cdot \left(-\unit{w} \times \vect{\cal X}_m(u,v)\right)\,. \label{eq:guiding_Hout}
\end{align}
\end{subequations}
The above construct guarantees that  $\matr{S}$ is unitary.
From here onwards, the $\omega$ dependence of quantities is suppressed.

\subsubsection{Maxwell's equations}
Consider two sets of electromagnetic fields inside $\Omega$: $\{\vect{\cal E}_p(\vr), \vect{\cal H}_p(\vr) \}$ and $\{\vect{\cal E}_q(\vr), \vect{\cal H}_q(\vr)\}$. 
The frequency derivative of Maxwell's equations for $\{\vect{\cal E}_p(\vr), \vect{\cal H}_p(\vr) \}$ reads
\begin{subequations}
\begin{align}
\nabla \times \vect{\cal H}_p'(\vr) &= j \varepsilon \vect{\cal E}_p(\vr) + j\omega \varepsilon \vect{\cal E}_p'(\vr)  \label{eq:WS_FO_7} \\
\nabla \times \vect{\cal E}_p'(\vr) &=  -j \mu \vect{\cal H}_p(\vr) - j\omega \mu \vect{\cal H}_p'(\vr)   \label{eq:WS_FO_8}
\end{align}
\end{subequations}
where $\varepsilon$ and $\mu$ are the permittivity and permeability of the medium inside $\Omega$. \revise{Here and throughout the remainder of the paper, $\varepsilon$ and $\mu$ are assumed frequency-independent.}
The conjugate of Maxwell's equations for $\{\vect{\cal E}_q(\vr), \vect{\cal H}_q(\vr)\}$ reads
\begin{subequations}
\begin{align}
\nabla \times \vect{\cal H}_q^*(\vr) &= -j\omega \varepsilon \vect{\cal E}_q^*(\vr) \label{eq:WS_FO_5} \\
\nabla \times \vect{\cal E}_q^*(\vr) &= j\omega \mu \vect{\cal H}_q^*(\vr) \,. \label{eq:WS_FO_6}
\end{align}
\end{subequations}
Adding the dot-product of~\eqref{eq:WS_FO_5} and $\frac{1}{2} \vect{\cal E}_p'(\vr)$ to the dot-product of~\eqref{eq:WS_FO_7} and $\frac{1}{2}\vect{\cal E}_q^*(\vr)$ yields
\begin{align}
\frac{1}{2}\vect{\cal E}_p'(\vr) \cdot \nabla \times \vect{\cal H}_q^*(\vr) + \frac{1}{2} \vect{\cal E}_q^*(\vr) &\cdot \nabla \times \vect{\cal H}_p'(\vr) \nonumber \\
&=   \frac{j}{2}\varepsilon \vect{\cal E}_q^*(\vr) \cdot \vect{\cal E}_p(\vr)\,.  \label{eq:WS_FO_13a}
\end{align}
Similarly, adding the dot-product of \eqref{eq:WS_FO_6} and $\frac{1}{2} \vect{\cal H}_p'(\vr)$ to the dot-product of~\eqref{eq:WS_FO_8} and $\frac{1}{2} \vect{\cal H}_q^*(\vr)$ results in
\begin{align}
\frac{1}{2} \vect{\cal H}_q^*(\vr) \cdot \nabla \times \vect{\cal E}_p'(\vr) &+ \frac{1}{2} \vect{\cal H}_p'(\vr) \cdot \nabla \times \vect{\cal E}_q^*(\vr) \nonumber \\
&=  - \frac{j}{2} \mu \vect{\cal H}_q^*(\vr) \cdot \vect{\cal H}_p(\vr)\,.  \label{eq:WS_FO_13}
\end{align}
Subtracting~\eqref{eq:WS_FO_13} from~\eqref{eq:WS_FO_13a} yields
\begin{align}
\frac{j}{2} \nabla \cdot &\left[  \vect{\cal E}_p'(\vr) \times \vect{\cal H}_q^*(\vr) +  \vect{\cal E}_q^*(\vr)   \times \vect{\cal H}_p'(\vr)  \right]   \nonumber \\
&\quad \quad = \frac{1}{2} \varepsilon \vect{\cal E}_q^*(\vr) \cdot \vect{\cal E}_p(\vr) + \frac{1}{2} \mu \vect{\cal H}_q^*(\vr) \cdot \vect{\cal H}_p(\vr)\,.
\label{eq:WS_FO_14}  
\end{align}
Finally, integrating the 
left and right hand sides (LHS and RHS)
 of~\eqref{eq:WS_FO_14} over $\Omega$,  applying the divergence theorem, and enforcing the boundary conditions on the network's PEC walls yields
\begin{align}
\frac{j}{2} &\int_{d\Omega} \unit{w} \cdot \left[  \vect{\cal E}_{q,\myparallel}^*(\vr)   \times \vect{\cal H}_{p,\myparallel}'(\vr)  + \vect{\cal E}_{p,\myparallel}'(\vr) \times \vect{\cal H}_{q,\myparallel}^*(\vr)   \right]   dS  \nonumber \\
&\quad = \frac{1}{2}  \int_{\Omega} \varepsilon \vect{\cal E}_q^*(\vr) \cdot \vect{\cal E}_p(\vr)   + \mu \vect{\cal H}_q^*(\vr) \cdot \vect{\cal H}_p(\vr)  dV \,.
\label{eq:WS_FO_17}
\end{align}


\subsubsection{WS Relationship}
The evaluation of the LHS of~\eqref{eq:WS_FO_17} requires expressions for $\vect{\cal E}_{p,\myparallel}'(\vr)$, $\vect{\cal H}_{p,\myparallel}'(\vr)$, $\vect{\cal E}_{q,\myparallel}^*(\vr)$, and $\vect{\cal H}_{q,\myparallel}^*(\vr)$  on $d\Omega$.
Substituting~\eqref{eq:guiding_Einc}--\eqref{eq:guiding_Hinc} and~\eqref{eq:guiding_Eout}--\eqref{eq:guiding_Hout} into~\eqref{eq:guiding_Etot}--\eqref{eq:guiding_Htot}, and differentiating w.r.t
frequency, yields
\begin{subequations}
\begin{align}
\vect{\cal E}_{p,\myparallel}'(\vr) &= \left(n_p e^{j \beta_p w} \right)' \vect{\cal X}_p(u,v) \label{eq:Enprime} \\
&\quad + \sum_{m = 1}^{M_g} \matr{S}_{mp}' n_m e^{-j \beta_m w} \vect{\cal X}_m(u,v) \nonumber \\
&\quad + \sum_{m = 1}^{M_g} \matr{S}_{mp} \left(n_m e^{-j \beta_m w}\right)' \vect{\cal X}_m(u,v) \nonumber \\
\vect{\cal H}_{p,\myparallel}'(\vr) &=  \left(\frac{ n_p}{Z_p} e^{j \beta_p w}\right)' \left[- \unit{w} \times \vect{\cal X}_p(u,v) \right] \label{eq:Hnprime} \\
&\quad - \sum_{m = 1}^{M_g} \matr{S}_{mp}' \frac{n_m}{Z_m} e^{-j \beta_m w} \left[- \unit{w} \times \vect{\cal X}_m(u,v) \right] \nonumber \\
&\quad  - \sum_{m = 1}^{M_g} \matr{S}_{mp} \left(\frac{n_m}{Z_m} e^{-j \beta_m w} \right)' \left[- \unit{w} \times \vect{\cal X}_m(u,v) \right]\,.
\nonumber
\end{align}
\end{subequations}
Likewise,  substituting   \eqref{eq:guiding_Einc}--\eqref{eq:guiding_Hinc} and \eqref{eq:guiding_Eout}--\eqref{eq:guiding_Hout} into \eqref{eq:guiding_Etot}--\eqref{eq:guiding_Htot} with $p\rightarrow q$, and 
complex conjugating the result yields
\begin{subequations}
\begin{align}
\vect{\cal E}_{q,\myparallel}^*(\vr) =& n_q e^{-j \beta_q w} \vect{\cal X}_q^*(u,v) \nonumber \\
&\quad + \sum_{m = 1}^{M_g} \matr{S}_{mq}^* n_m e^{j \beta_m w} \vect{\cal X}_m^*(u,v) 
\label{eq:Emconj}
\\
\vect{\cal H}_{q,\myparallel}^*(\vr) =& \frac{ n_q}{Z_q} e^{-j \beta_q w} \left(-\unit{w} \times \vect{\cal X}_q^*(u,v)\right) \nonumber \\
& - \sum_{m = 1}^{M_g} \matr{S}_{mq}^* \frac{n_m}{Z_m} e^{j \beta_m w} \left(-\unit{w} \times \vect{\cal X}_m^*(u,v)\right) \,.
\label{eq:Hmconj}
\end{align}
\end{subequations}
Substituting~\eqref{eq:Emconj}--\eqref{eq:Hmconj} and \eqref{eq:Enprime}--\eqref{eq:Hnprime} into the LHS of~\eqref{eq:WS_FO_17}, and manipulating the resulting equation produces
\begin{align}
\frac{j}{2} &\int_{d\Omega} \unit{w} \cdot  \bigg[\vect{\cal E}_{p,\myparallel}'(\vr) \times \vect{\cal H}_{q,\myparallel}^*(\vr) + \vect{\cal E}_{q,\myparallel}^*(\vr) \times \vect{\cal H}_{p,\myparallel}'(\vr) \bigg] dS \nonumber \\
&= j \sum_{m=1}^{M_g} \matr{S}_{mq}^* \matr{S}_{mp}' - \frac{j}{2} \matr{S}_{pq}^* \bfrac{1}{Z_p}' \left[ n_p  \right]^2     \nonumber \\
& \quad + \frac{j}{2} \matr{S}_{qp}  \bfrac{1}{Z_q}' \left[ n_q \right]^2 \,.  
\label{eq:WS1_closed}
\end{align}
Therefore, the surface integral on the LHS of~\eqref{eq:WS_FO_17} can be computed given knowledge of scattering matrix and its frequency derivative. 
Next, denote the RHS of~\eqref{eq:WS_FO_17} as 
\begin{align}
\widetilde{\matr{{Q}}}_{qp} =  \frac{1}{2}  \int_{\Omega}\left[ \varepsilon \vect{\cal E}_q^*(\vr) \cdot \vect{\cal E}_p(\vr) + \mu  \vect{\cal H}_q^*(\vr) \cdot \vect{\cal H}_p(\vr) \right] dV \,. \label{eq:Qc_tilde_closed}
\end{align}
Substituting~\eqref{eq:WS1_closed} and~\eqref{eq:Qc_tilde_closed} into~\eqref{eq:WS_FO_17} yields
\begin{align}
\widetilde{\matr{{Q}}}_{qp} &+ \frac{j}{2} \matr{S}_{qp}^\dag \bfrac{1}{Z_p}' \left[n_p\right]^2 - \frac{j}{2} \matr{S}_{qp}  \bfrac{1}{Z_q}'  \left[n_q\right]^2\nonumber \\
&
=  j \sum_{m=1}^{M_g} \matr{S}_{qm}^\dag \matr{S}_{mp}'  \,.  
\label{eq:WS_eqn_1_closed}
\end{align}
In matrix form,~\eqref{eq:WS_eqn_1_closed} reads
\begin{align}
{\matr{Q}}^g = j \matr{S}^{\dag} \matr{S}'
\label{eq:WS_guiding_final}
\end{align}
where
\begin{align}
\matr{Q}^g = \widetilde{\matr{Q}} - \frac{j}{2} \matr{D}^g \matr{S} + \frac{j}{2}  \matr{S}^{\dag} \matr{D}^g  
\label{eq:WS_guiding_final_2}
\end{align}
is the WS time delay matrix for guiding systems, and
$\matr{D}^g$ is a diagonal matrix with $(p,p)$-th entry
\begin{align}
\matr{D}^g_{pp} = \bfrac{1}{Z_p}' \left[n_p\right]^2\,.
\label{eq:Dg_entry}
\end{align}
Equations~\eqref{eq:WS_guiding_final}--\eqref{eq:Dg_entry} show that the entries of the WS time delay matrix for guiding systems are energy-like overlap integrals of the electric and magnetic fields that exist in $\Omega$ upon excitation of the systems' ports (diagonal elements \emph{are} energies).
Note that if all propagating modes are TEM, then $\matr{Q}^g = \widetilde{\matr{Q}}$ because $Z_p$ becomes frequency independent.
Knowledge of the fields at frequency $\omega$ throughout $\Omega$ for all possible port excitations therefore allows for the computation of both $\matr{S}$ and $\matr{Q}$, and, via~\eqref{eq:WS_guiding_final}, $\matr{S}$'s frequency derivative.


\input{graphics/open_system.tex}

\subsection{Scattering Systems}
\label{sec:scattering_systems}
\subsubsection{Setup}

Consider a lossless scatterer that resides in free space and is circumscribed by a sphere of radius $a$ centered at the origin (Fig.~\ref{fig:sample_scatterer}). 
Let $\Omega$ and $d\Omega$ denote the volume and surface of a concentric sphere of radius $R \gg a$, respectively.
On $d\Omega$, fields interacting with the scatterer can be described in terms of $M_s = {\cal O}((ka)^2)$ propagating TEM modes~\cite{Harrington_2001,Bucci_1989}. 
Assume that the scatterer is excited by a  \revise{\emph{radially incoming}} unit power field with (naturally transverse) electric and magnetic components
\begin{subequations}
\begin{align}
\vect{\cal E}_{ p,\myparallel}^i(\vr, \omega) &= n  \frac{e^{jk(\omega) r}}{r} \vect{\cal X}_p(\theta,\phi) \label{eq:scatter_Einc} \\    
\vect{\cal H}_{p,\myparallel}^i(\vr, \omega) &=  \frac{n}{Z} \frac{e^{jk(\omega)r}}{ r} \left( -\unit{r} \times \vect{\cal X}_p(\theta,\phi) \right) \label{eq:scatter_Hinc} \,.
\end{align}
\end{subequations}
Here $\vr = (r,\theta, \phi)$, $\unit{r}$ is the radial unit  vector, and $1 \le p \le M_s$ denotes the index of a mode with transverse mode-profile $\vect{\cal X}_p(\theta,\phi)$. 
In~\eqref{eq:scatter_Einc}-\eqref{eq:scatter_Hinc}, 
$Z = \sqrt{\mu_o/\varepsilon_o}$,  
$n = \sqrt{Z}$, and $k(\omega) = \omega \sqrt{\mu_o \varepsilon_o}$ are the mode-independent impedance, power normalization factor, and wave number; $\mu_o$ and  $\varepsilon_o$ are the free-space permeability and permittivity.
The mode profiles are assumed orthonormal, i.e.
\begin{align}
    \int_{0}^{2\pi} \int_{0}^{\pi} \vect{\cal X}_p(\theta,\phi) \cdot \vect{\cal X}_q^*(\theta,\phi) \sin\theta d\theta d\phi = \delta_{pq}\,.
\end{align}
There exists many possible choices for $\vect{\cal X}_p(\theta,\phi)$. A specific realization in terms of vector spherical harmonics is outlined in Appendix~\ref{app:radiation_ports}.
Note that \eqref{eq:scatter_Einc}-\eqref{eq:scatter_Hinc} should not be confused with the ``incident field'' in scattering computations as the latter carries no net energy across $d\Omega$.

Next, let $\vect{\cal E}_p(\vr)$ and $\vect{\cal H}_p(\vr)$ denote the fields throughout $\Omega$ due to excitation \eqref{eq:scatter_Einc}--\eqref{eq:scatter_Hinc}.  Near $d\Omega$, these fields can be expressed as
\begin{subequations}
\begin{align}
\vect{\cal E}_{ p,\myparallel}(\vr, \omega) &= \vect{\cal E}_{ p,\myparallel}^i(\vr, \omega ) + \vect{\cal E}_{ p,\myparallel}^o(\vr, \omega) \label{eq:scatter_Etot} \\
\vect{\cal H}_{p,\myparallel}(\vr, \omega) &= \vect{\cal H}_{p,\myparallel}^i(\vr, \omega) + \vect{\cal H}_{p,\myparallel}^o(\vr, \omega) \,, \label{eq:scatter_Htot}
\end{align}
\end{subequations}
where the \emph{outgoing} (automatically transverse) fields near $d\Omega$ are
\begin{subequations}
\begin{align}
\vect{\cal E}_{ p,\myparallel}^o(\vr, \omega) &= \sum_{m=1}^{M_s} \matr{S}_{mp}(\omega) n \frac{e^{-jk(\omega)r}}{r} \vect{\cal X}_{m}^*(\theta, \phi) \label{eq:scatter_Eout}\\
\vect{\cal H}_{p,\myparallel}^o(\vr, \omega) &= -\sum_{m=1}^{M_s} \matr{S}_{mp}(\omega) \frac{n}{Z} \frac{e^{-jk(\omega)r}}{r} \left(-\unit{r} \times \vect{\cal X}_{m}^*(\theta,\phi) \right)\,. \label{eq:scatter_Hout}
\end{align}
\end{subequations}
The above construct guarantees that $\matr{S}$ is independent of $R$ and that~\eqref{eq:SSI} holds, i.e. that $\matr{S}$ is unitary.

\subsubsection{WS Relationship}

To derive the WS relationship for scattering systems, once again consider two sets of fields: $\{\vect{\cal E}_p(\vr), \vect{\cal H}_p(\vr) \}$ and $\{\vect{\cal E}_q(\vr), \vect{\cal H}_q(\vr) \}$.
The above derivation \eqref{eq:WS_FO_7}-\eqref{eq:WS_FO_17} for guiding systems continues to hold true with $\unit{w} \rightarrow \unit{r}$.
The evaluation of the LHS of~\eqref{eq:WS_FO_17} requires expressions for $\vect{\cal E}_{p,\myparallel}'(\vr)$, $\vect{\cal H}_{p,\myparallel}'(\vr)$, $\vect{\cal E}_{q,\myparallel}^*(\vr)$, and $\vect{\cal H}_{q,\myparallel}^*(\vr)$  on $d\Omega$.
Substituting  \eqref{eq:scatter_Einc}--\eqref{eq:scatter_Hinc} and \eqref{eq:scatter_Eout}--\eqref{eq:scatter_Hout} into~\eqref{eq:scatter_Etot}--\eqref{eq:scatter_Htot}, and differentiating w.r.t. frequency  yields
\begin{subequations}
\begin{align}
\vect{\cal E}_{p,\myparallel}'(\vr) &= n\left(  \frac{e^{jkr}}{r}\right)' \vect{\cal X}_p(\theta,\phi) + \sum_{m=1}^{M_s} \matr{S}_{mp}' n \frac{e^{-jkr}}{ r} \vect{\cal X}_{m}^*(\theta,\phi) \nonumber \\
& \quad \quad + \sum_{m=1}^{M_s} \matr{S}_{mp} n \left( \frac{e^{-jkr}}{r}\right)' \vect{\cal X}_{m}^*(\theta,\phi)
  \label{eq:dEp_Open}\\
\vect{\cal H}_{p,\myparallel}'(\vr) &=  \frac{n}{Z} \left( \frac{e^{jkr}}{r}\right)' \left( -\unit{r} \times \vect{\cal X}_p(\theta,\phi) \right)  \nonumber \\
& \quad -  \sum_{m=1}^{M_s} \matr{S}_{mp}' \frac{n}{Z} \frac{e^{-jkr}}{r} \left(-\unit{r} \times \vect{\cal X}_{m}^*(\theta,\phi) \right) \nonumber \\
&\quad -  \sum_{m=1}^{M_s} \matr{S}_{mp} \frac{n}{Z} \left( \frac{e^{-jkr}}{r}\right)' \left(-\unit{r} \times \vect{\cal X}_{m}^*(\theta,\phi) \right)\,. \label{eq:dHp_Open}
\end{align}
\end{subequations}
Likewise,
substituting  \eqref{eq:scatter_Einc}--\eqref{eq:scatter_Hinc} and \eqref{eq:scatter_Eout}--\eqref{eq:scatter_Hout} into~\eqref{eq:scatter_Etot}--\eqref{eq:scatter_Htot} with $p \rightarrow q$, and complex conjugating the result yields
\begin{subequations}
\begin{align}
\vect{\cal E}_{q,\myparallel}^*(\vr) &= n  \frac{e^{-jkr}}{r} \vect{\cal X}_q^*(\theta,\phi)  + \sum_{m=1}^{M_s} \matr{S}_{mq}^* n \frac{e^{jkr}}{r} \vect{\cal X}_{m}(\theta,\phi)  \label{eq:Eq_Open}\\
\vect{\cal H}_{q,\myparallel}^*(\vr) &=  \frac{n}{Z} \frac{e^{-jkr}}{r} \left( -\unit{r} \times \vect{\cal X}_q^*(\theta,\phi) \right) \nonumber \\
& \quad \quad -  \sum_{m=1}^{M_s} \matr{S}_{mq}^* \frac{n}{Z} \frac{e^{jkr}}{r} \left(-\unit{r} \times \vect{\cal X}_{m}(\theta,\phi) \right) \,. \label{eq:Hq_Open}
\end{align}
\end{subequations}

Substituting \eqref{eq:dEp_Open}--\eqref{eq:dHp_Open} and \eqref{eq:Eq_Open}--\eqref{eq:Hq_Open}  into the LHS of~\eqref{eq:WS_FO_17} with $\unit{w} \rightarrow \unit{r}$, and simplifying the result yields
\begin{align}
\widetilde{\matr{{Q}}}_{qp} =  j \sum_{m=1}^{M_s} \matr{S}_{qm}^\dag \matr{S}_{mp}' + 2\sqrt{\mu_o\varepsilon_o} R \delta_{qp}\,,
\label{eq:open_Qtilde}
\end{align}
where $\widetilde{\matr{Q}}_{qp}$ is still given by~\eqref{eq:Qc_tilde_closed}. 
To arrive at a WS relationship that is independent of $R$, consider  the quantity $\widetilde{\matr{{Q}}}_{qp,\infty}^s$
obtained by replacing in ~\eqref{eq:Qc_tilde_closed} 
$\{ \vect{\cal E}_p(\vr), \vect{\cal H}_p(\vr) \}$ and $\{ \vect{\cal E}_q(\vr), \vect{\cal H}_q(\vr) \}$
by
$\{ \vect{\cal E}_{p,\myparallel}(\vr), \vect{\cal H}_{p,\myparallel}(\vr) \}$ and $\{ \vect{\cal E}_{q,\myparallel}(\vr), \vect{\cal H}_{q,\myparallel}(\vr) \}$, i.e.
\begin{align}
\widetilde{\matr{{Q}}}_{qp,\infty}^s =&  \frac{1}{2} \int_{\Omega} \big[  \varepsilon \vect{\cal E}_{q,\myparallel}^*(\vr) \cdot \vect{\cal E}_{p,\myparallel}(\vr)  \nonumber \\
 & \quad +  \mu \vect{\cal H}_{q,\myparallel}^*(\vr) \cdot \vect{\cal H}_{p, \myparallel}(\vr) \big] dV \,. \label{eq:Q_inf1}
\end{align}
The quantities in the integrand  in~\eqref{eq:Q_inf1} are \emph{not} the transverse to $\unit{r}$ components of
$\{ \vect{\cal E}_p(\vr), \vect{\cal H}_p(\vr) \}$ and $\{ \vect{\cal E}_q(\vr), \vect{\cal H}_q(\vr) \}$.
Instead, they are the quantities in~\eqref{eq:scatter_Etot}--\eqref{eq:scatter_Htot} evaluated for arbitrary $\vr \in \Omega$.
Substituting~\eqref{eq:scatter_Einc}-\eqref{eq:scatter_Hinc} and~\eqref{eq:scatter_Eout}-\eqref{eq:scatter_Hout} into~\eqref{eq:scatter_Etot}-\eqref{eq:scatter_Htot}, and then evaluating~\eqref{eq:Q_inf1} yields
\begin{align}
\widetilde{\matr{Q}}_{qp,\infty}^{s} &= 2 \sqrt{\mu_o\varepsilon_o} R \delta_{qp}\,.
\label{eq:Qinf_Openonly}
\end{align}
Subtracting~\eqref{eq:Qinf_Openonly} from both sides of~\eqref{eq:open_Qtilde} yields \begin{align}
{\matr{Q}}_{qp}^s  &=  j \sum_{m=1}^{M_s} \matr{S}_{qm}^\dag \matr{S}_{mp}' 
\label{eq:WS_closed_final_sum}
\end{align}
where ${\matr{Q}}_{qp}^s = \widetilde{\matr{Q}}_{qp} - \widetilde{\matr{Q}}_{qp,\infty}^s$.
In matrix form~\eqref{eq:WS_closed_final_sum} reads
\begin{align}
{\matr{Q}}^s  &=  j \matr{S}^\dag \matr{S}' 
\label{eq:Ws_Sdag_Sprime}
\end{align}
with
\begin{align}
{\matr{Q}}_{qp}^s &= \frac{1}{2}\varepsilon   \int_{R^3} \left[   \vect{\cal E}_q^*(\vr) \cdot \vect{\cal E}_p(\vr) - \vect{\cal E}_{q,\myparallel}^*(\vr) \cdot \vect{\cal E}_{p,\myparallel}(\vr)
  \right] dV   \label{eq:Qqp_s} \\
 + &\frac{1}{2} \mu \int_{R^3} \left[  \vect{\cal H}_q^*(\vr) \cdot \vect{\cal H}_p(\vr)-   \vect{\cal H}_{q,\myparallel}^*(\vr) \cdot \vect{\cal H}_{p, \myparallel}(\vr) \right] dV \nonumber
\end{align}
where the domain of integration was changed from $\Omega$ to $R^3$ because the bracketed integrands converge rapidly.
The above renormalization procedure is different from Smith's~\cite{Smith_1960}, who used an averaging scheme to render all integrals convergent.  
Instead, the procedure resembles that used in~\cite{VDB_2010,Gustafsson_2015} for expressing the energy stored in antenna fields.
Just like for guiding systems, \eqref{eq:Ws_Sdag_Sprime} and \eqref{eq:Qqp_s} show that entries of the WS time delay matrix for scattering systems are energy-like overlap integrals of fields that exist throughout $\Omega$ for all possible port excitations.  
The renormalization procedure in \eqref{eq:Qqp_s} extracts the time delay caused by the scattering process from the total time waves naturally dwell within $\Omega$ (which tends to infinity as $R \rightarrow \infty$).  Evaluation of $\matr{S}$ and $\matr{Q}$ at frequency $\omega$ once again permits the computation of $\matr{S}$'s frequency derivative via~\eqref{eq:Ws_Sdag_Sprime}.

\input{graphics/mixed_systems.tex}

\subsection{Radiating Systems}

\subsubsection{Setup}
Radiating systems are hybrids of the guiding and scattering systems considered in Secs.~\ref{sec:guiding_systems} and~\ref{sec:scattering_systems}.
Consider a radiating system composed of lossless antennas that are fed by PEC waveguides of uniform cross section (Fig.~\ref{fig:mixed_scatterer}). 
The antennas and their feeds reside in free space and are circumscribed by a sphere of radius $a$. 
Let $\Omega$ denote the volume of a concentric sphere of radius $R \gg a$, and let   $d\Omega$ denote the union of the concentric sphere's surface and the waveguides' physical ports.
On $d\Omega$, fields interacting with the system can be characterized in terms of $M_r = M_g + M_s$ propagating modes. 


Assume that the system is excited by an \emph{incoming} unit-power field with  transverse electric and magnetic components $\vect{\cal E}_{p,\myparallel}^{i}(\vr, \omega)$
and
$\vect{\cal H}_{p,\myparallel}^{i}(\vr, \omega)$ near $d\Omega$.
If $p \le M_g$, then $\vect{\cal E}_{p,\myparallel}^{i}(\vr, \omega)$ and $\vect{\cal H}_{p,\myparallel}^{i}(\vr, \omega)$ are given by~\eqref{eq:guiding_Einc}--\eqref{eq:guiding_Hinc}.
Likewise, if $p > M_g$, then $\vect{\cal E}_{p,\myparallel}^{i}(\vr, \omega)$ and $\vect{\cal H}_{p,\myparallel}^{i}(\vr, \omega)$  are given by~\eqref{eq:scatter_Einc}--\eqref{eq:scatter_Hinc}.

Let $\vect{\cal E}_{p}(\vr, \omega)$ and $\vect{\cal H}_{p}(\vr, \omega)$ denote the electric and magnetic fields throughout $\Omega$ due to $\vect{\cal E}_{p,\myparallel}^{i}(\vr, \omega)$ and $\vect{\cal H}_{p,\myparallel}^{i}(\vr, \omega)$.
Near $d\Omega$, the total transverse electric and magnetic fields $\vect{\cal E}_{p,\myparallel}(\vr,\omega)$ and $\vect{\cal H}_{p,\myparallel}(\vr,\omega)$ are given by~\eqref{eq:guiding_Etot}--\eqref{eq:guiding_Htot} with the transverse \emph{outgoing} fields given by
\begin{subequations}
\begin{align}
    \vect{\cal E}_{p,\myparallel}^o(\vr,\omega) &= \sum_{m = 1}^{M_g} \matr{S}_{mp}(\omega) n_m(\omega) e^{-j \beta_m(\omega) w} \vect{\cal X}_m(u,v) \nonumber \\ 
    &\quad + \sum_{m=M_g + 1}^{M_r} \matr{S}_{mp}(\omega) n \frac{e^{-jk(\omega)r}}{r} \vect{\cal X}_{m}^*(\theta, \phi) \label{eq:rad_guide_Eout} \\
    \vect{\cal H}_{p,\myparallel}^o(\vr,\omega) &= - \sum_{m = 1}^{M_g} \matr{S}_{mp}(\omega) \frac{n_m(\omega)}{Z_m(\omega)} e^{-j \beta_m(\omega) w} \label{eq:rad_guide_Hout}     \\
    &\quad \quad \quad \cdot \left(-\unit{w} \times \vect{\cal X}_m(u,v)\right) \nonumber \\
    &-\sum_{m=M_g+1}^{M_r} \matr{S}_{mp}(\omega) \frac{n}{Z} \frac{e^{-jk(\omega)r}}{r} \left(-\unit{r} \times \vect{\cal X}_{m}^*(\theta,\phi) \right) \,. \nonumber 
\end{align}
\end{subequations}
All modal quantities in~\eqref{eq:rad_guide_Eout}-\eqref{eq:rad_guide_Hout} were  defined in Sec.~\ref{sec:guiding_systems} and Sec.~\ref{sec:scattering_systems}.

\subsubsection{The WS Relationship}

The WS relationship can be derived by following the same procedure as in Secs.~\ref{sec:guiding_systems} and~\ref{sec:scattering_systems}.
First, consider two sets of fields: $\{ \vect{\cal E}_p(\vr), \vect{\cal H}_p(\vr) \}$ and $\{ \vect{\cal E}_q(\vr), \vect{\cal H}_q(\vr) \}$. 
The derivation in~\eqref{eq:WS_FO_7}-\eqref{eq:WS_FO_17} continues to hold true with $\unit{w} \rightarrow \unit{r}$ if $\vr$ is on the spherical surface of radius $R$. 
Expressions for $\vect{\cal E}_{p,\myparallel}'(\vr)$, $\vect{\cal H}_{p,\myparallel}'(\vr)$, $\vect{\cal E}_{q,\myparallel}^*(\vr)$, and $\vect{\cal H}_{q,\myparallel}^*(\vr)$  on $d\Omega$ can be derived using~\eqref{eq:guiding_Etot}--\eqref{eq:guiding_Htot},  \eqref{eq:guiding_Einc}--\eqref{eq:guiding_Hinc}, \eqref{eq:scatter_Einc}--\eqref{eq:scatter_Hinc}, and~\eqref{eq:rad_guide_Eout}--\eqref{eq:rad_guide_Hout}.
Substituting these expressions into~\eqref{eq:WS_FO_17} and simplifying the result yields
\begin{align}
\widetilde{\matr{Q}}_{qp} =& j \sum_{m=1}^{M_r} \matr{S}_{qm}^\dag \matr{S}_{mp}'
 -\frac{j}{2} \matr{S}_{qp}^\dag \bfrac{1}{Z_p}' (n_p)^2 \hat{\delta}_{p \le M_g} \nonumber \\
&+\frac{j}{2}  \matr{S}_{qp} \bfrac{1}{Z_q}' (n_q)^2  \hat{\delta}_{q \le M_g}   \nonumber  \\
& 
+ R\sqrt{\mu_o \varepsilon_o}  \left({\delta}_{qp}   \hat{\delta}_{q > M_g}   +    \sum_{m=M_g+1}^{ M_r} \matr{S}_{mp} \matr{S}_{qm}^\dag\right)  \label{eq:WS_hybrid}
\end{align}
where $\widetilde{\matr{Q}}_{qp}$ is still given by~\eqref{eq:Qc_tilde_closed}, and $\hat{\delta}_{f}=1$ if $f$ is true and is $0$ otherwise.
Note that the second and third terms on the RHS of~\eqref{eq:WS_hybrid} are due to non-TEM waveguide modes and resemble the second and third terms on the LHS of~\eqref{eq:WS_eqn_1_closed}.
The final term on the RHS of~\eqref{eq:WS_hybrid} is proportional to $R$ and resembles the second term on the RHS of~\eqref{eq:open_Qtilde}. 
As in Sec.~\ref{sec:scattering_systems}, a WS relationship that is independent of $R$ is obtained by introducing  $\widetilde{\matr{Q}}_{qp,\infty}^r$, which is still given by~\eqref{eq:Q_inf1}.
Substituting $\vect{\cal E}_{p,\myparallel}(\vr)$ and $\vect{\cal H}_{p,\myparallel}(\vr)$ into~\eqref{eq:Q_inf1} and evaluating the resulting integral yields
\begin{align}
 \widetilde{\matr{Q}}_{qp,\infty}^r =    R\sqrt{\mu_o \varepsilon_o}  \left(\delta_{qp}   \hat{\delta}_{q>M_g}   +    \sum_{m=M_g+1}^{ M_r} \matr{S}_{mp} \matr{S}_{qm}^\dag\right) \,,
 \label{eq:Qinf_rad}
\end{align}
which is identical to~\eqref{eq:Qinf_Openonly} for the case of $M_g = 0$.
Using \eqref{eq:Qinf_rad}  and \eqref{eq:WS_hybrid} results in 
\begin{align}
   \widetilde{\matr{Q}}_{qp}^r &+ \frac{j}{2} \matr{S}_{qp}^\dag \bfrac{1}{Z_p}' (n_p)^2 \hat{\delta}_{p \le M_g} \nonumber \\
   &-\frac{j}{2}   \matr{S}_{qp} \bfrac{1}{Z_q}' (n_q)^2  \hat{\delta}_{q \le M_g}  = j \sum_{m=1}^{M_s + M_g} \matr{S}_{qm}^\dag \matr{S}_{mp}'
  \label{eq:WS_hybrid2} 
\end{align}
where $\widetilde{\matr{Q}}_{qp}^r$ is given by the same as expression as $\matr{Q}_{qp}^{s}$ in~\eqref{eq:Qqp_s}.
In matrix form, \eqref{eq:WS_hybrid2} reads
\begin{align}
\matr{Q}^r = j \matr{S}^\dag \matr{S}'
\label{eq:rad_Qe_Sdag_Sprime}
\end{align}
where
\begin{align}
\matr{Q}^r &= \widetilde{\matr{Q}}^r +  \frac{j}{2} \begin{bmatrix} \matr{S}^{gg} & \matr{S}^{gs} \\ 
\matr{S}^{sg} & \matr{S}^{ss}
\end{bmatrix}^\dag 
\begin{bmatrix}
\matr{D}^{g} & \matr{0} \\
\matr{0} & \matr{0}
\end{bmatrix}\nonumber \\
&\quad - \frac{j}{2} \begin{bmatrix}
\matr{D}^{g} & \matr{0} \\
\matr{0} & \matr{0}
\end{bmatrix} 
\begin{bmatrix} \matr{S}^{gg} & \matr{S}^{gs} \\ 
\matr{S}^{sg} & \matr{S}^{ss}
\end{bmatrix}
\label{eq:rad_final_block_eqn}
\end{align}
with the $M_g \times M_g$ diagonal matrix $\matr{D}^g$ still given by~\eqref{eq:Dg_entry}.
In~\eqref{eq:rad_final_block_eqn},
the scattering matrix $\matr{S}$ was decomposed into four blocks that separate the waveguide and free-space ports.
Equations \eqref{eq:rad_Qe_Sdag_Sprime} and~\eqref{eq:rad_final_block_eqn} show that the WS time delay matrix for radiating systems can be computed from knowledge of the fields throughout $\Omega$ due to all possible port excitations.  Once again, knowledge of $\matr{S}$ and $\matr{Q}$ at frequency $\omega$ permits the computation of $\matr{S}'$, which in turn can be used to compute the frequency derivative of the antenna impedance matrix (see Sec.~\ref{sec:impedance_formulation} below).

\subsection{Alternative Expressions for $\matr{Q}$}

In the previous sections, the entries of the WS time delay matrix for guiding, scattering, and radiating systems were
expressed in terms of integrals of both electric \emph{and} magnetic fields over $\Omega$.
By manipulating Maxwell's equations~\eqref{eq:WS_FO_7}--\eqref{eq:WS_FO_8} and their frequency derivatives~\eqref{eq:WS_FO_5}--\eqref{eq:WS_FO_6}, the following alternatives to~\eqref{eq:WS_FO_17} may be derived: 
\begin{subequations}
\begin{align}
\int_{d\Omega} \unit{w} \cdot &\left[ \vect{\cal E}_p'(\vr) \times \left[ \nabla \times \vect{\cal E}_q^*(\vr) \right]  - \vect{\cal E}_q^*(\vr) \times \left[ \nabla \times \vect{\cal E}_p'(\vr) \right] \right] dS \nonumber \\
&= \frac{2k^2}{\omega} \int_{\Omega}  \vect{\cal E}_p(\vr) \cdot \vect{\cal E}_q^*(\vr) dV 
\label{eq:WS_div2_E}\\
\int_{d\Omega} \unit{w} &\cdot \big[ \vect{\cal H}_p'(\vr)  \times \left[ \nabla \times \vect{\cal H}_q^*(\vr)\right] \nonumber \\
&\quad \quad - \vect{\cal H}_q^*(\vr) \times \left[ \nabla \times \vect{\cal H}_p'(\vr) \right] \big] dS  \nonumber \\ 
&\quad \quad \quad \quad \quad =\frac{2k^2}{\omega} \int_{\Omega}  \vect{\cal H}_p(\vr) \cdot \vect{\cal H}_q^*(\vr) dV\,.
\label{eq:WS_div2_H}
\end{align}
\end{subequations}
The above expressions only require integration of electric or magnetic fields.

Using~\eqref{eq:WS_div2_E} instead of~\eqref{eq:WS_FO_17} to derive the WS relationship for guiding system results in
\begin{align}
    \matr{Q}^{g,e} = j \matr{S}^\dag \matr{S}'
\end{align}
where
\begin{align}
\matr{Q}^{g,e} &= \widetilde{\matr{Q}}^e - \frac{j}{2} \left( \matr{D}^g + \frac{1}{\omega} \matr{I}_{M_g} \right) \matr{S} \nonumber \\
&\quad + \frac{j}{2}  \matr{S}^{\dag} \left( \matr{D}^g + \frac{1}{\omega} \matr{I}_{M_g} \right) 
\label{eq:AlternateWS_E_Closed}
\end{align}
and 
\begin{align}
\widetilde{\matr{{Q}}}_{qp}^e &= \varepsilon \int_{\Omega} \vect{\cal E}_p(\vr) \cdot \vect{\cal E}_q^*(\vr) dV\,.
\end{align}

Likewise, using~\eqref{eq:WS_div2_H} to derive the WS relationship for guiding systems yields
\begin{align}
\matr{Q}^{g,h} = j \matr{S}^{\dag} \matr{S}'  
\end{align}
where
\begin{align}
\matr{Q}^{g,h} &= \widetilde{\matr{Q}}^h - \frac{j}{2} \left( \matr{D}^g - \frac{1}{\omega} \matr{I}_{M_g} \right) \matr{S} \nonumber \\
&\quad + \frac{j}{2}  \matr{S}^{\dag} \left( \matr{D}^g 
 - \frac{1}{\omega} \matr{I}_{M_g} \right)
\label{eq:AlternateWS_H_Closed}
\end{align}
and
\begin{align}
\widetilde{\matr{{Q}}}_{qp}^h &= \mu \int_{\Omega} \vect{\cal H}_p(\vr) \cdot \vect{\cal H}_q^*(\vr) dV\,.
\end{align}
Note that~\eqref{eq:WS_eqn_1_closed} can be retrieved by adding~\eqref{eq:AlternateWS_E_Closed} and \eqref{eq:AlternateWS_H_Closed}.
Similar WS relations involving $\widetilde{\matr{Q}}^e$ and $\widetilde{\matr{Q}}^h$ can be derived for scattering and radiating systems starting from~\eqref{eq:AlternateWS_E_Closed} and~\eqref{eq:AlternateWS_H_Closed}.
Expressions for $\matr{Q}$ involving only electric or magnetic fields are useful in many computational electromagnetics settings that model only one field type. 

\subsection{Impedance Formulation}
\label{sec:impedance_formulation}

WS relationship~\eqref{eq:WS_2} can be used to obtain the frequency derivative of a system's impedance matrix.
\revise{In the past, expressions for the frequency derivative of the impedance of a single port antenna system were derived in \cite{Best_2005, VDB_2010}. Evaluation of these expressions however requires knowledge of the frequency derivative of current density ($\vect{\cal J}'$) or far-fields ($\vect{\cal E}_{\myparallel}'$ and $\vect{\cal H}_{\myparallel}'$), which can only be obtained for canonical problems or by numerically solving the electromagnetic problem at two different frequencies.
In this section, the WS relationship is used to compute the frequency derivative of the impedance matrix of a multiport system in terms of energy-like quantities present in $\matr{Q}$. We note that, alternatively, the frequency derivative of the antenna impedance may also be computed using the frequency derivative of the method of moment impedance matrix \cite{gustafsson2014q}.  
}

Recall that the $M\times M$ scattering matrix relates the amplitudes of incoming and outgoing waves $\vect{a}$ and $\vect{b}$ as
\begin{align}
\vect{b}  = \matr{S} \vect{a}\,.  \label{eq:Sab}
\end{align}
The impedance matrix, on the other hand, relates ports voltages and currents $\vect{v}$ and $\vect{i}$ 
\begin{align}
\vect{v} &=  \matr{Z} \vect{i}
\label{eq:VZI}
\end{align}
where
\begin{subequations}
\begin{align}
\vect{v} &= \matr{N} \left( \vect{a} + \vect{b} \right)  \label{eq:Veq1} \\
\vect{i} &= \matr{N} \matr{Y} \left( \vect{a} - \vect{b} \right) \,.\label{eq:Ieq1}
\end{align}
\end{subequations}
Here, $\matr{N}$ and $\matr{Y}$ are diagonal matrices whose $(p,p)$-th entries are mode power normalization constants $n_p$ (or $n$) and admittances $Z_p^{-1}$ (or $Z^{-1}$), respectively.

Defining $
\widehat{\matr{Z}} = \matr{N}^{-1} \matr{Z} \matr{N}  \matr{Y}$, it follows from~\eqref{eq:Sab}--\eqref{eq:Ieq1} that
\begin{align}
\widehat{\matr{Z}} = \left( \matr{I}_M + \matr{S} \right)  \left( \matr{I}_M - \matr{S} \right)^{-1}\,.
\end{align}
Alternatively, $\matr{S}$ may be written in terms of $\widehat{\matr{Z}}$ as
\begin{align}
{\matr{S}} = \left( \widehat{\matr{Z}} + \matr{I}_M \right)^{-1}  \left( \widehat{\matr{Z}} - \matr{I}_M \right)\,. \label{eq:ZandS1}
\end{align}
Taking the frequency derivative of~\eqref{eq:ZandS1} by applying the chain rule yields
\begin{align}
\widehat{\matr{Z}} '  &= 
\left( \widehat{\matr{Z}} + \matr{I}_M \right)  \matr{S}' \left( \matr{I}_M -\matr{S}\right)^{-1} \,.
\label{eq:Zhat_prime}
\end{align}
Finally, using the WS relationship~\eqref{eq:WS_2} in~\eqref{eq:Zhat_prime} yields the frequency derivative of the impedance matrix in terms of the WS time delay matrix $\matr{Q}$ and the scattering matrix $\matr{S}$
\begin{align}
\widehat{\matr{Z}} '  &= -j \left( \widehat{\matr{Z}} + \matr{I}_M \right)  \matr{S} \matr{Q} \left( \matr{I}_M -\matr{S}\right)^{-1} 
\end{align}
or
\begin{align}
    {\matr{Z}} '  &= -j \matr{N} \left( \widehat{\matr{Z}} + \matr{I}_M \right)^{-1}  \matr{S} \matr{Q} \left( \matr{I}_M -\matr{S}\right)^{-1} \matr{Y}^{-1} \matr{N}^{-1} \,.
    \label{eq:Zprime_multiport}
\end{align} 
For radiating systems, the frequency derivative of the antenna impedance matrix is easily extracted from~\eqref{eq:Zprime_multiport}.

%% file: graphics/TM_Line.tex
    

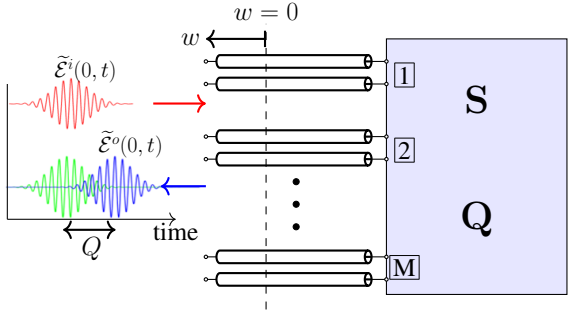
\begin{figure}
\centering
    \begin{circuitikz}[scale = 0.4, every node/.style={scale=0.4}]
    \draw [fill = blue!10] (6,-0.5) rectangle (12,8);
    \begin{scope}[shift={(0, 4)}]
    \draw (0,0)  to[TL, o-o] (6,0);
    \draw (0,0.75)  to[TL, o-o] (6,0.75);
    \end{scope}
    
    \begin{scope}[shift={(0, 6.5)}]
    \draw (0,0)  to[TL, o-o] (6,0);
    \draw (0,0.75)  to[TL, o-o] (6,0.75);
    \end{scope}
    
    \draw (0,0)  to[TL, o-o] (6,0);
    \draw (0,0.75)  to[TL, o-o] (6,0.75);
    
    \node at (9,6) {\scalebox{1.5}{\Huge $\matr{S}$}};
    
    \node at (9,2) {\scalebox{1.5}{\Huge $\matr{Q}$}};
    
    \node at (6.6,6.8) {\Huge \boxed{$1$}};
    
    \node at (6.6,4.3) {\Huge \boxed{$2$}};
    
    \draw[fill = black] (3,1.75) circle (0.1); 
    
    \draw[fill = black] (3,2.5) circle (0.1); 
    
    \draw[fill = black] (3,3.25) circle (0.1); 
    
    \begin{scope}[shift={(-1,-2.15)}]
    
    \node at (-3.5,8) {\includegraphics[scale = 0.99]{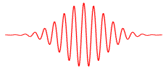}};
    \draw [->,thick, red] (-0.75,8) -- (1,8);
    
    \node at (-2.5,5.25) {\includegraphics[scale = 0.99]{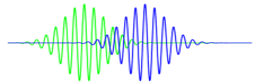}};
    \draw [<-,thick, blue] (-0.5,5.25) -- (1,5.25);
    \draw[<->, thick] (-3.75,3.8) -- (-2, 3.8);
    \node at (-2.8,3.7) [below] {\Huge $Q$};
    \end{scope}
    
    \node at (6.6,0.4) {\Huge \boxed{$M$}};
    
    \draw [dashed] (2,-1) -- (2,8.5);
    \draw [<-|, thick] (0, 8.0) -- (2.0,8.0);
    \node at (2,8.5 )[above] {\Huge $w=0$};
    \node at (-0.5,8) {\Huge $w$};
    
    \draw (-6.6,2) -- (-6.6,6.5);
    \draw[->] (-6.6,2) -- (-1,2);
    \node at (-1, 2) [below] {\Huge time};
    \node at (-4,7) {\huge $\widetilde{{\cal E}}^i(0,t)$};
    
    \node at (-2.5,4.5) {\huge $\widetilde{{\cal E}}^o(0,t)$};
    
    \end{circuitikz}
    \caption{A generic $M$-port linear, time-invariant, lossless, and reciprocal system. }
    \label{fig:TM_Line}
\end{figure}

%% file: graphics/closed_system.tex
\begin{figure}
\begin{center}
\begin{tikzpicture}
\node (p1) at (0,0) { \includegraphics[width=0.9\columnwidth]{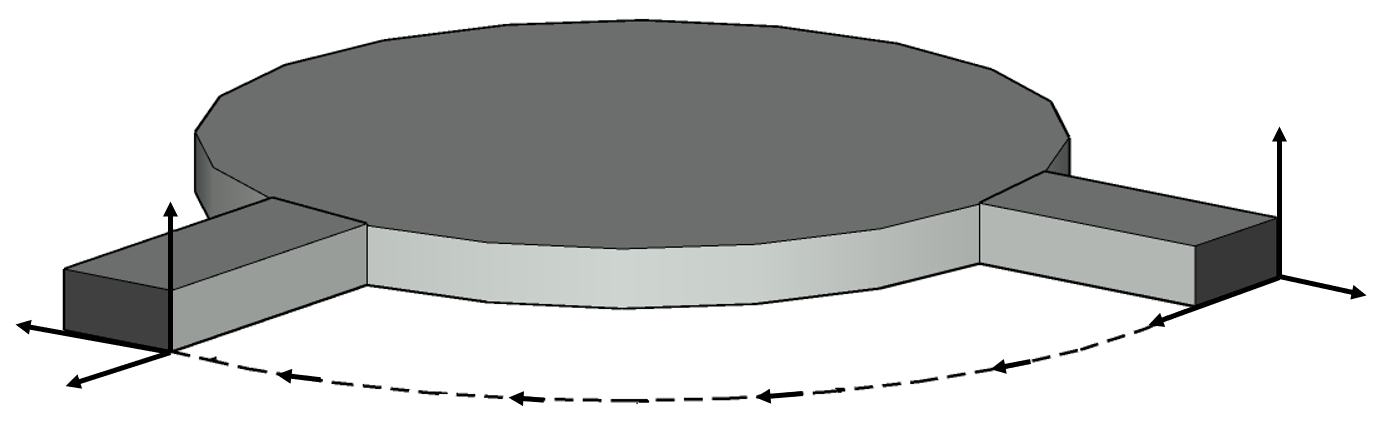}};
\draw[->] (-3.6,0.4) -- (-3.3,-0.55);
\node at (-3.6,0.2) [above] {\begin{tabular}{c} physical\\ port 1\end{tabular}};

\draw[->] (2.75,0.65) -- (3.2,-0.3);
\node at (2.7,0.6) [above] {\begin{tabular}{c} physical \\ port 2 \end{tabular}};



\node at (-3.05,0.25) {$\unit{u}$};
\node at  (-3.8,-0.55) [left] {$\unit{v}$};
\node at  (-3.75,-0.85) [below] {$\unit{w}$};

\node at (2.3,-0.85) [right] {$\unit{v}$};
\node at (3.7,-0.55) [right] {$\unit{w}$};
\node at (3.1,0.75) [right] {$\unit{u}$};

\node at (-0.1,1.3) {$d\Omega$};
\node at (-0.25,0.35) {$\Omega$};
\node at (3.7,-0.2) {$d\Omega_p$};
\end{tikzpicture}
\end{center}
\caption{Guiding system.  The global coordinate system covers the entire system; one or more propagating modes exist in each physical port.}
\label{fig:GuidingStructure_1}
\end{figure}
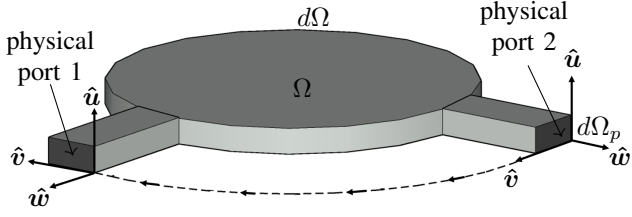

%% file: graphics/open_system.tex
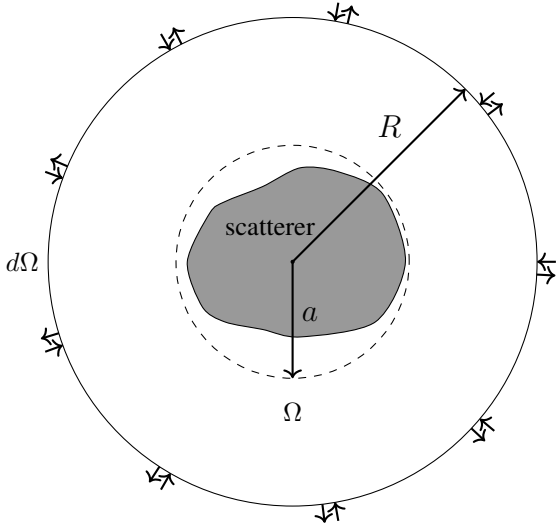
\begin{figure}
\centering
\begin{tikzpicture}
\draw (0,0) circle (3.25);
\begin{scope}[shift={(-0.4,0)}, scale=1,scale=1]
\draw [black, fill = black!40] plot [smooth cycle] coordinates {(1.9,0)  (1.6, 0.9) (1.0, 1.2) (0.5, 1.25) (0,1) (-0.5,0.8) (-0.7, 0.65) (-1,0) (-0.8, -0.6) (-0.5,-0.8) (0,-0.9) (0.5, -1)  (1.5, -0.8)};
\end{scope}
\draw[fill=black] (0,0) circle (0.02);
\draw[->,thick] (0,0) -- (45:3.25);
\node at (50:2) [above] {\large $R$};
\node at (0,-2) {$\Omega$};
\node at (-0.3,0.45) {scatterer};
\node at (180:3.25) [left] {$d\Omega$};
\foreach \th in {40, 80, 120, ..., 360}
{
\draw [->,thick] (\th:3.5) -- (\th:3.25);
\draw [->,thick] (\th-3:3.25) -- (\th-3:3.5);
}

\draw[dashed] (0,0) circle (1.55);
\draw[->,thick] (0,0) -- (270:1.55);
\node at (0, -0.7) [right] {\large $a$};
\end{tikzpicture}
\caption{Scattering system excited through free-space port defined on sphere of radius $R$.}
\label{fig:sample_scatterer}
\end{figure}

%% file: graphics/mixed_systems.tex
\begin{figure}
\centering
\begin{tikzpicture}
\node (p1) at (0,0.75) { \includegraphics[width=0.3\columnwidth]{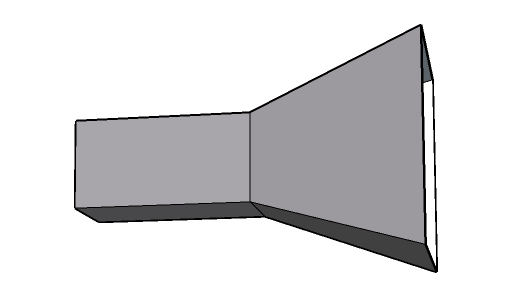}};
\node (p1) at (0,-0.75) { \includegraphics[width=0.3\columnwidth]{graphics/horn.png}};
\draw (0,0) circle (3.25);
\draw[dashed] (0,0) circle (1.7);

\draw[fill=black] (0,0) circle (0.02);
\draw[->] (0,0) -- (45:3.25);
\node at (50:2.5) [above] {$R$};
\node at (0,-2) {$\Omega$};
\foreach \th in {40, 80, 120, ..., 360}
{
\draw [->,thick] (\th:3.5) -- (\th:3.25);
\draw [->,thick] (\th-3:3.25) -- (\th-3:3.5);
}
\node at (140:2) [above] {$d\Omega$};
\draw[->] (140:2.0) -- (160:3.2);
\draw[->] (140:2) -- (-1,0.78);
\draw[->] (140:2) -- (-1,-0.7);

\draw[->] (0,0) -- (340:1.7);
\node at (340:1.5) [above] {$a$};
\node at (-2.5,0) {\begin{tabular}{c} physical\\ ports \end{tabular}};
\draw[->] (-2.3,0.4) -- (-1,0.6);
\draw[->] (-2.3,-0.4) -- (-1,-0.8);


\begin{scope}[shift={(0,-1.5)}]
\node at (-0.95,1.05) [above] {$\unit{u}$};
\node at (-0.57,0.20) [right] {$\unit{v}$};
\node at (-1.4,0.42) [below] {$\unit{w}$};

\draw[->, thick] (-0.95,0.45) -- (-0.95,1.1);
\draw[->, thick] (-0.95,0.45) -- (-0.5,0.23);
\draw[->,thick] (-0.95,0.45) -- (-1.5,0.42); 
\end{scope}

\end{tikzpicture}
\caption{Radiating system excited by through waveguide and free-space ports.}
\label{fig:mixed_scatterer}
\end{figure}

%% file: results.tex
\section{Illustrative Examples}

This section applies WS methods to the characterization of fields with well-defined time delays in guiding and scattering systems.  
It also demonstrates the use of equations \eqref{eq:rad_Qe_Sdag_Sprime} and \eqref{eq:Zprime_multiport} to compute the frequency derivative of antenna impedance matrices.  
The examples in this sections are merely illustrative in nature.  
While WS methods have applications in many branches of electromagnetics (see Sec.~\ref{sec:conclusions}), their treatment is beyond the scope of this paper.

\subsection{Guiding Systems}
 
\subsubsection{A PEC-Terminated Rectangular Waveguide}

This first didactic example verifies the WS relationship \eqref{eq:WS_guiding_final} for an air-filled rectangular waveguide with dimensions $a \times b$ and length $l$ that is terminated in a short \revise{(at $w = -l$)}.   Since both $\matr{S}$ and $\matr{S}'$ are diagonal, so is $\matr{Q}^g$.  Assume the waveguide is excited by a unit-power $\mathrm{TM}_{mn}$ mode.   
The total electric and magnetic fields inside the waveguide are 
\begin{subequations}
\begin{align}
    \vect{\cal E}_p(u,v,w) &= \vect{\cal E}_{p,\myparallel}(u,v,w) + {\cal E}_{p,w}(u,v,w) \unit{w}   \label{eq:waveguide_PEC_2} \\
    \vect{\cal H}_p(u,v,w) &= \vect{\cal H}_{p,\myparallel}(u,v,w) \,,
    \label{eq:waveguide_PEC_3}
\end{align}
\end{subequations}
where $\vect{\cal E}_{p,\myparallel}(u,v,w)$ and $\vect{\cal H}_{p,\myparallel}(u,v,w)$ are given by \eqref{eq:guiding_Etot} and \eqref{eq:guiding_Htot} with  mode profile
\begin{align}
     \vect{\cal X}_{p}(u,v) =&  \frac{1}{\sqrt{k_c^2 ab/ 4}}\bigg[ \left(\frac{m\pi}{a} \right) \cos \left( \frac{m \pi}{a} u\right) \sin \left(\frac{n \pi}{b}v \right) \unit{u} \nonumber \\
& +  \left(\frac{n\pi}{b} \right) \sin\left( \frac{m \pi}{a} u\right) \cos \left(\frac{n \pi}{b}v \right) \unit{v} \bigg] \,,
 \end{align}
wave impedance $Z_p = \beta_p Z / k$, propagation constant
$\beta_p = \sqrt{k^2 - k_c^2}$, 
cutoff wave number $k_c =  \sqrt{(m\pi/a)^2 + (n\pi/b)^2}$, and a diagonal scattering matrix $\matr{S}$ with $(p,p)$-th entry 
\begin{align}
    \matr{S}_{pp} = -e^{-2j\beta_p l}\,.
    \label{eq:waveguide_PEC_1}
\end{align}
In~\eqref{eq:waveguide_PEC_2}, the longitudinal component of $\vect{\cal E}_p(\vr)$ reads 
\begin{align}
{\cal E}_{p,w}(u,v,w) &=  \frac{jn_p k_c}{\beta_p}  \frac{2}{\sqrt{ab}} \sin \left( \frac{m \pi}{a} u \right) \sin \left( \frac{n \pi}{b} v \right)  \label{eq:waveguide_PEC_int3}\\
    &\quad \left[ e^{j\beta_p w} - \matr{S}_{pp} e^{-j\beta_p w} \right] \,. \nonumber
\end{align}
Substituting~\eqref{eq:waveguide_PEC_2}--\eqref{eq:waveguide_PEC_3} into \eqref{eq:Qc_tilde_closed} and evaluating the resulting integral yields
\begin{align}
    \widetilde{\matr{Q}}^g_{pp} = \frac{2l \sqrt{\mu_o \varepsilon_o}}{ \cos \theta_p} + \frac{1}{\omega} \tan^2 \theta_p \sin (2 \beta_p l)\,
    \label{eq:waveguide_PEC_Qtilde}
\end{align}
where $\cos \theta_p = \beta_p / k$.
The last two terms on the LHS of~\eqref{eq:WS_eqn_1_closed} 
are easily shown to equal the negative of the second term on the RHS of~\eqref{eq:waveguide_PEC_Qtilde}, yielding
\begin{align}
    \matr{Q}^g_{pp} &= \frac{2 l \sqrt{\mu_o\varepsilon_o}}{\cos \theta_p} \,.
    \label{eq:waveguide_PEC_Qg}
\end{align}
Using \eqref{eq:waveguide_PEC_Qg} and the fact that 
\begin{align}
\matr{S}'_{pp} = -j \frac{ 2l \sqrt{\mu_o \varepsilon_o}}{ \cos \theta_p} \matr{S}_{pp}\,,
\end{align}
which follows from \eqref{eq:waveguide_PEC_1}, it is easily verified that the WS relationship \eqref{eq:WS_guiding_final} holds true, i.e. that $\matr{Q}^g_{pp} = j \matr{S}^\dag_{pp} \matr{S}_{pp}'$.
Not surprisingly, the time delay $\matr{Q}^g_{pp}$ in \eqref{eq:waveguide_PEC_Qg} equals the time needed for light to travel the length of the zig-zag ray path from the port to the short and back, traditionally
associated with the $\mathrm{TM}_{mn}$ mode~\revise{\cite{Pozar_2005}}.
The above analysis can be repeated for $\mathrm{TE}_{mn}$ modes with identical results.

\begin{figure}[t]
\centering
\input{graphics/results/waveguide2/transient}
\caption{Envelopes of incoming ({\Large \color{blue} ---}) and outgoing waves for the air- ({\Large \color{red} -- --}) and inhomogeneously-filled
({\Large \color{black} -\,-\,-\,-}) waveguides; computed group delays for the air-filled waveguide are: (a. maximum of envelope) $89.99 \mathrm{ns}$, (b. WS $\matr{Q}$ matrix) $89.897 \mathrm{ns}$, (c. frequency derivative of phase of reflection coefficient) $89.942 \mathrm{ns}$; computed group delays for the inhomogeneously-filled waveguide are (a) $178.221 \mathrm{ns}$, (b) $178.703 \mathrm{ns}$, (c) $178.785 \mathrm{ns}$.}
\label{fig:transient_plot}
\end{figure}
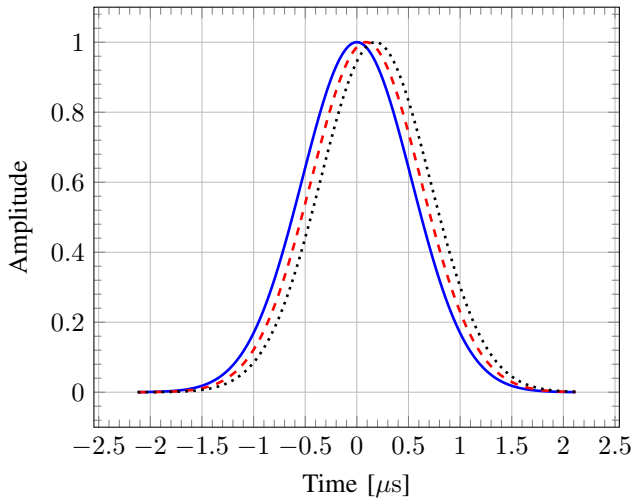

\revise{
Next, assume that the waveguide has dimensions $l = 40\mathrm{cm}$, $a = 3\mathrm{cm}$, and $b=1.5\mathrm{cm}$, and is excited
by the incoming transient $\text{TE}_{10}$ pulse
\begin{align}
    \vect{\cal E}_p(u,v,w=l) = \vect{\cal X}_{10}(u,v) T(t)
\end{align}
where $T(t) = e^{-\frac{1}{2} \left(\frac{t}{\sigma} \right)^2} \cos (\omega_o t)$ is a modulated Gaussian with $\sigma = 10^3/\omega_o$, $\omega_o = 2\pi f_o$, and $f_o = 7.5 \mathrm{GHz}$.
In addition to the air-filled waveguide, an inhomogeneously filled one with permittivity 
\begin{align}
    \varepsilon(u,v,w) = \left( 1 + 30e^{-(w/3)^2} \right) \varepsilon_o
\end{align}
is studied. 
For both structures, the frequency domain $\text{TE}_{10}$ reflection coefficient at $z = l$ for $\omega_o - 3/\sigma < \omega < \omega_o + 3/\sigma$ (this band contains 99.99\% of the pulse's energy) is approximated by complex exponentials using the Prony method \cite{Prony}, resulting in a semi-analytical expression for the outgoing transient field in terms of time- and phase-shifted, and possibly decaying modulated Gaussians. 
For the air-filled waveguide, the frequency-domain reflection coefficient is analytically available (eqn.~\eqref{eq:waveguide_PEC_1}), while for the inhomogeneously filled one it is obtained by solving a Riccatti equation \cite{Chew1995waves}. Fig.~\ref{fig:transient_plot} shows the time-domain signatures of the incoming and outgoing fields at $z=l$ (for clarity, only the envelopes are shown as all waveforms contain thousands of oscillations). The outgoing fields' group delays were computed (a) by numerically locating the maximum in their envelopes, (b) by computing the (sole) entry of the WS time delay matrix at $\omega = \omega_o$ via \eqref{eq:WS_guiding_final_2}, and (c) by numerical differentiation of the phase of reflection coefficient at $\omega = \omega_o$ via \eqref{eq:Oneport_Narrowband_qp}. For both waveguides, these group-delay estimates are reported in the caption of Fig.~\ref{fig:transient_plot} and seen to agree to within 0.3\%; for the air-filled waveguide they also agree with the analytical result in \eqref{eq:waveguide_PEC_Qg}.

}

\begin{figure}[t]
    \centering
    \includegraphics[width=0.95\columnwidth]{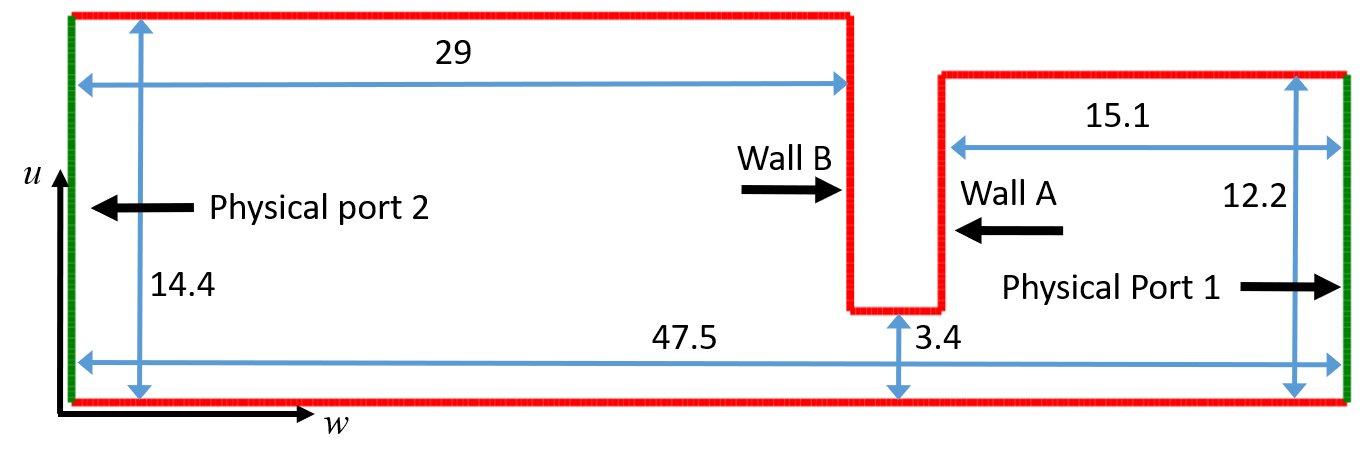}
    \caption{Notched waveguide.  Red: PEC walls; Green: physical apertures.  Dimensions are shown in $\mathrm{cm}$.}
    \label{fig:waveguide2_geometry}
\end{figure}

\begin{figure}[t]
    \centering
    \includegraphics[width=0.92\columnwidth, height = 0.27\columnwidth]{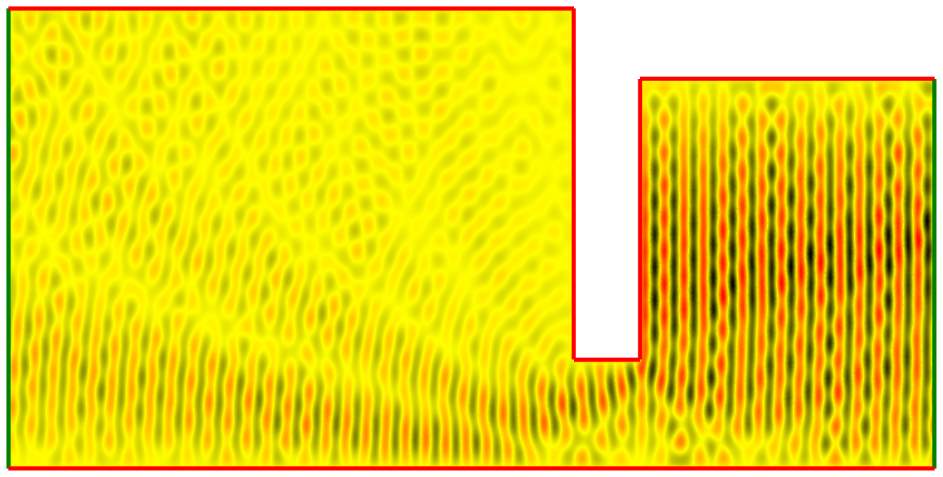}
    \caption{Electric field distribution ${\cal E}_{v,1}(u,w)$. }
    \label{fig:waveguide2_regfield}
\end{figure}

\subsubsection{Notched Waveguide}
\label{sec:notched_waveguide}

\begin{figure*}[t]
\null \hfill
    \subfloat[WS mode \#1 \label{fig:waveguide_WS1}]{\includegraphics[width=0.3\textwidth, height=0.09\textwidth]{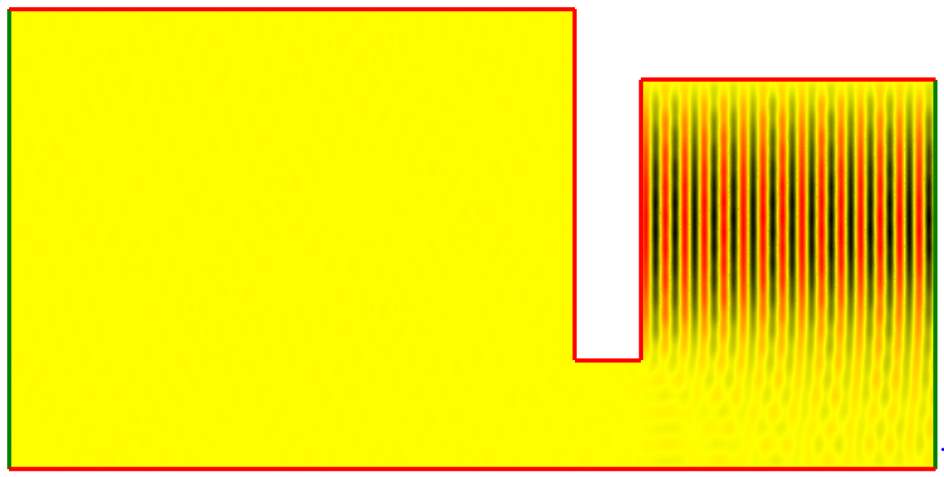}}   \hfill
    \subfloat[WS mode \#3]{\includegraphics[width=0.3\textwidth, height=0.09\textwidth]{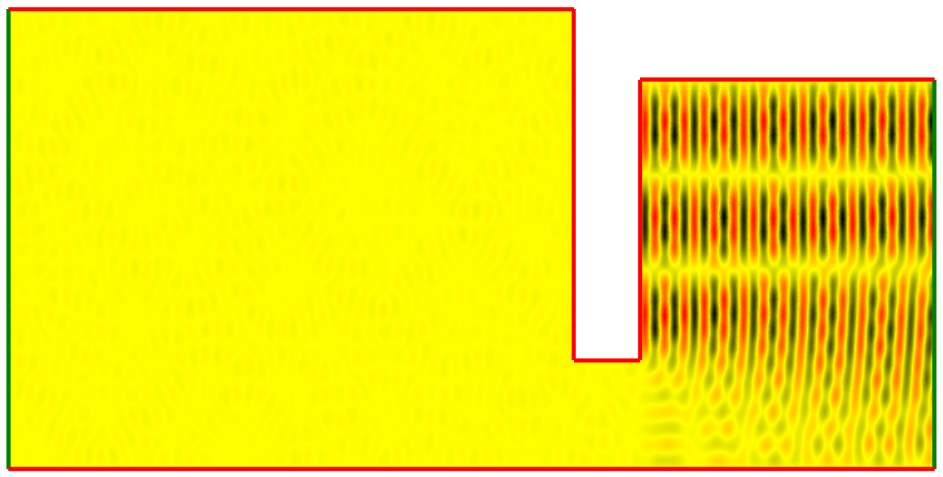}} 
    \hfill \subfloat[WS mode \#5]{\includegraphics[width=0.3\textwidth, height=0.09\textwidth]{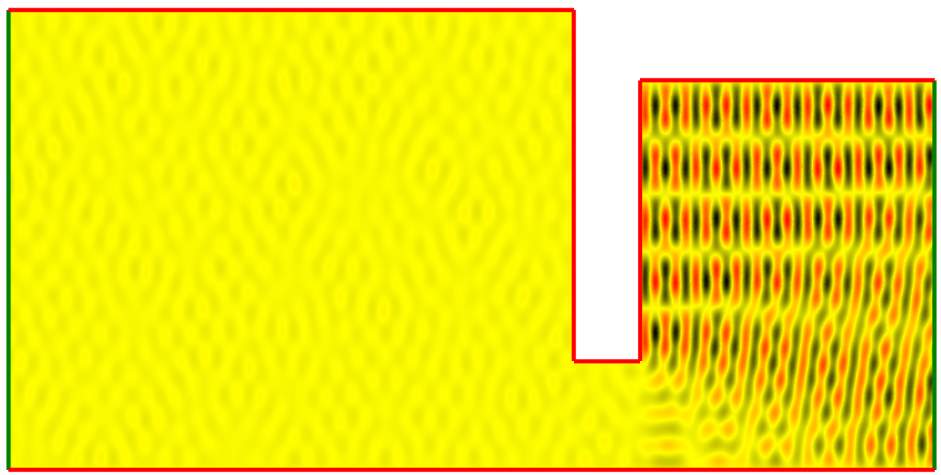}} \hfill \null  \\
    \null \hfill
    \subfloat[WS mode \#15 \label{fig:waveguide_WS15}]{\includegraphics[width=0.3\textwidth, height=0.09\textwidth]{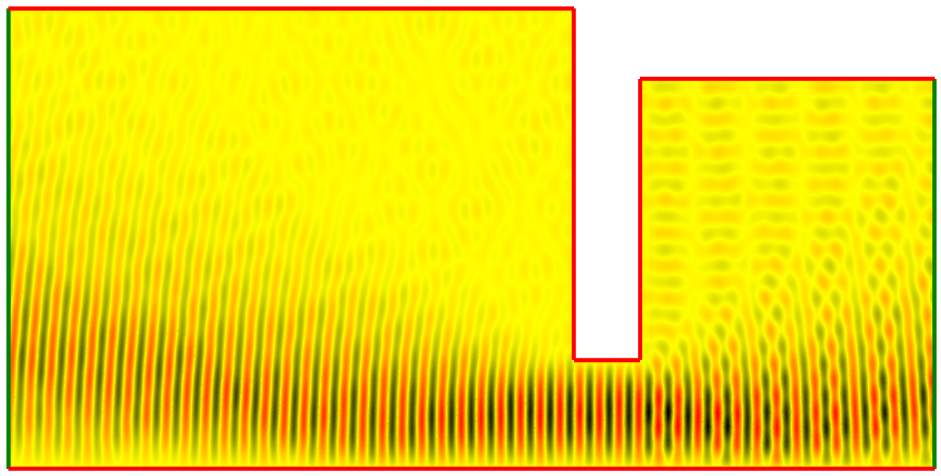}}   \hfill
    \subfloat[WS mode \#18]{\includegraphics[width=0.3\textwidth, height=0.09\textwidth]{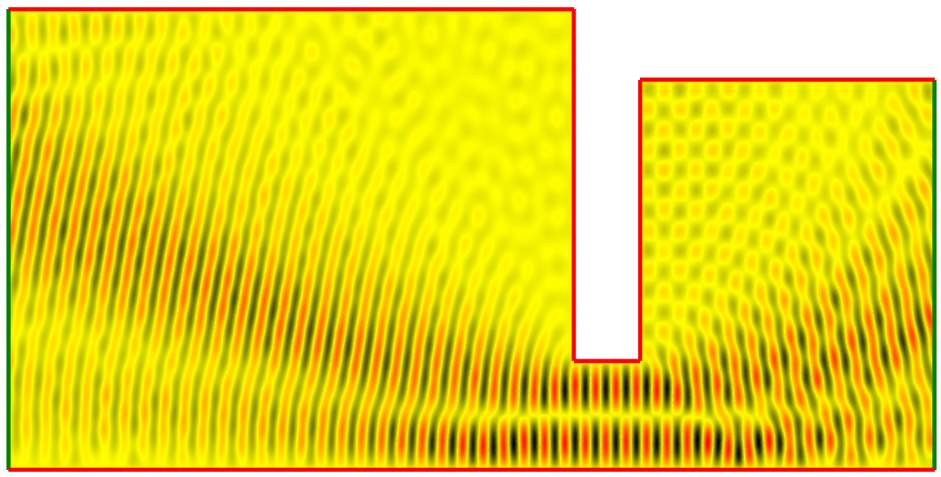}} 
    \hfill \subfloat[WS mode \#23]{\includegraphics[width=0.3\textwidth, height=0.09\textwidth]{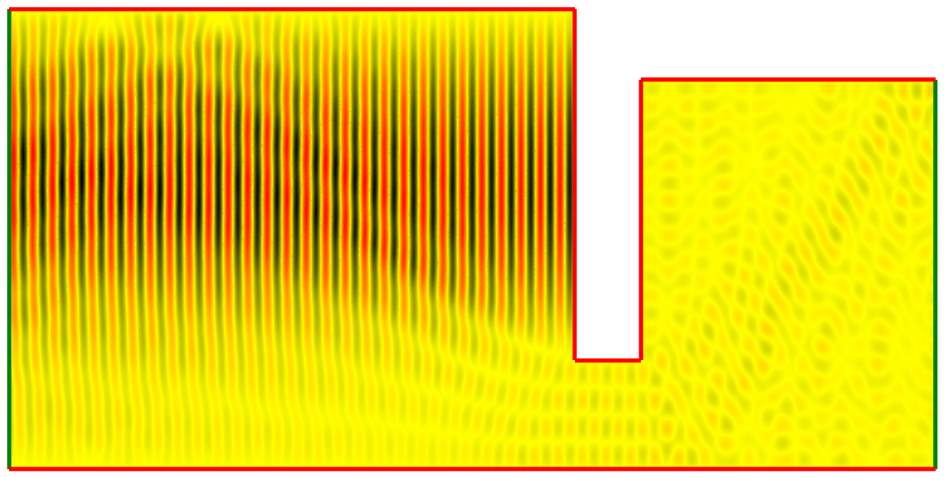}} \hfill \null  \\
    \null \hfill
    \subfloat[WS mode \#25]{\includegraphics[width=0.3\textwidth, height=0.09\textwidth]{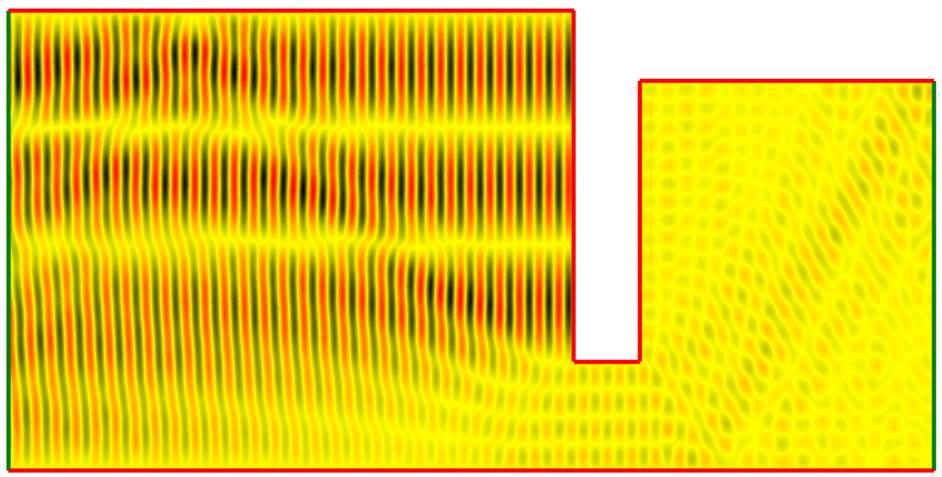}}   \hfill
    \subfloat[WS mode \#44]{\includegraphics[width=0.3\textwidth, height=0.09\textwidth]{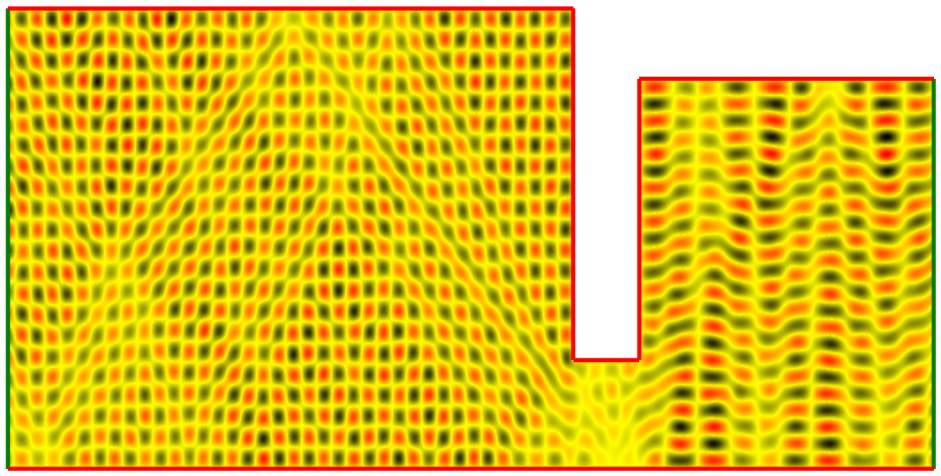}} \hfill
    \subfloat[WS mode \#52]{\includegraphics[width=0.3\textwidth, height=0.09\textwidth]{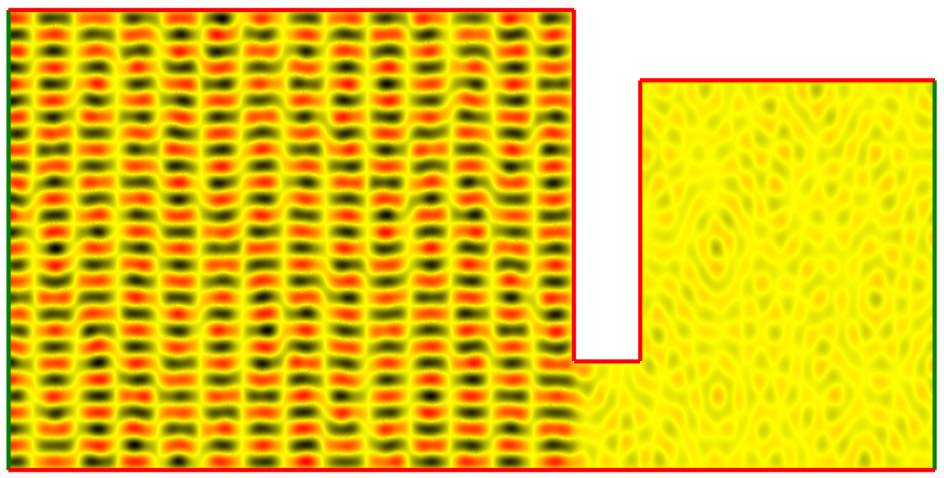}}
     \hfill \null 
     \caption{Field distribution with select WS modes in notched waveguide.}
    \label{fig:waveguide2_fields}
\end{figure*}

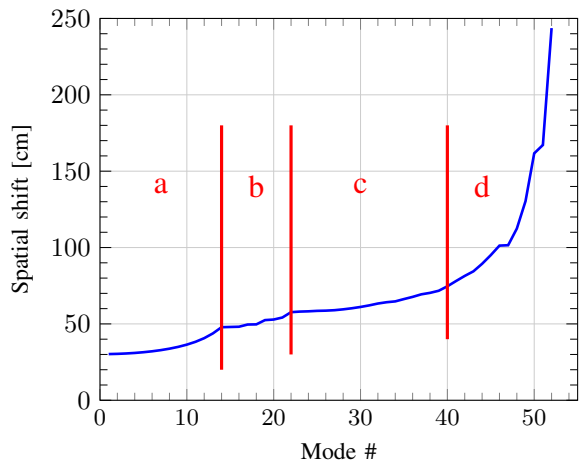
\begin{figure}[tbh]
    \centering
    \input{graphics/results/waveguide2/delays}
    \caption{Spatial shifts (delays) of WS modes for notched waveguide.}
    \label{fig:waveguide2_spatialshifts}
\end{figure}

Consider the waveguide shown in Fig.~\ref{fig:waveguide2_geometry}; due to the excitation described below, the structure’s height along $v$ is immaterial. 
The system is excited by $v$-polarized electric fields with mode profiles
\begin{align}
    \vect{\cal X}_m(u) = \sqrt{\frac{2}{a_\mathrm{port}}} \sin \left(\frac{m \pi u}{a_\mathrm{port}} \right) \unit{v}
    \label{eq:Waveguide2_1}
\end{align}
and frequency $f = 30~\mathrm{GHz}$. Physical ports 1 and 2 support $24$ and $28$ propagating modes, respectively.
In contrast to the previous example, all modes couple; $\matr{S}$ and $\matr{Q}$ therefore are dense $M_g \times M_g$ matrices with $M_g = 52$. 
\revise{Matrices} $\matr{S}$ and $\matr{Q}$ are computed using a third order accurate finite element method, and the WS relationship holds true to $7$ digits $\left( \abs{\matr{Q} - j \matr{S}^\dag \matr{S}'}/\abs{\matr{Q}} \approx 10^{-7} \right)$. 
Fig.~\ref{fig:waveguide2_regfield} shows the electric field distribution  ${\cal E}_{v,1}(u,w)$  throughout $\Omega$ due to excitation of physical port 1 by an incoming field with mode profile~\eqref{eq:Waveguide2_1} with $m = 1$. Many different scattering phenomena are in play, causing the field to be devoid of obvious \revise{modal} structure. 

Next, the WS time delay matrix is diagonalized (see~\eqref{eq:Q_diagonal}). 
Fields associated with select WS modes (port excitations specified by columns of $\matr{W}$  (see~\eqref{eq:Q_diagonal}) are shown in Fig.~\ref{fig:waveguide2_fields}\revise{)}. 
The structured nature of these fields (relative to that in Fig.~\ref{fig:waveguide2_regfield}) is immediately apparent. 
Time delays ($\widebar{\matr{Q}}$ ’s diagonal elements) are converted to equivalent spatial shifts measured in centimeter (100 times 
the eigenvalue multiplied by the
free-space speed of light) (Fig.~\ref{fig:waveguide2_spatialshifts}). Four different delay/shift regimes are observed (a-d).
\begin{enumerate}[a., leftmargin=* ]
    \item	WS modes 1-14.  \emph{Fields originating in, and reflecting back to physical port 1}.  Consider WS mode \#1 shown in Fig.~\ref{fig:waveguide_WS1}; this mode is characterized by a shift of $30.2~\mathrm{cm}$, or just over twice the distance from physical port 1 to wall A, representing the shortest possible path a field originating from either aperture can take before exiting the system.  Other modes in this category, e.g. WS modes \#3 and \#5, share similar characteristics but experience slightly larger delays/shifts than WS mode \#1 as they travel at a (small) angle w.r.t. the $w$-axis.
    \item WS modes 15-22. \emph{Fields originating in physical port 1 and exiting though physical port 2 (and vice versa)}.  Consider WS mode \#15 shown in Fig.~\ref{fig:waveguide_WS15}; this mode is characterized by a shift of $48.0~\mathrm{cm}$, or $0.5~\mathrm{cm}$ more than the structure’s physical length of $47.5~\mathrm{cm}$.  After all modes originating in, and reflecting back to, physical aperture 1 have been exhausted, the shortest possible time a signal can dwell in the system is by traversing the entire cavity (from physical port 1 to 2, or vice versa), avoiding contact with walls A and B. Other modes in this category, e.g. WS mode \#18, share similar characteristics but experience slightly larger delays than WS mode \#15 as they travel at a (small) angle w.r.t. the $w$-axis.
    \item WS modes 23-40. \emph{Fields originating in, and reflecting back to, physical port 2}.  These modes are similar to WS modes 1-14, but originate in physical aperture 2 and bounce off wall B.  The shift of WS mode \#23 is 58.1 cm, or slightly higher than twice the distance from physical port 2 to wall B and back.
    \item WS modes 41-52. \emph{Highly resonant fields, traveling at steep angles w.r.t. the $w$-axis}.  These modes originate either in physical aperture 1 or 2 and are characterized by large time delays. 
\end{enumerate}

\subsection{Scattering Systems}

\begin{figure*}[t]
\null \hfill
    \subfloat[WS mode \#1 \label{fig:strip_WS1}]{\includegraphics[width=0.3\textwidth, height=0.3\textwidth]{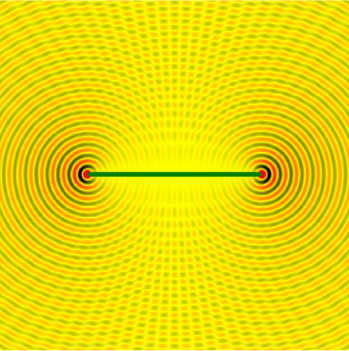}}   \hfill
    \subfloat[WS mode \#3 \label{fig:strip_WS3}]{\includegraphics[width=0.3\textwidth, height=0.3\textwidth]{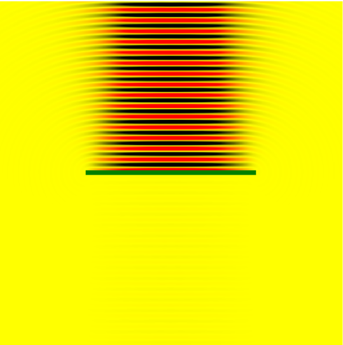}} 
    \hfill \subfloat[WS mode \#16 \label{fig:strip_WS16}]{\includegraphics[width=0.3\textwidth, height=0.3\textwidth]{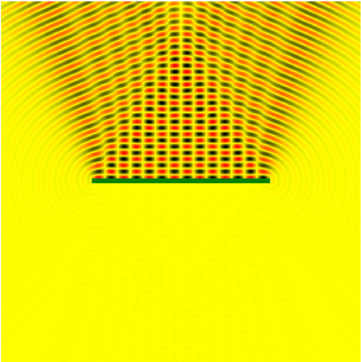}} \hfill \null \\
    \null \hfill
    \subfloat[WS mode \#1 \label{fig:strip_WS1_current}]{\input{graphics/results/strip/curr_ws_1.tex}}  \hfill 
    \subfloat[WS mode \#3\label{fig:strip_WS3_current}]{\input{graphics/results/strip/curr_ws_3.tex}} \hfill
    \subfloat[WS mode \#16 \label{fig:strip_WS16_current}]{\input{graphics/results/strip/curr_ws_16.tex}} 
    \hfill \null
    \caption{Field distribution (Figs. (a)--(c)) and magnitude of current (Figs. (d)--(f)) for select WS modes of the PEC strip.  Distance is measured from the left edge of the strip.}
    \label{fig:PEC_Strip_Mode_Distibution}
\end{figure*}

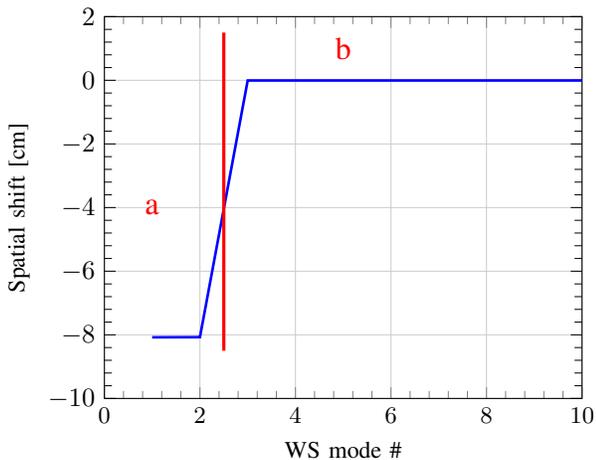
\begin{figure}[t]
    \centering
    \input{graphics/results/strip/delays}
    \caption{Spatial shifts (delays) of WS modes for strip. Shifts are only shown for the first 10 modes as all others are zero.}
    \label{fig:strip_delays}
\end{figure}

\subsubsection{PEC Strip}

Consider the strip shown in Fig.~\ref{fig:strip_WS1}. The strip is $8~\mathrm{cm}$ wide and centered about the origin (the location of the strip relative to the origin affects the WS time delays, as they are derived from scattering matrices defined on spheres/cylinders centered at the origin). 
The strip is illuminated by $\mathrm{TM}_z$ fields at $f = 30~\mathrm{GHz}$. The strip’s dense $\matr{S}$ and $\matr{Q}$ matrices are computed using an integral equation code considering $M_s = 100$ excitations by cylindrical harmonics; because $100 \gg 2 k \text{ (width of strip)}/2 \cong 50$, these excitations adequately resolve all the system’s degrees of freedom.  The WS relationship \eqref{eq:Ws_Sdag_Sprime} is found to hold true to 5 digits $\left( \abs{\matr{Q} - j \matr{S}^\dag \matr{S}'}/\abs{\matr{Q}} \approx 10^{-5} \right)$. Very much like the fields due to a waveguide port excitation in Fig.~\ref{fig:waveguide2_regfield}, all cylindrical harmonics excite both surface and edge scattering phenomena, causing each of them to experience a different delay (as defined by~\eqref{eq:Multiport_Narrowband_qp}).  

Next, the WS time delay matrix is diagonalized (see~\eqref{eq:Q_diagonal}).  \emph{Total} fields (i.e. sums of incident and scattered fields) and the strip’s currents for select WS modes (port excitations specified by columns of $\matr{W}$ (see \eqref{eq:Q_diagonal}) are shown in Figs.~\ref{fig:strip_WS1}--\ref{fig:strip_WS16_current}. 
Time delays ($\widebar{\matr{Q}}$'s diagonal elements) are converted to equivalent spatial shifts just as was done for the notched waveguide in Sec.~\ref{sec:notched_waveguide} (Fig.~\ref{fig:strip_delays}). Two different delay/shift regimes are observed.

\begin{enumerate}[a., leftmargin=* ]
\item \emph{WS edge modes.}   The total field and current of WS mode \#1 are shown in Figs.~\ref{fig:strip_WS1} and~\ref{fig:strip_WS1_current}; this mode is characterized by a (negative!) shift of $-8.1~\mathrm{cm}$, or just over twice the distance from the strip’s edge to the origin.  
The total field and currents are concentrated near the strip’s edges; the center of the strip is virtually quiescent (the same to a large extent holds true for the incident and scattered field (not shown)).  
The total fields associated with this mode are concentrated along the strip’s axis; the two ``beams'' (coming in predominantly from the $\pm x$-axis) that excite the strip reflect back upon hitting the edge, causing them to travel (roughly) $8~\mathrm{cm}$ less (from the edge to the origin and back) than a field that does not interact with the strip.  WS mode \#1 is characterized by an even current distribution on the strip.  WS mode \#2 (not shown) is very similar to mode \#1, except that it supports odd currents and fields.
\item \emph{WS Geomeric Optics (GO)-like modes.} The fields and current distributions of WS modes \#3 and \#16 are shown in Figs.~\ref{fig:strip_WS3}--\ref{fig:strip_WS16} and~\ref{fig:strip_WS3_current}--\ref{fig:strip_WS16_current}. 
These modes are GO-like in nature, and consist of beam-like incident fields that avoid the strip’s edges while exciting quasi-periodic currents on the strip, 
producing beams that specularly reflect away from the strip.
These modes are characterized by very small time delays/spatial shifts as the GO ray’s round trip time is virtually identical to that experienced by a wave that does not interact with the strip.  
For each WS mode hitting the strip from the top, there is one hitting it from the bottom (not shown; while in principle these modes are degenerate, the symmetry was broken here by positioning the strip a very small distance above the origin). 
Note that the WS GO fields extremize group time delays (see Sec. \ref{sec:SimultaneousDiagonalization}), in accordance with the generalized Fermat principle for optics and high-frequency electromagnetic fields~\cite{Born2013}.
\end{enumerate}

\begin{figure*}[t]
\null \hfill
    \subfloat[WS mode \#1 \label{fig:cavity_WS1}]{\includegraphics[width=0.22\textwidth, height=0.22\textwidth]{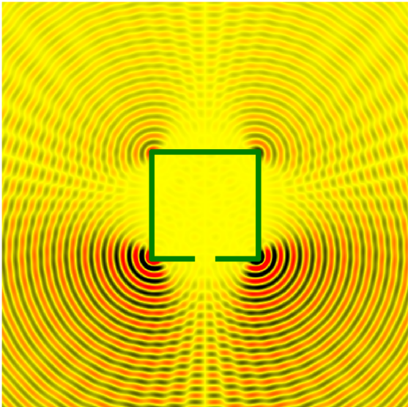}}   \hfill
    \subfloat[WS mode \#5 \label{fig:cavity_WS5}]{\includegraphics[width=0.22\textwidth, height=0.22\textwidth]{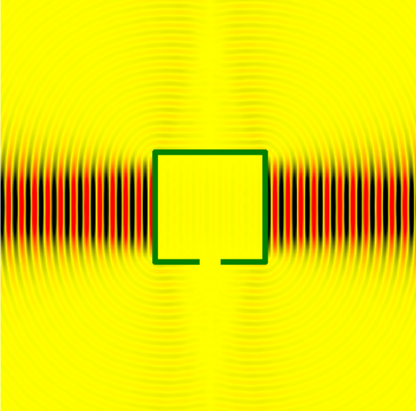}} 
    \hfill 
 \subfloat[WS mode \#15 \label{fig:cavity_WS15}]{\includegraphics[width=0.22\textwidth, height=0.22\textwidth]{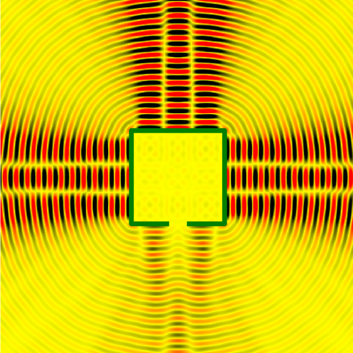}} \hfill 
 \subfloat[WS mode \#120 \label{fig:cavity_WS190}]{\includegraphics[width=0.22\textwidth, height=0.22\textwidth]{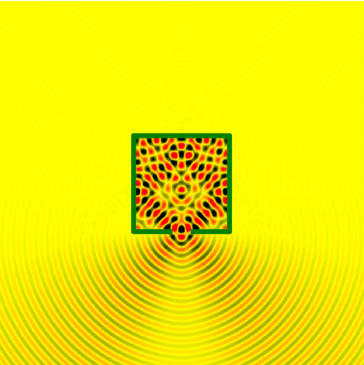}} \hfill
    \null \\
    \null \hfill
    \subfloat[WS mode \#1 \label{fig:cavity_WS1_current}]{\input{graphics/results/cavity/curr_ws_1.tex}} \hfill 
    \subfloat[WS mode \#5 \label{fig:cavity_WS5_current}]{\input{graphics/results/cavity/curr_ws_5.tex}} \hfill
     \subfloat[WS mode \#15 \label{fig:cavity_WS15_current}]{\input{graphics/results/cavity/curr_ws_15.tex}} \hfill
      \subfloat[WS mode \#120 \label{fig:cavity_WS190_current}]{\input{graphics/results/cavity/curr_ws_190.tex}} \hfill \null
      \caption{Field distribution (Figs.(a)-(d)) and current on the cavity walls (Figs. (e)--(h)) for select WS modes. Distance is measured by traversing the cavity walls counterclockwise starting from the right edge of the aperture.}
\end{figure*}

\subsubsection{PEC Cavity}

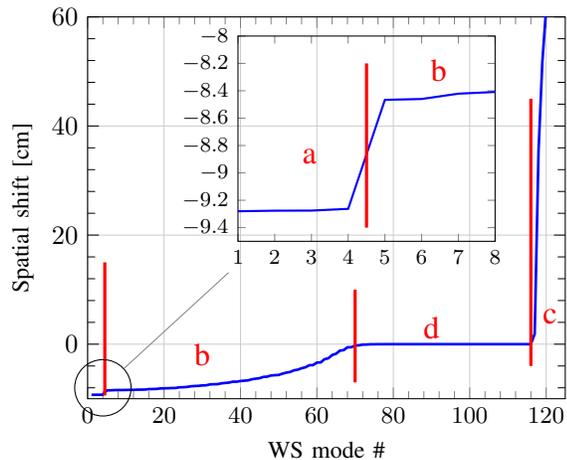
\begin{figure}[t]
    \centering
    \input{graphics/results/cavity/delays}
    \caption{Spatial shifts (delays) of WS modes for cavity.}
    \label{fig:cavity_delays}
\end{figure}

 \begin{figure*}[t]
 \null \hfill
 \subfloat[$\matr{Z}_{11}$ \label{fig:Twodipole_Z11}]{\includegraphics[width=0.3\textwidth]{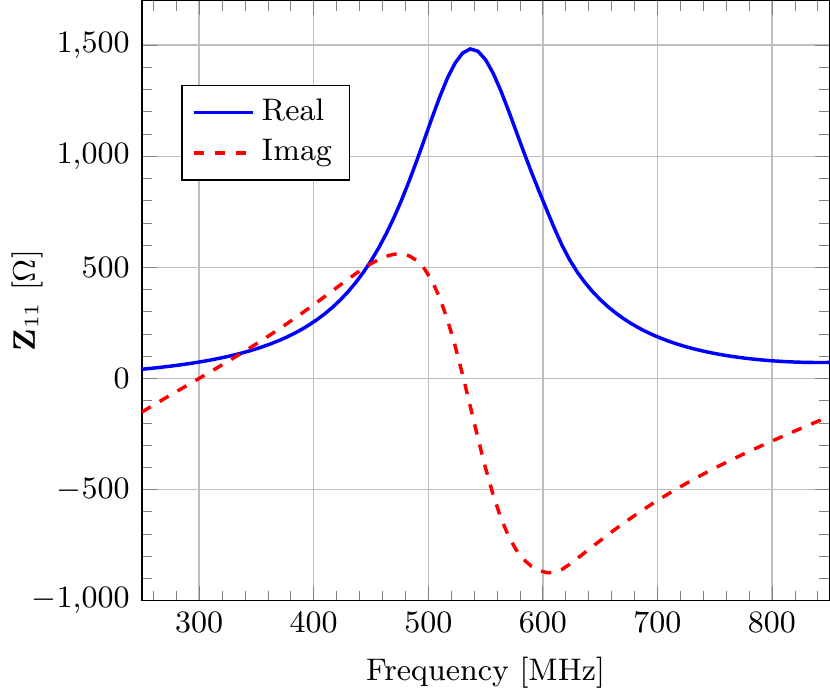}}
 \hfill
 \subfloat[$\matr{Z}_{12}$ \label{fig:Twodipole_Z12}]{\includegraphics[width=0.3\textwidth]{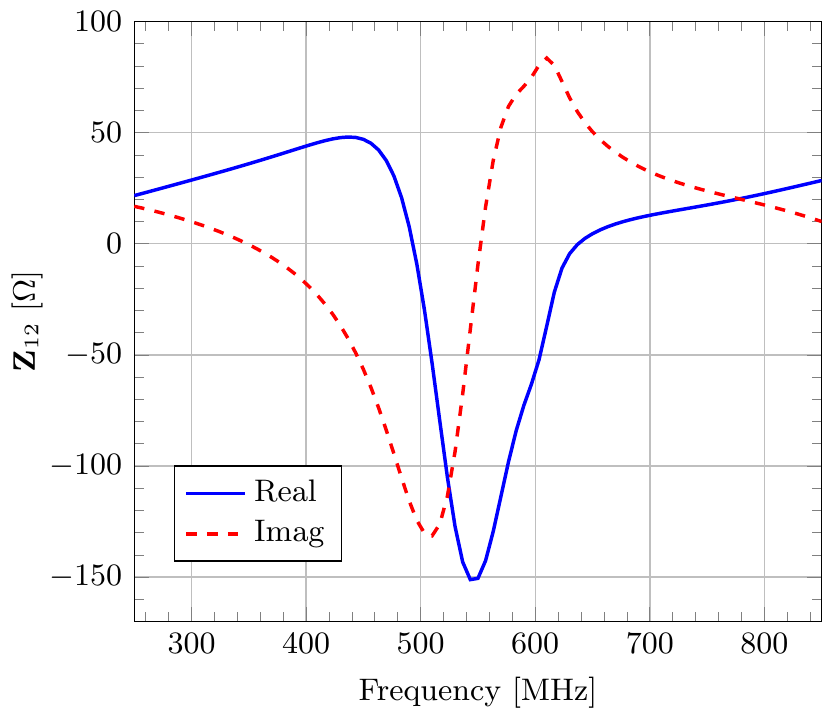}}
 \hfill \subfloat[$\matr{Z}_{11}'$\label{fig:Twodipole_dZ11}]{\includegraphics[width=0.3\textwidth]{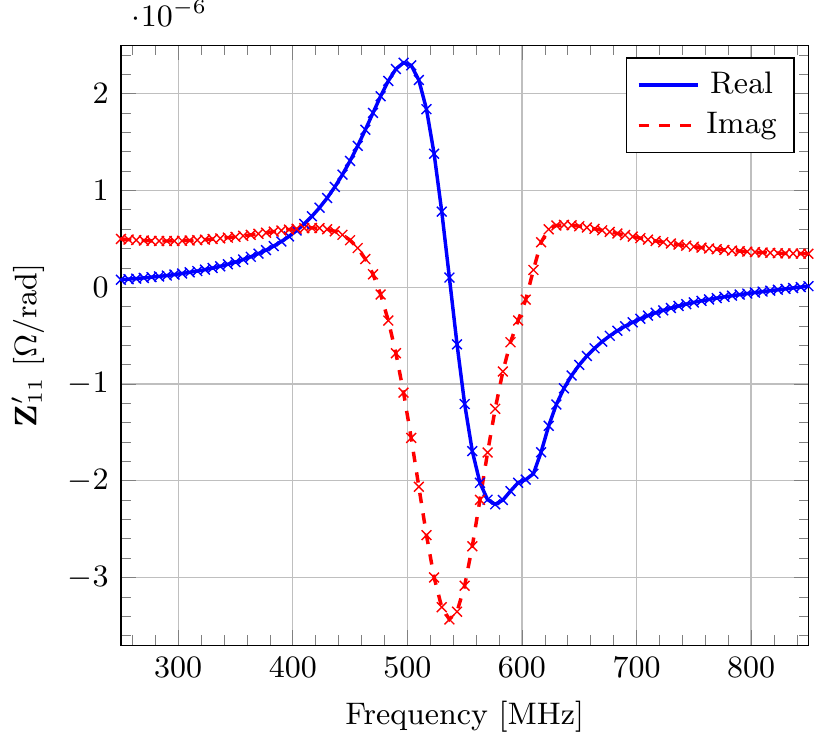}} \hfill \null \\
 \null \hfill
 \subfloat[$\matr{Z}_{12}'$\label{fig:Twodipole_dZ12}]{\includegraphics[width=0.3\textwidth]{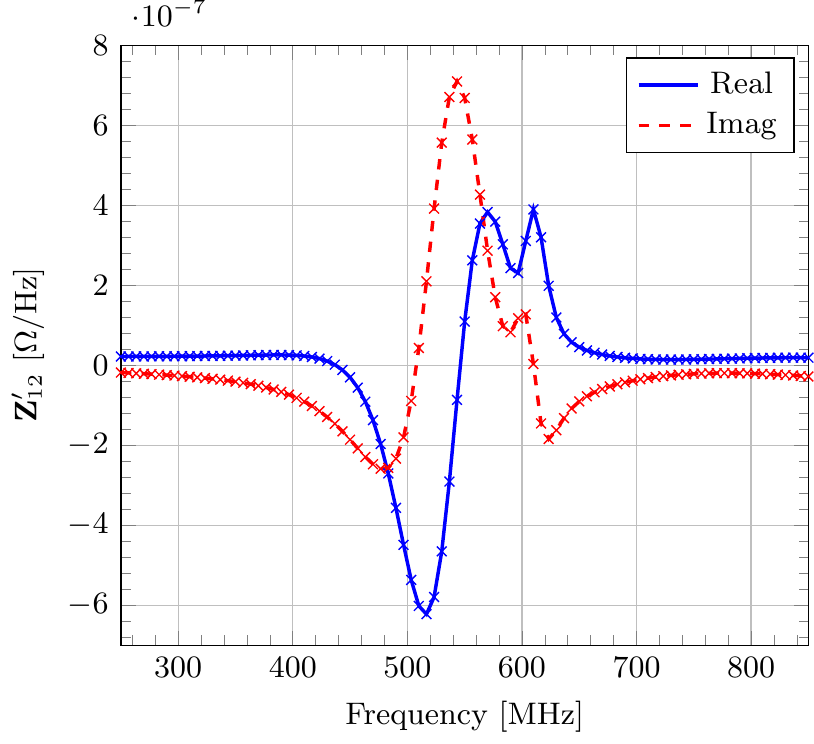}} \hfill 
 \subfloat[Scattered field \label{fig:Twodipole_Rad1}]{\includegraphics[width=0.3\textwidth]{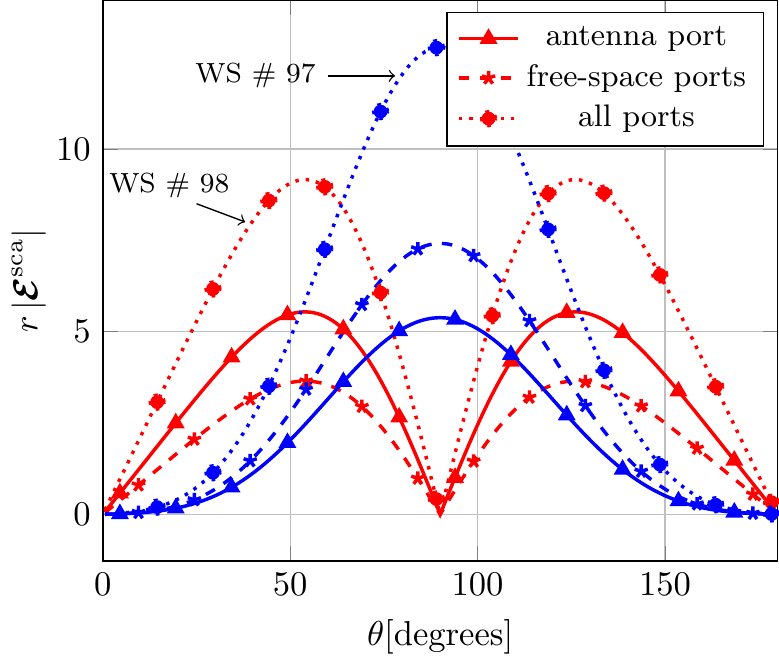}} \hfill 
 \subfloat[Scattered field \label{fig:Twodipole_Rad2}]{\includegraphics[width=0.3\textwidth]{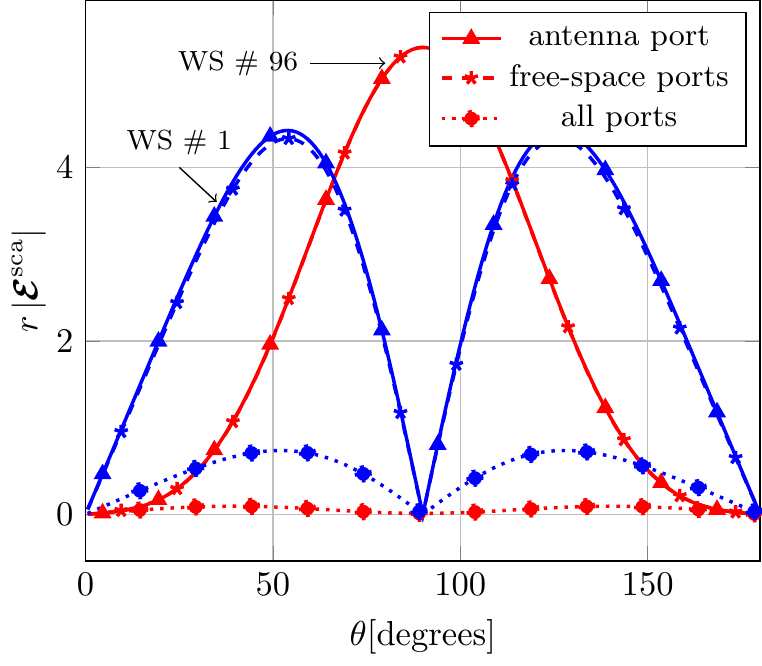}} \hfill \null
     \caption{(a)-(b) Selected entries of $2 \times 2$ input impedance matrix of the two-element array of strip dipoles; (c)-(d) selected entries of frequency derivative of the input impedance matrix computed with the WS (---) and the FD method ($\times$); (e)-(f) scattered fields due to antenna and free-space ports excitation.}
     \label{fig:twodipole_result}
 \end{figure*}

 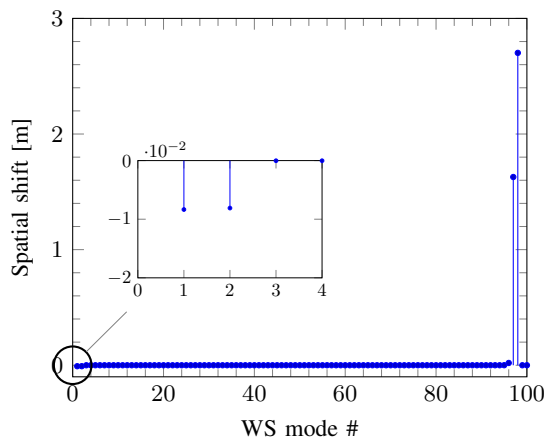
\begin{figure}[t]
\centering
\input{Plots/TwoDipole_delays}
\caption{Spatial shifts (delays) of WS modes for two-elements array of strip dipoles.}
\label{fig:TwoDipoles_Shift}
\end{figure}

Consider the square cavity shown in Fig.~\ref{fig:cavity_WS1}. 
The cavity’s base is $8.24~\mathrm{cm}$ long and contains a hole of width $1.6~\mathrm{cm}$.  
The cavity is centered about the origin and illuminated by $\mathrm{TM}_z$ fields at $f= 30~\mathrm{GHz}$. 
The cavity’s dense $\matr{S}$ and $\matr{Q}$ matrices are computed using an integral equation code using $M_s = 120$ cylindrical harmonic excitations, which adequately resolve all of the system’s degrees of freedom.  The WS relationship \eqref{eq:Ws_Sdag_Sprime} is found to hold true to 4 digits $\left( \abs{\matr{Q} - j \matr{S}^\dag \matr{S}'}/\abs{\matr{Q}} \approx 10^{-4} \right)$.   The WS time delay matrix is diagonalized (see \eqref{eq:Q_diagonal}) and \emph{total} fields and currents for select WS modes are shown in Figs.~\ref{fig:cavity_WS1}--\ref{fig:cavity_WS190_current}.  Spatial shifts computed from $\matr{Q}$’s eigenvalues are shown in (Fig.~\ref{fig:cavity_delays}). Three different delay/shift regimes are observed.
\begin{enumerate}[a., leftmargin=* ]
\item \emph{WS corner modes.}  The total field and current of WS mode \#1 are shown in Figs.~\ref{fig:cavity_WS1} and~\ref{fig:cavity_WS1_current}; just like for the strip, this mode is characterized by a negative spatial shift.  
The total field and currents are concentrated near the cavity’s four corners; 
the cavity’s facets and aperture are virtually quiescent. 
The cavity supports a total of four corner modes (inset in Fig.~\ref{fig:cavity_delays}) with different current and field symmetries across the origin.

\item \emph{WS GO-like modes.}  The fields and current distributions of WS modes \#5 and \#15 are shown in Figs.~\ref{fig:cavity_WS5}--\ref{fig:cavity_WS15} and~\ref{fig:cavity_WS5_current}--\ref{fig:cavity_WS15_current}. 
These modes are GO-like in nature and consist of beam-like incident fields that avoid the cavity’s corners and aperture while exciting quasi-periodic currents on, and causing specular reflections from, its facets.  
These modes spend less time in the system than free space fields that do not interact with the system, causing them to be characterized by negative time delays/shifts.
\item \emph{WS cavity modes.} The total field and current of WS mode \#120 are shown in Figs.~\ref{fig:cavity_WS190} and~\ref{fig:cavity_WS190_current}.  
This mode is excited by a beam focused on the cavity’s aperture, exciting a strong quasi-resonant field in its interior. 
This field bounces back and forth many times between the cavity’s walls, causing it to experience a large positive time delay before escaping. 
The cavity supports a total of four such modes, each characterized by a different aperture field distribution.  All cavity modes are characterized by large currents near the aperture and virtually quiescent total fields on the cavity’s exterior walls away from the aperture. 
\end{enumerate}
The WS modes in region d are composed of fields that do not interact with the cavity. Indeed, because $M_s = 120 > 2 k\sqrt{2} (\text{width of cavity})/2$, many cylindrical harmonics do not reach the cavity and their WS combinations do not experience any time delay.

\subsection{Radiating Systems}
\label{sec:twodipole}

\revise{
Consider a co-linear array of two center-fed $z$-directed strip dipoles of length $l = 0.476\mathrm{m}$ and width $w=4\mathrm{mm}$. The dipoles' centers are spaced $0.5 \mathrm{m}$ apart along the $z$-axis.
Figs.~\ref{fig:Twodipole_Z11}--\ref{fig:Twodipole_Z12} show the self and mutual impedances of the array in the frequency band $250-850~\mathrm{MHz}$.
The frequency derivative of the antenna impedance matrix is obtained in two ways:
(i) using the WS relationship ($\matr{Q}$ is computed from the field in the system for all port excitations using \eqref{eq:rad_final_block_eqn}, next $\matr{S}'$ is computed using \eqref{eq:rad_Qe_Sdag_Sprime}, and finally $\matr{Z}'$ is computed using \eqref{eq:Zprime_multiport}; the latter contains a $2 \times 2$ block with the frequency derivatives of antenna self and mutual impedances), and
(ii) using a finite difference in frequency approximation. 
The WS method is observed to accurately predict the frequency derivative of the array's input impedance matrix (Figs.~\ref{fig:Twodipole_dZ11}--\ref{fig:Twodipole_dZ12}).  Specifically, the WS relationship was found to hold true to 4 digits over the entire frequency range $\left(\abs{\matr{Q} - j \matr{S}^\dag \matr{S}'}/\abs{\matr{Q}} \approx 10^{-4}\right)$. 

Next, $\matr{Q}$ is diagonalized at strip dipole's resonant frequency of $f = 300 \mathrm{MHz}$.
Fig.~\ref{fig:TwoDipoles_Shift} shows the WS modes' spatial shift. Only five WS modes are observed to meaningfully interact with the antennas; all others consist of free-space port excitations producing fields that by and large bypass the antenna and hence experience negligible time delays (the WS time-delay matrix $\matr{Q}$ was computed using two guided and 96 free-space excitations, many with higher-order azimuthal dependencies that do not interact with the nearly axisymmetric strip dipole).
 The WS modes with non-zero spatial shifts can be categorized as follows:
\begin{itemize}
    \item WS modes \# 97-98: these modes experience the largest spatial shifts, i.e. store the most energy. These modes excite both the antenna and free-space ports. Fig.~\ref{fig:Twodipole_Rad1} shows the scattered field when the antenna and free-space ports are simultaneously excited using WS modes \# 97 and 98's eigenvectors. The antenna and free-space ports excite the antenna ``(quasi-) symmetrically'', producing induced currents that add constructively, resulting in strong radiation. Each of these modes produces a beam pointing to a different angle w.r.t. the array axis.
    \item WS modes \# 1-2 and 96: these modes can be considered as the asymmetric counterpart of WS modes \# 97-98.  Fig.~\ref{fig:Twodipole_Rad2} shows that while both the antenna and free space ports are excited, they produce currents that add destructively, resulting in little far-field radiation and a small spatial shift, i.e little energy storage.
\end{itemize}
}

%% file: graphics/results/waveguide2/transient.tex
\begin{tikzpicture}[thick,scale=1, every node/.style={scale=1}]
\centering
\begin{axis}
[
width=2.75in,
 height=2.2in,
 compat=newest,
 minor tick num=4,
scale only axis,
xlabel near ticks,
ylabel near ticks,
xtick distance = 0.5,
xlabel={Time [$\mu \mathrm{s}$]},
ylabel={Amplitude},
axis background/.style={fill=white},
grid=major,
]

\addplot[forget plot, color=blue, line width=1pt]
  table[x=t, y=data1, col sep=comma]{graphics/results/waveguide2/transient_data.csv};

\addplot[forget plot, color=red, dashed, line width=1pt]
  table[x=t, y=data2, col sep=comma]{graphics/results/waveguide2/transient_data.csv};

\addplot[forget plot, color=black, dotted, line width=1pt]
  table[x=t, y=data3, col sep=comma]{graphics/results/waveguide2/transient_data.csv};

\end{axis}
\end{tikzpicture}

%% file: graphics/results/waveguide2/delays.tex
\begin{tikzpicture}
\centering
\begin{axis}[width=2.5in,
height=2in,
xlabel near ticks,
ylabel near ticks,
yticklabel style = {font=\small},
xticklabel style = {font=\small},
at={(0.758in,0.481in)},
scale only axis,
grid style={line width=.1pt, draw=gray!20},major grid style={line width=.2pt,draw=gray!40},
minor tick num=4,
grid = major,
xmin=0,
xmax=55,
xlabel={\small Mode \#},
every outer y axis line/.append style={black},
every y tick label/.append style={font=\color{black}},
every y tick/.append style={black},
ymin=0,
ymax=250,
ylabel={\small Spatial shift [cm]},
legend style={legend cell align=left, align=left, draw=black,at={(0.18,0.7)},anchor=south}
]

\addplot[color=blue, line width=1pt]
  table[row sep=crcr]
  {
1 30.245907675944668 \\
2 30.390359060403 \\
3 30.63551867367133 \\
4 30.988147539902837 \\
5 31.459983342089053 \\
6 32.0656770294652 \\
7 32.82862441527828 \\
8 33.77900707994825 \\
9 34.95833608262327 \\
10 36.429607157238216 \\
11 38.284224869140004 \\
12 40.66179738349786 \\
13 43.78230885272667 \\
14 47.84293960352561 \\
15 47.99875706000922 \\
16 48.14888680751052 \\
17 49.593231699149854 \\
18 49.69552559211831 \\
19 52.48939519198427 \\
20 52.86315916303667 \\
21 54.18285832560063 \\
22 57.68472236059042 \\
23 58.05971034806293 \\
24 58.24162557148877 \\
25 58.5436784941552 \\
26 58.68027830311645 \\
27 58.97558681507196 \\
28 59.533991317430214 \\
29 60.23736054360277 \\
30 61.09742079060934 \\
31 62.12734187681109 \\
32 63.343116568484724 \\
33 64.19666286851952 \\
34 64.74767758336614 \\
35 66.26093284132975 \\
36 67.69325433006213 \\
37 69.36921232302814 \\
38 70.32322618321042 \\
39 71.69958152109194 \\
40 74.62371804156273 \\
41 78.07725951146865 \\
42 81.4181429881076 \\
43 84.4293187269094 \\
44 89.36116344978268 \\
45 95.03756574060739 \\
46 101.2721962738003 \\
47 101.51998879397695 \\
48 112.42393734011969 \\
49 130.37808104852888 \\
50 161.67960852889914 \\
51 167.13975403680004 \\
52 243.67756634104146 \\
};
\draw[red, very thick] (axis cs: 14, 20) -- (axis cs: 14, 180);
\draw[red, very thick] (axis cs: 22, 30) -- (axis cs: 22, 180);
\draw[red, very thick] (axis cs: 40, 40) -- (axis cs: 40, 180);
\node at (axis cs: 7, 140) {\large {\color{red} a}};
\node at (axis cs: 18, 140) {\large {\color{red} b}};
\node at (axis cs: 30, 140) {\large {\color{red} c}};
\node at (axis cs: 44, 140) {\large {\color{red} d}};
\end{axis}
\end{tikzpicture}

%% file: graphics/results/strip/curr_ws_1.tex
\begin{tikzpicture}
\centering
\begin{axis}[width=1.85in,
height=1.25in, ylabel near ticks, xlabel near ticks,
scale only axis,
grid style={line width=.1pt, draw=gray!20},major grid style={line width=.2pt,draw=gray!40},
minor tick num=4,
grid = major,
yticklabel style = {font=\small},
xticklabel style = {font=\small},
xmin=0,
xmax=8,
xlabel={\small Distance [$\mathrm{cm}$]},
every outer y axis line/.append style={black},
every y tick label/.append style={font=\color{black}, font=\small},
every y tick/.append style={black},
ymin=0,
ymax=0.8,
legend style={legend cell align=left, align=left, draw=black,at={(0.5,0.7)},anchor=south, font=\small}
]

\addplot[color=blue, line width=1pt]
  table[col sep= comma]
  {
0.01,0.5829683585
0.02,0.2204490125
0.03,0.1787769699
0.04,0.1473578059
0.05,0.1269026348
0.06,0.1115038455
0.07,0.09910952521
0.08,0.08864482571
0.09,0.07951038651
0.1,0.07134590479
0.11,0.06392400145
0.12,0.05709615123
0.13,0.05076304211
0.14,0.04485728217
0.15,0.03933278357
0.16,0.03415796005
0.17,0.02931120572
0.18,0.02477779429
0.19,0.02054769445
0.2,0.01661399565
0.21,0.01297175263
0.22,0.009617125421
0.23,0.006546733537
0.24,0.00375716948
0.25,0.001244633669
0.26,0.0009953358531
0.27,0.002968058147
0.28,0.00467984266
0.29,0.006138071725
0.3,0.007351240545
0.31,0.008328961203
0.32,0.00908193611
0.33,0.009621905414
0.34,0.009961572311
0.35,0.01011450977
0.36,0.01009505182
0.37,0.009918172401
0.38,0.009599354421
0.39,0.009154451704
0.4,0.008599546212
0.41,0.007950802891
0.42,0.00722432432
0.43,0.006436007239
0.44,0.005601402885
0.45,0.004735582938
0.46,0.003853012725
0.47,0.002967433166
0.48,0.002091752808
0.49,0.001237951091
0.5,0.000416993829
0.51,0.0003612382984
0.52,0.001088007627
0.53,0.001755763352
0.54,0.002358171087
0.55,0.002890128104
0.56,0.003347764374
0.57,0.00372842972
0.58,0.004030667562
0.59,0.004254175876
0.6,0.004399756154
0.61,0.004469251264
0.62,0.00446547325
0.63,0.004392122191
0.64,0.004253697342
0.65,0.004055401846
0.66,0.003803042337
0.67,0.003502924831
0.68,0.003161748259
0.69,0.002786497046
0.7,0.002384334089
0.71,0.00196249545
0.72,0.001528188044
0.73,0.001088491516
0.74,0.0006502654264
0.75,0.000220062763
0.76,0.000195949302
0.77,0.0005920606128
0.78,0.0009630805144
0.79,0.001304391076
0.8,0.00161199064
0.81,0.001882527431
0.82,0.002113323076
0.83,0.00230238605
0.84,0.002448415161
0.85,0.002550793353
0.86,0.002609572195
0.87,0.002625447559
0.88,0.002599727082
0.89,0.002534290105
0.9,0.002431540877
0.91,0.002294355854
0.92,0.002126026009
0.93,0.001930195094
0.94,0.001710794844
0.95,0.001471978091
0.96,0.001218050819
0.97,0.0009534041086
0.98,0.0006824469585
0.99,0.0004095408881
1,0.000138937203
1.01,0.0001252822668
1.02,0.000379260212
1.03,0.0006194140073
1.04,0.0008424804144
1.05,0.001045554337
1.06,0.001226121234
1.07,0.001382082909
1.08,0.00151177647
1.09,0.001613986387
1.1,0.001687949651
1.11,0.001733354142
1.12,0.001750330422
1.13,0.001739437226
1.14,0.001701641057
1.15,0.001638290297
1.16,0.001551084395
1.17,0.001442038676
1.18,0.001313445417
1.19,0.001167831846
1.2,0.00100791576
1.21,0.0008365594788
1.22,0.0006567228403
1.23,0.0004714159649
1.24,0.0002836524853
1.25,9.640391891e-05
1.26,8.744416949e-05
1.27,0.0002651336118
1.28,0.0004340721323
1.29,0.0005918683963
1.3,0.0007363631744
1.31,0.0008656562548
1.32,0.0009781288025
1.33,0.001072460944
1.34,0.001147644425
1.35,0.001202990273
1.36,0.001238131465
1.37,0.001253020671
1.38,0.001247923235
1.39,0.001223405596
1.4,0.001180319435
1.41,0.001119781871
1.42,0.001043152118
1.43,0.0009520050004
1.44,0.0008481018282
1.45,0.0007333591027
1.46,0.0006098155906
1.47,0.0004795982888
1.48,0.0003448878217
1.49,0.0002078838036
1.5,7.077069078e-05
1.51,6.431537137e-05
1.52,0.0001953182281
1.53,0.0003202914731
1.54,0.0004374258764
1.55,0.0005450741166
1.56,0.0006417725008
1.57,0.000726259407
1.58,0.0007974902387
1.59,0.0008546487383
1.6,0.0008971545656
1.61,0.0009246671027
1.62,0.0009370855091
1.63,0.0009345451016
1.64,0.0009174101939
1.65,0.0008862635757
1.66,0.0008418928627
1.67,0.0007852739907
1.68,0.0007175521628
1.69,0.0006400205947
1.7,0.0005540974265
1.71,0.0004613011927
1.72,0.0003632252537
1.73,0.0002615116019
1.74,0.0001578244563
1.75,5.382405399e-05
1.76,4.885896292e-05
1.77,0.0001486481845
1.78,0.0002440447352
1.79,0.00033364891
1.8,0.0004161798454
1.81,0.0004904929601
1.82,0.0005555949441
1.83,0.0006106561102
1.84,0.0006550199727
1.85,0.0006882099571
1.86,0.0007099331958
1.87,0.000720081406
1.88,0.0007187288943
1.89,0.000706127773
1.9,0.0006827005169
1.91,0.0006490300262
1.92,0.0006058473965
1.93,0.0005540176295
1.94,0.0004945235447
1.95,0.0004284481744
1.96,0.0003569559441
1.97,0.000281272953
1.98,0.0002026666754
1.99,0.0001224254104
2,4.183780231e-05
2.01,3.782725294e-05
2.02,0.000115340009
2.03,0.0001895282914
2.04,0.0002592946733
2.05,0.0003236321599
2.06,0.0003816381675
2.07,0.0004325266109
2.08,0.0004756379455
2.09,0.0005104470484
2.1,0.0005365688539
2.11,0.0005537616977
2.12,0.0005619283611
2.13,0.0005611148409
2.14,0.000551506906
2.15,0.0005334245359
2.16,0.0005073143663
2.17,0.0004737402969
2.18,0.0004333724411
2.19,0.0003869746224
2.2,0.0003353906369
2.21,0.0002795295217
2.22,0.0002203500763
2.23,0.0001588448931
2.24,9.602415681e-05
2.25,3.289946852e-05
2.26,2.95320504e-05
2.27,9.030313227e-05
2.28,0.0001484909829
2.29,0.0002032309319
2.3,0.000253728917
2.31,0.0002992726282
2.32,0.0003392411607
2.33,0.0003731130525
2.34,0.0004004726098
2.35,0.0004210144528
2.36,0.0004345462405
2.37,0.0004409895686
2.38,0.0004403790554
2.39,0.000432859667
2.4,0.0004186823532
2.41,0.0003981980949
2.42,0.0003718504861
2.43,0.0003401669929
2.44,0.0003037490529
2.45,0.0002632611915
2.46,0.0002194193453
2.47,0.0001729785912
2.48,0.0001247204865
2.49,7.544022627e-05
2.5,2.59338258e-05
2.51,2.301447087e-05
2.52,7.064436025e-05
2.53,0.0001162310419
2.54,0.000159096002
2.55,0.0001986168615
2.56,0.0002342361493
2.57,0.0002654688821
2.58,0.0002919088513
2.59,0.000313233542
2.6,0.0003292076301
2.61,0.00033968503
2.62,0.0003446094844
2.63,0.0003440137168
2.64,0.0003380171844
2.65,0.0003268224948
2.66,0.0003107105677
2.67,0.0002900346424
2.68,0.000265213248
2.69,0.0002367222683
2.7,0.0002050862449
2.71,0.0001708690718
2.72,0.0001346642413
2.73,9.708480701e-05
2.74,5.875322835e-05
2.75,2.029126262e-05
2.76,1.768993451e-05
2.77,5.459934555e-05
2.78,8.987511822e-05
2.79,0.0001229929479
2.8,0.0001534736933
2.81,0.0001808901166
2.82,0.0002048726584
2.83,0.0002251141703
2.84,0.0002413735514
2.85,0.0002534782501
2.86,0.0002613256126
2.87,0.0002648830794
2.88,0.0002641872487
2.89,0.0002593418442
2.9,0.000250514641
2.91,0.000237933421
2.92,0.0002218810413
2.93,0.0002026897136
2.94,0.0001807346036
2.95,0.0001564268673
2.96,0.0001302062483
2.97,0.000102533367
2.98,7.388183142e-05
2.99,4.473030357e-05
3,1.555465094e-05
3.01,1.317969028e-05
3.02,4.102501945e-05
3.03,6.75582306e-05
3.04,9.238712305e-05
3.05,0.0001151560713
3.06,0.0001355509741
3.07,0.000153303417
3.08,0.0001681939944
3.09,0.000180054757
3.1,0.0001887707586
3.11,0.0001942806995
3.12,0.0001965766718
3.13,0.000195703032
3.14,0.0001917544367
3.15,0.0001848730928
3.16,0.0001752452841
3.17,0.0001630972476
3.18,0.0001486904831
3.19,0.000132316586
3.2,0.0001142916989
3.21,9.495068636e-05
3.22,7.464113274e-05
3.23,5.371727053e-05
3.24,3.253394148e-05
3.25,1.144069193e-05
3.26,9.223901131e-06
3.27,2.913758267e-05
3.28,4.799936727e-05
3.29,6.553400732e-05
3.3,8.149591658e-05
3.31,9.567249473e-05
3.32,0.0001078868099
3.33,0.0001179996083
3.34,0.0001259106315
3.35,0.0001315592355
3.36,0.0001349243165
3.37,0.0001360235612
3.38,0.0001349120506
3.39,0.0001316802559
3.4,0.000126451476
3.41,0.0001193787754
3.42,0.0001106414856
3.43,0.0001004413433
3.44,8.899834172e-05
3.45,7.65463738e-05
3.46,6.332874962e-05
3.47,4.959367073e-05
3.48,3.55897419e-05
3.49,2.15615997e-05
3.5,7.745732679e-06
3.51,5.633436517e-06
3.52,1.836714338e-05
3.53,3.026549129e-05
3.54,4.116023022e-05
3.55,5.090706461e-05
3.56,5.938746004e-05
3.57,6.650992785e-05
3.58,7.22107772e-05
3.59,7.645433372e-05
3.6,7.923263369e-05
3.61,8.05646125e-05
3.62,8.049481459e-05
3.63,7.909166083e-05
3.64,7.644531655e-05
3.65,7.266520998e-05
3.66,6.787725666e-05
3.67,6.222084945e-05
3.68,5.584567732e-05
3.69,4.890843788e-05
3.7,4.156950961e-05
3.71,3.398964895e-05
3.72,2.632677605e-05
3.73,1.873290992e-05
3.74,1.135130965e-05
3.75,4.313873639e-06
3.76,2.261157403e-06
3.77,8.271154431e-06
3.78,1.363067926e-05
3.79,1.827266499e-05
3.8,2.2149182e-05
3.81,2.523178211e-05
3.82,2.751142222e-05
3.83,2.899797685e-05
3.84,2.971935714e-05
3.85,2.97202616e-05
3.86,2.906059084e-05
3.87,2.781356479e-05
3.88,2.606358682e-05
3.89,2.390390337e-05
3.9,2.143411147e-05
3.91,1.875756937e-05
3.92,1.597876649e-05
3.93,1.320070957e-05
3.94,1.052238105e-05
3.95,8.036323682e-06
3.96,5.826402578e-06
3.97,3.965791832e-06
3.98,2.515228028e-06
3.99,1.521567157e-06
4,1.016675155e-06
4.01,1.016675155e-06
4.02,1.521567156e-06
4.03,2.515228027e-06
4.04,3.965791831e-06
4.05,5.826402575e-06
4.06,8.03632368e-06
4.07,1.052238105e-05
4.08,1.320070957e-05
4.09,1.597876648e-05
4.1,1.875756937e-05
4.11,2.143411146e-05
4.12,2.390390336e-05
4.13,2.606358682e-05
4.14,2.781356478e-05
4.15,2.906059083e-05
4.16,2.97202616e-05
4.17,2.971935713e-05
4.18,2.899797684e-05
4.19,2.751142222e-05
4.2,2.523178211e-05
4.21,2.2149182e-05
4.22,1.827266499e-05
4.23,1.363067926e-05
4.24,8.27115443e-06
4.25,2.261157403e-06
4.26,4.313873639e-06
4.27,1.135130965e-05
4.28,1.873290992e-05
4.29,2.632677605e-05
4.3,3.398964894e-05
4.31,4.156950961e-05
4.32,4.890843788e-05
4.33,5.584567731e-05
4.34,6.222084944e-05
4.35,6.787725665e-05
4.36,7.266520997e-05
4.37,7.644531654e-05
4.38,7.909166083e-05
4.39,8.049481459e-05
4.4,8.05646125e-05
4.41,7.923263369e-05
4.42,7.645433372e-05
4.43,7.22107772e-05
4.44,6.650992784e-05
4.45,5.938746004e-05
4.46,5.090706461e-05
4.47,4.116023021e-05
4.48,3.026549129e-05
4.49,1.836714338e-05
4.5,5.633436518e-06
4.51,7.745732679e-06
4.52,2.15615997e-05
4.53,3.55897419e-05
4.54,4.959367072e-05
4.55,6.332874962e-05
4.56,7.654637379e-05
4.57,8.899834172e-05
4.58,0.0001004413433
4.59,0.0001106414856
4.6,0.0001193787754
4.61,0.000126451476
4.62,0.0001316802559
4.63,0.0001349120506
4.64,0.0001360235612
4.65,0.0001349243165
4.66,0.0001315592355
4.67,0.0001259106315
4.68,0.0001179996083
4.69,0.0001078868099
4.7,9.567249473e-05
4.71,8.149591658e-05
4.72,6.553400732e-05
4.73,4.799936726e-05
4.74,2.913758267e-05
4.75,9.223901131e-06
4.76,1.144069193e-05
4.77,3.253394148e-05
4.78,5.371727053e-05
4.79,7.464113274e-05
4.8,9.495068635e-05
4.81,0.0001142916989
4.82,0.0001323165859
4.83,0.0001486904831
4.84,0.0001630972476
4.85,0.0001752452841
4.86,0.0001848730928
4.87,0.0001917544367
4.88,0.000195703032
4.89,0.0001965766718
4.9,0.0001942806995
4.91,0.0001887707586
4.92,0.000180054757
4.93,0.0001681939944
4.94,0.000153303417
4.95,0.0001355509741
4.96,0.0001151560713
4.97,9.238712305e-05
4.98,6.755823059e-05
4.99,4.102501944e-05
5,1.317969028e-05
5.01,1.555465094e-05
5.02,4.473030357e-05
5.03,7.388183142e-05
5.04,0.000102533367
5.05,0.0001302062483
5.06,0.0001564268673
5.07,0.0001807346036
5.08,0.0002026897136
5.09,0.0002218810413
5.1,0.000237933421
5.11,0.000250514641
5.12,0.0002593418442
5.13,0.0002641872487
5.14,0.0002648830794
5.15,0.0002613256126
5.16,0.0002534782501
5.17,0.0002413735514
5.18,0.0002251141703
5.19,0.0002048726584
5.2,0.0001808901166
5.21,0.0001534736933
5.22,0.0001229929479
5.23,8.987511822e-05
5.24,5.459934555e-05
5.25,1.768993451e-05
5.26,2.029126262e-05
5.27,5.875322835e-05
5.28,9.708480701e-05
5.29,0.0001346642413
5.3,0.0001708690718
5.31,0.0002050862449
5.32,0.0002367222683
5.33,0.000265213248
5.34,0.0002900346424
5.35,0.0003107105677
5.36,0.0003268224948
5.37,0.0003380171844
5.38,0.0003440137168
5.39,0.0003446094844
5.4,0.00033968503
5.41,0.0003292076301
5.42,0.000313233542
5.43,0.0002919088513
5.44,0.0002654688821
5.45,0.0002342361493
5.46,0.0001986168615
5.47,0.000159096002
5.48,0.0001162310419
5.49,7.064436025e-05
5.5,2.301447087e-05
5.51,2.59338258e-05
5.52,7.544022627e-05
5.53,0.0001247204865
5.54,0.0001729785912
5.55,0.0002194193453
5.56,0.0002632611915
5.57,0.0003037490529
5.58,0.0003401669929
5.59,0.0003718504861
5.6,0.0003981980949
5.61,0.0004186823532
5.62,0.000432859667
5.63,0.0004403790554
5.64,0.0004409895686
5.65,0.0004345462405
5.66,0.0004210144528
5.67,0.0004004726098
5.68,0.0003731130524
5.69,0.0003392411607
5.7,0.0002992726282
5.71,0.000253728917
5.72,0.0002032309319
5.73,0.0001484909829
5.74,9.030313227e-05
5.75,2.95320504e-05
5.76,3.289946852e-05
5.77,9.602415681e-05
5.78,0.0001588448931
5.79,0.0002203500763
5.8,0.0002795295217
5.81,0.0003353906369
5.82,0.0003869746224
5.83,0.0004333724411
5.84,0.0004737402968
5.85,0.0005073143663
5.86,0.0005334245359
5.87,0.000551506906
5.88,0.0005611148408
5.89,0.0005619283611
5.9,0.0005537616977
5.91,0.0005365688539
5.92,0.0005104470484
5.93,0.0004756379454
5.94,0.0004325266109
5.95,0.0003816381675
5.96,0.0003236321599
5.97,0.0002592946733
5.98,0.0001895282914
5.99,0.000115340009
6,3.782725294e-05
6.01,4.183780231e-05
6.02,0.0001224254104
6.03,0.0002026666754
6.04,0.000281272953
6.05,0.0003569559441
6.06,0.0004284481744
6.07,0.0004945235447
6.08,0.0005540176295
6.09,0.0006058473965
6.1,0.0006490300262
6.11,0.0006827005169
6.12,0.0007061277729
6.13,0.0007187288943
6.14,0.000720081406
6.15,0.0007099331958
6.16,0.0006882099571
6.17,0.0006550199727
6.18,0.0006106561102
6.19,0.0005555949441
6.2,0.0004904929601
6.21,0.0004161798453
6.22,0.00033364891
6.23,0.0002440447352
6.24,0.0001486481845
6.25,4.885896292e-05
6.26,5.382405399e-05
6.27,0.0001578244563
6.28,0.0002615116019
6.29,0.0003632252537
6.3,0.0004613011927
6.31,0.0005540974265
6.32,0.0006400205947
6.33,0.0007175521628
6.34,0.0007852739907
6.35,0.0008418928627
6.36,0.0008862635757
6.37,0.0009174101939
6.38,0.0009345451016
6.39,0.0009370855091
6.4,0.0009246671027
6.41,0.0008971545656
6.42,0.0008546487383
6.43,0.0007974902387
6.44,0.000726259407
6.45,0.0006417725008
6.46,0.0005450741166
6.47,0.0004374258764
6.48,0.0003202914731
6.49,0.0001953182281
6.5,6.431537137e-05
6.51,7.077069078e-05
6.52,0.0002078838035
6.53,0.0003448878217
6.54,0.0004795982888
6.55,0.0006098155906
6.56,0.0007333591027
6.57,0.0008481018282
6.58,0.0009520050003
6.59,0.001043152118
6.6,0.001119781871
6.61,0.001180319435
6.62,0.001223405596
6.63,0.001247923235
6.64,0.001253020671
6.65,0.001238131465
6.66,0.001202990273
6.67,0.001147644425
6.68,0.001072460944
6.69,0.0009781288025
6.7,0.0008656562548
6.71,0.0007363631744
6.72,0.0005918683963
6.73,0.0004340721323
6.74,0.0002651336118
6.75,8.744416949e-05
6.76,9.640391891e-05
6.77,0.0002836524853
6.78,0.0004714159649
6.79,0.0006567228403
6.8,0.0008365594788
6.81,0.00100791576
6.82,0.001167831846
6.83,0.001313445417
6.84,0.001442038676
6.85,0.001551084395
6.86,0.001638290297
6.87,0.001701641057
6.88,0.001739437226
6.89,0.001750330422
6.9,0.001733354142
6.91,0.001687949651
6.92,0.001613986387
6.93,0.00151177647
6.94,0.001382082909
6.95,0.001226121234
6.96,0.001045554337
6.97,0.0008424804144
6.98,0.0006194140073
6.99,0.000379260212
7,0.0001252822668
7.01,0.000138937203
7.02,0.0004095408881
7.03,0.0006824469585
7.04,0.0009534041086
7.05,0.001218050819
7.06,0.001471978091
7.07,0.001710794844
7.08,0.001930195094
7.09,0.002126026009
7.1,0.002294355854
7.11,0.002431540877
7.12,0.002534290105
7.13,0.002599727082
7.14,0.002625447559
7.15,0.002609572195
7.16,0.002550793353
7.17,0.002448415161
7.18,0.00230238605
7.19,0.002113323076
7.2,0.001882527431
7.21,0.00161199064
7.22,0.001304391076
7.23,0.0009630805144
7.24,0.0005920606128
7.25,0.000195949302
7.26,0.000220062763
7.27,0.0006502654263
7.28,0.001088491516
7.29,0.001528188044
7.3,0.00196249545
7.31,0.002384334089
7.32,0.002786497046
7.33,0.003161748259
7.34,0.003502924831
7.35,0.003803042337
7.36,0.004055401846
7.37,0.004253697342
7.38,0.004392122191
7.39,0.00446547325
7.4,0.004469251264
7.41,0.004399756154
7.42,0.004254175876
7.43,0.004030667562
7.44,0.00372842972
7.45,0.003347764374
7.46,0.002890128104
7.47,0.002358171087
7.48,0.001755763352
7.49,0.001088007627
7.5,0.0003612382984
7.51,0.000416993829
7.52,0.001237951091
7.53,0.002091752808
7.54,0.002967433166
7.55,0.003853012725
7.56,0.004735582938
7.57,0.005601402885
7.58,0.006436007239
7.59,0.00722432432
7.6,0.007950802891
7.61,0.008599546212
7.62,0.009154451704
7.63,0.009599354421
7.64,0.009918172401
7.65,0.01009505182
7.66,0.01011450977
7.67,0.009961572311
7.68,0.009621905414
7.69,0.00908193611
7.7,0.008328961203
7.71,0.007351240545
7.72,0.006138071725
7.73,0.00467984266
7.74,0.002968058147
7.75,0.0009953358531
7.76,0.001244633669
7.77,0.00375716948
7.78,0.006546733537
7.79,0.009617125421
7.8,0.01297175263
7.81,0.01661399565
7.82,0.02054769445
7.83,0.02477779429
7.84,0.02931120572
7.85,0.03415796005
7.86,0.03933278357
7.87,0.04485728217
7.88,0.05076304211
7.89,0.05709615123
7.9,0.06392400145
7.91,0.07134590479
7.92,0.07951038651
7.93,0.08864482571
7.94,0.09910952521
7.95,0.1115038455
7.96,0.1269026348
7.97,0.1473578059
7.98,0.1787769699
7.99,0.2204490125
8,0.5829683585
};\addlegendentry{Current};

\end{axis}
\end{tikzpicture}

%% file: graphics/results/strip/curr_ws_3.tex
\begin{tikzpicture}
\centering
\begin{axis}[
width=1.85in,
height=1.25in,
scale only axis,
xlabel near ticks, ylabel near ticks,
grid style={line width=.1pt, draw=gray!20},major grid style={line width=.2pt,draw=gray!40},
minor tick num=4,
yticklabel style = {font=\small},
xticklabel style = {font=\small},
grid = major,
xmin=0,
xmax=8,
xlabel={\small Distance [$\mathrm{cm}$]},
every outer y axis line/.append style={black},
every y tick label/.append style={font=\color{black}},
every y tick/.append style={black},
ymin=0,
ymax=0.06,
legend style={legend cell align=left, align=left, draw=black,at={(0.5,0.7)},anchor=south, font=\small}
]

\addplot[color=blue, line width=1pt]
  table[col sep= comma]
  {
0.01,6.92681681e-05
0.02,0.0002726258885
0.03,0.0003859961238
0.04,0.0004936629792
0.05,0.0005986757787
0.06,0.0007039423182
0.07,0.000810727393
0.08,0.0009197085656
0.09,0.001031251593
0.1,0.001145543313
0.11,0.001262658459
0.12,0.001382596976
0.13,0.001505306487
0.14,0.001630696729
0.15,0.001758649392
0.16,0.001889025189
0.17,0.002021669189
0.18,0.002156415003
0.19,0.002293088212
0.2,0.002431509238
0.21,0.002571495819
0.22,0.002712865177
0.23,0.002855435931
0.24,0.002999029807
0.25,0.003143473156
0.26,0.003288598315
0.27,0.003434244797
0.28,0.003580260341
0.29,0.003726501816
0.3,0.003872835978
0.31,0.004019140091
0.32,0.004165302411
0.33,0.004311222541
0.34,0.004456811648
0.35,0.004601992563
0.36,0.004746699752
0.37,0.00489087918
0.38,0.005034488061
0.39,0.005177494504
0.4,0.005319877071
0.41,0.005461624248
0.42,0.005602733832
0.43,0.005743212263
0.44,0.005883073888
0.45,0.006022340182
0.46,0.006161038937
0.47,0.006299203413
0.48,0.006436871483
0.49,0.006574084769
0.5,0.00671088778
0.51,0.006847327064
0.52,0.006983450386
0.53,0.00711930593
0.54,0.007254941548
0.55,0.007390404046
0.56,0.007525738532
0.57,0.007660987814
0.58,0.007796191866
0.59,0.007931387354
0.6,0.008066607238
0.61,0.00820188044
0.62,0.008337231591
0.63,0.008472680838
0.64,0.008608243737
0.65,0.008743931202
0.66,0.008879749533
0.67,0.009015700502
0.68,0.009151781498
0.69,0.00928798573
0.7,0.009424302485
0.71,0.009560717421
0.72,0.009697212909
0.73,0.0098337684
0.74,0.009970360829
0.75,0.01010696503
0.76,0.01024355415
0.77,0.01038010013
0.78,0.0105165741
0.79,0.01065294683
0.8,0.01078918916
0.81,0.0109252724
0.82,0.01106116869
0.83,0.0111968514
0.84,0.01133229543
0.85,0.0114674775
0.86,0.01160237641
0.87,0.01173697325
0.88,0.0118712516
0.89,0.01200519762
0.9,0.01213880017
0.91,0.01227205084
0.92,0.01240494397
0.93,0.01253747658
0.94,0.01266964835
0.95,0.01280146146
0.96,0.01293292045
0.97,0.01306403207
0.98,0.01319480507
0.99,0.01332524994
1,0.01345537873
1.01,0.01358520471
1.02,0.01371474217
1.03,0.01384400609
1.04,0.0139730119
1.05,0.01410177518
1.06,0.01423031139
1.07,0.0143586356
1.08,0.01448676227
1.09,0.01461470499
1.1,0.01474247626
1.11,0.01487008734
1.12,0.01499754805
1.13,0.01512486663
1.14,0.01525204963
1.15,0.01537910186
1.16,0.01550602628
1.17,0.015632824
1.18,0.01575949428
1.19,0.01588603459
1.2,0.01601244059
1.21,0.01613870629
1.22,0.01626482411
1.23,0.01639078504
1.24,0.01651657874
1.25,0.01664219373
1.26,0.01676761756
1.27,0.01689283701
1.28,0.01701783823
1.29,0.01714260701
1.3,0.01726712891
1.31,0.01739138949
1.32,0.0175153745
1.33,0.01763907005
1.34,0.01776246282
1.35,0.01788554015
1.36,0.01800829027
1.37,0.01813070237
1.38,0.01825276675
1.39,0.0183744749
1.4,0.01849581957
1.41,0.01861679482
1.42,0.01873739608
1.43,0.01885762013
1.44,0.01897746511
1.45,0.01909693046
1.46,0.01921601694
1.47,0.01933472647
1.48,0.01945306211
1.49,0.01957102796
1.5,0.01968862902
1.51,0.01980587106
1.52,0.01992276055
1.53,0.02003930443
1.54,0.02015551004
1.55,0.02027138494
1.56,0.02038693675
1.57,0.02050217305
1.58,0.02061710119
1.59,0.02073172817
1.6,0.02084606054
1.61,0.02096010425
1.62,0.02107386454
1.63,0.02118734588
1.64,0.02130055186
1.65,0.02141348515
1.66,0.02152614742
1.67,0.02163853934
1.68,0.02175066054
1.69,0.02186250961
1.7,0.02197408413
1.71,0.0220853807
1.72,0.02219639495
1.73,0.02230712164
1.74,0.02241755472
1.75,0.02252768737
1.76,0.02263751217
1.77,0.02274702112
1.78,0.02285620581
1.79,0.02296505747
1.8,0.02307356713
1.81,0.02318172574
1.82,0.02328952422
1.83,0.02339695364
1.84,0.02350400532
1.85,0.02361067088
1.86,0.02371694241
1.87,0.02382281248
1.88,0.0239282743
1.89,0.02403332173
1.9,0.02413794934
1.91,0.0242421525
1.92,0.02434592735
1.93,0.02444927087
1.94,0.02455218088
1.95,0.024654656
1.96,0.02475669567
1.97,0.02485830011
1.98,0.02495947028
1.99,0.02506020782
2,0.02516051501
2.01,0.02526039466
2.02,0.02535985009
2.03,0.02545888502
2.04,0.02555750348
2.05,0.02565570971
2.06,0.02575350811
2.07,0.02585090312
2.08,0.0259478991
2.09,0.02604450032
2.1,0.0261407108
2.11,0.02623653426
2.12,0.02633197405
2.13,0.02642703308
2.14,0.02652171376
2.15,0.02661601794
2.16,0.02670994686
2.17,0.02680350118
2.18,0.02689668086
2.19,0.02698948524
2.2,0.02708191297
2.21,0.02717396206
2.22,0.0272656299
2.23,0.02735691324
2.24,0.02744780829
2.25,0.02753831073
2.26,0.02762841575
2.27,0.02771811813
2.28,0.02780741232
2.29,0.02789629247
2.3,0.0279847525
2.31,0.02807278622
2.32,0.02816038738
2.33,0.02824754971
2.34,0.02833426705
2.35,0.0284205334
2.36,0.02850634296
2.37,0.02859169023
2.38,0.02867657007
2.39,0.02876097772
2.4,0.02884490888
2.41,0.0289283597
2.42,0.0290113269
2.43,0.02909380768
2.44,0.02917579984
2.45,0.02925730171
2.46,0.0293383122
2.47,0.02941883073
2.48,0.02949885729
2.49,0.02957839234
2.5,0.02965743685
2.51,0.02973599218
2.52,0.0298140601
2.53,0.02989164272
2.54,0.02996874242
2.55,0.03004536181
2.56,0.03012150369
2.57,0.03019717092
2.58,0.03027236644
2.59,0.03034709315
2.6,0.03042135389
2.61,0.03049515133
2.62,0.030568488
2.63,0.03064136612
2.64,0.03071378768
2.65,0.03078575429
2.66,0.03085726721
2.67,0.03092832729
2.68,0.03099893498
2.69,0.03106909024
2.7,0.03113879263
2.71,0.03120804122
2.72,0.03127683463
2.73,0.03134517105
2.74,0.03141304823
2.75,0.03148046354
2.76,0.03154741394
2.77,0.03161389608
2.78,0.0316799063
2.79,0.03174544067
2.8,0.03181049506
2.81,0.03187506516
2.82,0.03193914657
2.83,0.0320027348
2.84,0.03206582538
2.85,0.03212841385
2.86,0.03219049587
2.87,0.03225206722
2.88,0.03231312388
2.89,0.03237366206
2.9,0.03243367823
2.91,0.03249316919
2.92,0.03255213204
2.93,0.03261056426
2.94,0.03266846371
2.95,0.03272582864
2.96,0.0327826577
2.97,0.03283894994
2.98,0.03289470483
2.99,0.0329499222
3,0.03300460228
3.01,0.03305874566
3.02,0.03311235323
3.03,0.03316542623
3.04,0.03321796612
3.05,0.03326997464
3.06,0.03332145368
3.07,0.03337240532
3.08,0.03342283173
3.09,0.03347273514
3.1,0.03352211782
3.11,0.03357098199
3.12,0.03361932983
3.13,0.03366716338
3.14,0.03371448457
3.15,0.0337612951
3.16,0.03380759649
3.17,0.03385338999
3.18,0.03389867657
3.19,0.03394345692
3.2,0.0339877314
3.21,0.03403150005
3.22,0.03407476258
3.23,0.03411751835
3.24,0.0341597664
3.25,0.03420150543
3.26,0.03424273385
3.27,0.03428344975
3.28,0.03432365094
3.29,0.034363335
3.3,0.03440249927
3.31,0.03444114091
3.32,0.0344792569
3.33,0.03451684413
3.34,0.03455389937
3.35,0.03459041938
3.36,0.03462640088
3.37,0.03466184065
3.38,0.03469673551
3.39,0.03473108242
3.4,0.03476487845
3.41,0.03479812087
3.42,0.03483080714
3.43,0.03486293496
3.44,0.03489450228
3.45,0.03492550734
3.46,0.03495594867
3.47,0.03498582508
3.48,0.03501513574
3.49,0.03504388008
3.5,0.0350720579
3.51,0.03509966928
3.52,0.0351267146
3.53,0.03515319455
3.54,0.03517911005
3.55,0.03520446232
3.56,0.03522925276
3.57,0.035253483
3.58,0.03527715482
3.59,0.03530027014
3.6,0.03532283099
3.61,0.03534483945
3.62,0.03536629767
3.63,0.03538720777
3.64,0.03540757183
3.65,0.03542739189
3.66,0.03544666986
3.67,0.03546540753
3.68,0.03548360652
3.69,0.03550126827
3.7,0.03551839401
3.71,0.03553498472
3.72,0.03555104115
3.73,0.03556656378
3.74,0.03558155282
3.75,0.03559600821
3.76,0.0356099296
3.77,0.03562331637
3.78,0.03563616764
3.79,0.03564848227
3.8,0.03566025886
3.81,0.0356714958
3.82,0.03568219128
3.83,0.03569234329
3.84,0.03570194967
3.85,0.03571100816
3.86,0.03571951636
3.87,0.03572747184
3.88,0.03573487212
3.89,0.03574171473
3.9,0.03574799723
3.91,0.03575371725
3.92,0.03575887252
3.93,0.03576346089
3.94,0.03576748038
3.95,0.03577092919
3.96,0.03577380573
3.97,0.03577610865
3.98,0.03577783682
3.99,0.03577898942
4,0.03577956586
4.01,0.03577956586
4.02,0.03577898942
4.03,0.03577783682
4.04,0.03577610865
4.05,0.03577380573
4.06,0.03577092919
4.07,0.03576748038
4.08,0.03576346089
4.09,0.03575887252
4.1,0.03575371725
4.11,0.03574799723
4.12,0.03574171473
4.13,0.03573487212
4.14,0.03572747184
4.15,0.03571951636
4.16,0.03571100816
4.17,0.03570194967
4.18,0.03569234329
4.19,0.03568219128
4.2,0.0356714958
4.21,0.03566025886
4.22,0.03564848227
4.23,0.03563616764
4.24,0.03562331637
4.25,0.0356099296
4.26,0.03559600821
4.27,0.03558155282
4.28,0.03556656378
4.29,0.03555104115
4.3,0.03553498472
4.31,0.03551839401
4.32,0.03550126827
4.33,0.03548360652
4.34,0.03546540753
4.35,0.03544666986
4.36,0.03542739189
4.37,0.03540757183
4.38,0.03538720777
4.39,0.03536629767
4.4,0.03534483945
4.41,0.03532283099
4.42,0.03530027014
4.43,0.03527715482
4.44,0.035253483
4.45,0.03522925276
4.46,0.03520446232
4.47,0.03517911005
4.48,0.03515319455
4.49,0.0351267146
4.5,0.03509966928
4.51,0.0350720579
4.52,0.03504388008
4.53,0.03501513574
4.54,0.03498582508
4.55,0.03495594867
4.56,0.03492550734
4.57,0.03489450228
4.58,0.03486293496
4.59,0.03483080714
4.6,0.03479812087
4.61,0.03476487845
4.62,0.03473108242
4.63,0.03469673551
4.64,0.03466184065
4.65,0.03462640088
4.66,0.03459041938
4.67,0.03455389937
4.68,0.03451684413
4.69,0.0344792569
4.7,0.03444114091
4.71,0.03440249927
4.72,0.034363335
4.73,0.03432365094
4.74,0.03428344975
4.75,0.03424273385
4.76,0.03420150543
4.77,0.0341597664
4.78,0.03411751835
4.79,0.03407476258
4.8,0.03403150005
4.81,0.0339877314
4.82,0.03394345692
4.83,0.03389867657
4.84,0.03385338999
4.85,0.03380759649
4.86,0.0337612951
4.87,0.03371448457
4.88,0.03366716338
4.89,0.03361932983
4.9,0.03357098199
4.91,0.03352211782
4.92,0.03347273514
4.93,0.03342283173
4.94,0.03337240532
4.95,0.03332145368
4.96,0.03326997464
4.97,0.03321796612
4.98,0.03316542623
4.99,0.03311235323
5,0.03305874566
5.01,0.03300460228
5.02,0.0329499222
5.03,0.03289470483
5.04,0.03283894994
5.05,0.0327826577
5.06,0.03272582864
5.07,0.03266846371
5.08,0.03261056426
5.09,0.03255213204
5.1,0.03249316919
5.11,0.03243367823
5.12,0.03237366206
5.13,0.03231312388
5.14,0.03225206722
5.15,0.03219049587
5.16,0.03212841385
5.17,0.03206582538
5.18,0.0320027348
5.19,0.03193914657
5.2,0.03187506516
5.21,0.03181049506
5.22,0.03174544067
5.23,0.0316799063
5.24,0.03161389608
5.25,0.03154741394
5.26,0.03148046354
5.27,0.03141304824
5.28,0.03134517105
5.29,0.03127683463
5.3,0.03120804122
5.31,0.03113879263
5.32,0.03106909024
5.33,0.03099893498
5.34,0.03092832729
5.35,0.03085726721
5.36,0.03078575429
5.37,0.03071378768
5.38,0.03064136612
5.39,0.030568488
5.4,0.03049515133
5.41,0.03042135389
5.42,0.03034709315
5.43,0.03027236644
5.44,0.03019717092
5.45,0.03012150369
5.46,0.03004536181
5.47,0.02996874242
5.48,0.02989164272
5.49,0.0298140601
5.5,0.02973599218
5.51,0.02965743685
5.52,0.02957839234
5.53,0.02949885729
5.54,0.02941883073
5.55,0.0293383122
5.56,0.02925730171
5.57,0.02917579984
5.58,0.02909380768
5.59,0.0290113269
5.6,0.0289283597
5.61,0.02884490888
5.62,0.02876097772
5.63,0.02867657007
5.64,0.02859169023
5.65,0.02850634296
5.66,0.0284205334
5.67,0.02833426705
5.68,0.02824754971
5.69,0.02816038738
5.7,0.02807278622
5.71,0.0279847525
5.72,0.02789629247
5.73,0.02780741232
5.74,0.02771811813
5.75,0.02762841575
5.76,0.02753831073
5.77,0.02744780829
5.78,0.02735691324
5.79,0.0272656299
5.8,0.02717396206
5.81,0.02708191297
5.82,0.02698948524
5.83,0.02689668086
5.84,0.02680350118
5.85,0.02670994686
5.86,0.02661601794
5.87,0.02652171376
5.88,0.02642703308
5.89,0.02633197405
5.9,0.02623653426
5.91,0.0261407108
5.92,0.02604450032
5.93,0.0259478991
5.94,0.02585090312
5.95,0.02575350811
5.96,0.02565570971
5.97,0.02555750348
5.98,0.02545888502
5.99,0.02535985009
6,0.02526039466
6.01,0.02516051501
6.02,0.02506020782
6.03,0.02495947028
6.04,0.02485830011
6.05,0.02475669567
6.06,0.024654656
6.07,0.02455218088
6.08,0.02444927087
6.09,0.02434592735
6.1,0.0242421525
6.11,0.02413794934
6.12,0.02403332173
6.13,0.0239282743
6.14,0.02382281248
6.15,0.02371694241
6.16,0.02361067088
6.17,0.02350400532
6.18,0.02339695364
6.19,0.02328952422
6.2,0.02318172574
6.21,0.02307356713
6.22,0.02296505747
6.23,0.02285620581
6.24,0.02274702112
6.25,0.02263751217
6.26,0.02252768737
6.27,0.02241755472
6.28,0.02230712164
6.29,0.02219639495
6.3,0.0220853807
6.31,0.02197408413
6.32,0.02186250961
6.33,0.02175066054
6.34,0.02163853934
6.35,0.02152614742
6.36,0.02141348515
6.37,0.02130055186
6.38,0.02118734588
6.39,0.02107386454
6.4,0.02096010425
6.41,0.02084606054
6.42,0.02073172817
6.43,0.02061710119
6.44,0.02050217305
6.45,0.02038693675
6.46,0.02027138494
6.47,0.02015551004
6.48,0.02003930443
6.49,0.01992276055
6.5,0.01980587106
6.51,0.01968862902
6.52,0.01957102796
6.53,0.01945306211
6.54,0.01933472647
6.55,0.01921601694
6.56,0.01909693046
6.57,0.01897746511
6.58,0.01885762013
6.59,0.01873739608
6.6,0.01861679482
6.61,0.01849581957
6.62,0.0183744749
6.63,0.01825276675
6.64,0.01813070237
6.65,0.01800829027
6.66,0.01788554015
6.67,0.01776246282
6.68,0.01763907005
6.69,0.0175153745
6.7,0.01739138949
6.71,0.01726712891
6.72,0.01714260701
6.73,0.01701783823
6.74,0.01689283701
6.75,0.01676761756
6.76,0.01664219373
6.77,0.01651657874
6.78,0.01639078504
6.79,0.01626482411
6.8,0.01613870629
6.81,0.01601244059
6.82,0.01588603459
6.83,0.01575949428
6.84,0.015632824
6.85,0.01550602628
6.86,0.01537910186
6.87,0.01525204963
6.88,0.01512486663
6.89,0.01499754805
6.9,0.01487008734
6.91,0.01474247626
6.92,0.01461470499
6.93,0.01448676227
6.94,0.0143586356
6.95,0.01423031139
6.96,0.01410177518
6.97,0.0139730119
6.98,0.01384400609
6.99,0.01371474217
7,0.01358520471
7.01,0.01345537873
7.02,0.01332524994
7.03,0.01319480507
7.04,0.01306403207
7.05,0.01293292045
7.06,0.01280146146
7.07,0.01266964835
7.08,0.01253747658
7.09,0.01240494397
7.1,0.01227205084
7.11,0.01213880017
7.12,0.01200519762
7.13,0.0118712516
7.14,0.01173697325
7.15,0.01160237641
7.16,0.0114674775
7.17,0.01133229543
7.18,0.0111968514
7.19,0.01106116869
7.2,0.0109252724
7.21,0.01078918916
7.22,0.01065294683
7.23,0.0105165741
7.24,0.01038010013
7.25,0.01024355415
7.26,0.01010696503
7.27,0.009970360829
7.28,0.0098337684
7.29,0.009697212909
7.3,0.009560717421
7.31,0.009424302485
7.32,0.00928798573
7.33,0.009151781498
7.34,0.009015700502
7.35,0.008879749533
7.36,0.008743931202
7.37,0.008608243737
7.38,0.008472680838
7.39,0.008337231591
7.4,0.00820188044
7.41,0.008066607238
7.42,0.007931387354
7.43,0.007796191866
7.44,0.007660987814
7.45,0.007525738532
7.46,0.007390404046
7.47,0.007254941548
7.48,0.00711930593
7.49,0.006983450386
7.5,0.006847327064
7.51,0.00671088778
7.52,0.006574084769
7.53,0.006436871483
7.54,0.006299203413
7.55,0.006161038937
7.56,0.006022340182
7.57,0.005883073888
7.58,0.005743212263
7.59,0.005602733832
7.6,0.005461624248
7.61,0.005319877071
7.62,0.005177494504
7.63,0.005034488061
7.64,0.00489087918
7.65,0.004746699752
7.66,0.004601992563
7.67,0.004456811648
7.68,0.004311222541
7.69,0.004165302411
7.7,0.004019140091
7.71,0.003872835978
7.72,0.003726501816
7.73,0.003580260341
7.74,0.003434244797
7.75,0.003288598315
7.76,0.003143473156
7.77,0.002999029807
7.78,0.002855435931
7.79,0.002712865177
7.8,0.002571495819
7.81,0.002431509238
7.82,0.002293088212
7.83,0.002156415003
7.84,0.002021669189
7.85,0.001889025189
7.86,0.001758649392
7.87,0.001630696729
7.88,0.001505306487
7.89,0.001382596976
7.9,0.001262658459
7.91,0.001145543313
7.92,0.001031251593
7.93,0.0009197085656
7.94,0.000810727393
7.95,0.0007039423182
7.96,0.0005986757787
7.97,0.0004936629792
7.98,0.0003859961238
7.99,0.0002726258885
8,6.92681681e-05
};\addlegendentry{Current};

\end{axis}
\end{tikzpicture}

%% file: graphics/results/strip/curr_ws_16.tex
\begin{tikzpicture}
\centering
\begin{axis}[
width=1.85in,
height=1.25in,
at={(0.758in,0.481in)},
scale only axis,
yticklabel style = {font=\small},
xticklabel style = {font=\small},
grid style={line width=.1pt, draw=gray!20},major grid style={line width=.2pt,draw=gray!40},
minor tick num=4,
grid = major,
xmin=0,
xmax=8,
xlabel={\small Distance [$\mathrm{cm}$]},
every outer y axis line/.append style={black},
every y tick label/.append style={font=\color{black}},
xlabel near ticks, ylabel near ticks,
every y tick/.append style={black},
ymin=0,
ymax=0.05,
legend style={legend cell align=left, align=left, draw=black,at={(0.5,0.7)},anchor=south, font=\small}
]

\addplot[color=blue, line width=1pt]
  table[col sep= comma]
  {
0.01,0.001018089891
0.02,0.0039790442
0.03,0.005574217075
0.04,0.007053313639
0.05,0.008459719898
0.06,0.009834068662
0.07,0.01119228315
0.08,0.01254136356
0.09,0.01388354115
0.1,0.01521829329
0.11,0.01654336704
0.12,0.01785535936
0.13,0.01915007389
0.14,0.02042275705
0.15,0.02166826492
0.16,0.02288118848
0.17,0.02405595258
0.18,0.02518689778
0.19,0.02626835041
0.2,0.02729468414
0.21,0.02826037519
0.22,0.02916005255
0.23,0.02998854378
0.24,0.03074091715
0.25,0.03141252017
0.26,0.03199901485
0.27,0.03249640962
0.28,0.03290108797
0.29,0.03320983391
0.3,0.03341985395
0.31,0.03352879593
0.32,0.0335347644
0.33,0.03343633267
0.34,0.0332325515
0.35,0.03292295452
0.36,0.03250756032
0.37,0.0319868714
0.38,0.03136186994
0.39,0.03063401063
0.4,0.02980521062
0.41,0.0288778367
0.42,0.02785469003
0.43,0.02673898835
0.44,0.02553434617
0.45,0.02424475291
0.46,0.02287454925
0.47,0.02142840206
0.48,0.01991127795
0.49,0.01832841578
0.5,0.01668529833
0.51,0.01498762336
0.52,0.01324127431
0.53,0.01145229079
0.54,0.009626839156
0.55,0.007771183334
0.56,0.005891656039
0.57,0.003994630638
0.58,0.002086493767
0.59,0.0001736188594
0.6,0.001737659276
0.61,0.003641068736
0.62,0.005530423553
0.63,0.0073996454
0.64,0.009242783923
0.65,0.01105403564
0.66,0.01282776142
0.67,0.01455850246
0.68,0.01624099484
0.69,0.01787018268
0.7,0.01944122982
0.71,0.02094953023
0.72,0.02239071708
0.73,0.0237606706
0.74,0.02505552474
0.75,0.02627167282
0.76,0.02740577208
0.77,0.02845474744
0.78,0.02941579428
0.79,0.03028638061
0.8,0.03106424847
0.81,0.03174741482
0.82,0.03233417189
0.83,0.03282308705
0.84,0.03321300242
0.85,0.033503034
0.86,0.03369257067
0.87,0.03378127289
0.88,0.03376907119
0.89,0.03365616453
0.9,0.03344301843
0.91,0.03313036298
0.92,0.03271919067
0.93,0.03221075404
0.94,0.03160656311
0.95,0.03090838259
0.96,0.03011822881
0.97,0.0292383664
0.98,0.02827130452
0.99,0.02721979275
1,0.02608681657
1.01,0.02487559225
1.02,0.02358956126
1.03,0.02223238402
1.04,0.0208079331
1.05,0.0193202856
1.06,0.01777371489
1.07,0.01617268157
1.08,0.0145218236
1.09,0.01282594566
1.1,0.01109000764
1.11,0.009319112369
1.12,0.007518492396
1.13,0.005693496091
1.14,0.003849572868
1.15,0.001992257679
1.16,0.0001271547816
1.17,0.001740079149
1.18,0.003603752529
1.19,0.005458156902
1.2,0.00729758545
1.21,0.009116351876
1.22,0.01090880958
1.23,0.01266937105
1.24,0.01439252731
1.25,0.01607286741
1.26,0.01770509776
1.27,0.01928406128
1.28,0.02080475624
1.29,0.0222623546
1.3,0.02365221986
1.31,0.02496992429
1.32,0.02621126538
1.33,0.02737228148
1.34,0.0284492665
1.35,0.02943878366
1.36,0.03033767807
1.37,0.03114308828
1.38,0.03185245655
1.39,0.0324635379
1.4,0.03297440783
1.41,0.0333834688
1.42,0.03368945529
1.43,0.03389143751
1.44,0.03398882381
1.45,0.03398136166
1.46,0.03386913737
1.47,0.03365257443
1.48,0.03333243059
1.49,0.03290979373
1.5,0.03238607646
1.51,0.03176300967
1.52,0.03104263489
1.53,0.03022729568
1.54,0.02931962806
1.55,0.02832255001
1.56,0.02723925019
1.57,0.02607317583
1.58,0.02482802004
1.59,0.02350770842
1.6,0.0221163852
1.61,0.02065839885
1.62,0.01913828733
1.63,0.01756076301
1.64,0.01593069724
1.65,0.01425310482
1.66,0.01253312821
1.67,0.01077602168
1.68,0.00898713538
1.69,0.007171899375
1.7,0.005335807738
1.71,0.003484402659
1.72,0.001623258649
1.73,0.0002420331415
1.74,0.002105880475
1.75,0.003962705416
1.76,0.005806959552
1.77,0.007633139019
1.78,0.009435799321
1.79,0.01120956993
1.8,0.01294916866
1.81,0.01464941578
1.82,0.01630524789
1.83,0.01791173152
1.84,0.01946407643
1.85,0.02095764863
1.86,0.02238798309
1.87,0.02375079609
1.88,0.02504199725
1.89,0.02625770112
1.9,0.02739423848
1.91,0.02844816708
1.92,0.029416282
1.93,0.03029562549
1.94,0.03108349633
1.95,0.03177745854
1.96,0.03237534955
1.97,0.03287528774
1.98,0.03327567924
1.99,0.03357522408
2,0.03377292155
2.01,0.03386807472
2.02,0.03386029418
2.03,0.03374950087
2.04,0.03353592803
2.05,0.03322012214
2.06,0.03280294295
2.07,0.03228556247
2.08,0.03166946295
2.09,0.03095643382
2.1,0.03014856759
2.11,0.02924825462
2.12,0.02825817695
2.13,0.02718130093
2.14,0.02602086892
2.15,0.02478038991
2.16,0.02346362909
2.17,0.02207459653
2.18,0.02061753485
2.19,0.01909690599
2.2,0.01751737716
2.21,0.01588380596
2.22,0.01420122473
2.23,0.01247482427
2.24,0.01070993694
2.25,0.008912019111
2.26,0.007086633303
2.27,0.005239429794
2.28,0.003376127981
2.29,0.001502497479
2.3,0.0003756609311
2.31,0.00225253443
2.32,0.00412231728
2.33,0.005979229862
2.34,0.007817537542
2.35,0.009631569305
2.36,0.01141573606
2.37,0.01316454858
2.38,0.01487263495
2.39,0.01653475754
2.4,0.01814582938
2.41,0.0197009299
2.42,0.02119531999
2.43,0.02262445633
2.44,0.02398400499
2.45,0.02526985413
2.46,0.026478126
2.47,0.02760518795
2.48,0.02864766266
2.49,0.02960243743
2.5,0.03046667255
2.51,0.03123780878
2.52,0.03191357393
2.53,0.03249198849
2.54,0.03297137034
2.55,0.03335033858
2.56,0.03362781647
2.57,0.03380303347
2.58,0.03387552636
2.59,0.0338451396
2.6,0.03371202477
2.61,0.03347663922
2.62,0.0331397439
2.63,0.03270240039
2.64,0.0321659672
2.65,0.03153209529
2.66,0.03080272282
2.67,0.02998006926
2.68,0.02906662873
2.69,0.02806516268
2.7,0.02697869194
2.71,0.02581048805
2.72,0.02456406403
2.73,0.02324316444
2.74,0.02185175495
2.75,0.02039401124
2.76,0.01887430733
2.77,0.01729720345
2.78,0.01566743325
2.79,0.0139898906
2.8,0.01226961584
2.81,0.01051178159
2.82,0.008721678104
2.83,0.006904698198
2.84,0.005066321822
2.85,0.003212100244
2.86,0.001347639923
2.87,0.000521413897
2.88,0.002389393845
2.89,0.004250627276
2.9,0.006099453288
2.91,0.007930239841
2.92,0.009737400933
2.93,0.01151541378
2.94,0.01325883597
2.95,0.01496232242
2.96,0.0166206423
2.97,0.01822869561
2.98,0.01978152951
2.99,0.02127435429
3,0.02270255894
3.01,0.02406172619
3.02,0.02534764706
3.03,0.02655633475
3.04,0.02768403793
3.05,0.02872725322
3.06,0.02968273698
3.07,0.0305475162
3.08,0.03131889851
3.09,0.03199448135
3.1,0.03257216002
3.11,0.03305013485
3.12,0.03342691728
3.13,0.03370133489
3.14,0.03387253526
3.15,0.03393998888
3.16,0.03390349078
3.17,0.03376316113
3.18,0.03351944469
3.19,0.03317310914
3.2,0.03272524229
3.21,0.03217724823
3.22,0.03153084236
3.23,0.03078804541
3.24,0.02995117644
3.25,0.02902284487
3.26,0.02800594152
3.27,0.02690362883
3.28,0.02571933016
3.29,0.02445671835
3.3,0.02311970346
3.31,0.02171241988
3.32,0.02023921279
3.33,0.01870462396
3.34,0.01711337715
3.35,0.01547036294
3.36,0.01378062317
3.37,0.01204933507
3.38,0.01028179506
3.39,0.008483402272
3.4,0.006659641978
3.41,0.004816068796
3.42,0.002958289855
3.43,0.001091947928
3.44,0.0007772954384
3.45,0.002643776717
3.46,0.004501847099
3.47,0.006345889048
3.48,0.008170332662
3.49,0.009969671821
3.5,0.01173848008
3.51,0.01347142624
3.52,0.01516328967
3.53,0.01680897518
3.54,0.01840352756
3.55,0.01994214573
3.56,0.02142019639
3.57,0.02283322727
3.58,0.02417697985
3.59,0.02544740152
3.6,0.02664065733
3.61,0.02775314103
3.62,0.0287814856
3.63,0.02972257314
3.64,0.03057354414
3.65,0.03133180606
3.66,0.0319950412
3.67,0.03256121397
3.68,0.03302857724
3.69,0.0333956781
3.7,0.03366136274
3.71,0.03382478055
3.72,0.03388538737
3.73,0.03384294796
3.74,0.03369753748
3.75,0.03344954218
3.76,0.03309965915
3.77,0.03264889519
3.78,0.03209856462
3.79,0.03145028633
3.8,0.03070597977
3.81,0.02986785999
3.82,0.02893843177
3.83,0.0279204828
3.84,0.0268170759
3.85,0.02563154033
3.86,0.02436746216
3.87,0.02302867382
3.88,0.02161924271
3.89,0.02014345901
3.9,0.0186058227
3.91,0.0170110298
3.92,0.01536395788
3.93,0.01366965094
3.94,0.01193330361
3.95,0.01016024484
3.96,0.008355921043
3.97,0.006525878779
3.98,0.004675747113
3.99,0.002811219594
4,0.0009380360139
4.01,0.000938036014
4.02,0.002811219594
4.03,0.004675747113
4.04,0.006525878779
4.05,0.008355921043
4.06,0.01016024484
4.07,0.01193330361
4.08,0.01366965094
4.09,0.01536395788
4.1,0.0170110298
4.11,0.0186058227
4.12,0.02014345901
4.13,0.02161924271
4.14,0.02302867382
4.15,0.02436746216
4.16,0.02563154033
4.17,0.0268170759
4.18,0.0279204828
4.19,0.02893843177
4.2,0.02986785999
4.21,0.03070597977
4.22,0.03145028633
4.23,0.03209856462
4.24,0.03264889519
4.25,0.03309965915
4.26,0.03344954218
4.27,0.03369753748
4.28,0.03384294796
4.29,0.03388538737
4.3,0.03382478055
4.31,0.03366136274
4.32,0.0333956781
4.33,0.03302857724
4.34,0.03256121397
4.35,0.0319950412
4.36,0.03133180606
4.37,0.03057354414
4.38,0.02972257314
4.39,0.0287814856
4.4,0.02775314103
4.41,0.02664065733
4.42,0.02544740152
4.43,0.02417697985
4.44,0.02283322727
4.45,0.02142019639
4.46,0.01994214573
4.47,0.01840352756
4.48,0.01680897518
4.49,0.01516328967
4.5,0.01347142624
4.51,0.01173848008
4.52,0.009969671821
4.53,0.008170332662
4.54,0.006345889048
4.55,0.004501847099
4.56,0.002643776717
4.57,0.0007772954383
4.58,0.001091947928
4.59,0.002958289855
4.6,0.004816068796
4.61,0.006659641978
4.62,0.008483402272
4.63,0.01028179506
4.64,0.01204933507
4.65,0.01378062317
4.66,0.01547036294
4.67,0.01711337715
4.68,0.01870462396
4.69,0.02023921279
4.7,0.02171241988
4.71,0.02311970346
4.72,0.02445671835
4.73,0.02571933016
4.74,0.02690362883
4.75,0.02800594152
4.76,0.02902284487
4.77,0.02995117644
4.78,0.03078804541
4.79,0.03153084236
4.8,0.03217724823
4.81,0.03272524229
4.82,0.03317310914
4.83,0.03351944469
4.84,0.03376316113
4.85,0.03390349078
4.86,0.03393998888
4.87,0.03387253526
4.88,0.03370133489
4.89,0.03342691728
4.9,0.03305013485
4.91,0.03257216002
4.92,0.03199448135
4.93,0.03131889851
4.94,0.0305475162
4.95,0.02968273698
4.96,0.02872725322
4.97,0.02768403793
4.98,0.02655633475
4.99,0.02534764706
5,0.02406172619
5.01,0.02270255894
5.02,0.02127435429
5.03,0.01978152951
5.04,0.01822869561
5.05,0.0166206423
5.06,0.01496232242
5.07,0.01325883597
5.08,0.01151541378
5.09,0.009737400933
5.1,0.007930239841
5.11,0.006099453288
5.12,0.004250627276
5.13,0.002389393845
5.14,0.0005214138971
5.15,0.001347639923
5.16,0.003212100244
5.17,0.005066321822
5.18,0.006904698198
5.19,0.008721678104
5.2,0.01051178159
5.21,0.01226961584
5.22,0.0139898906
5.23,0.01566743325
5.24,0.01729720345
5.25,0.01887430733
5.26,0.02039401124
5.27,0.02185175495
5.28,0.02324316444
5.29,0.02456406403
5.3,0.02581048805
5.31,0.02697869194
5.32,0.02806516268
5.33,0.02906662873
5.34,0.02998006926
5.35,0.03080272282
5.36,0.03153209529
5.37,0.0321659672
5.38,0.03270240039
5.39,0.0331397439
5.4,0.03347663922
5.41,0.03371202477
5.42,0.0338451396
5.43,0.03387552636
5.44,0.03380303347
5.45,0.03362781647
5.46,0.03335033858
5.47,0.03297137034
5.48,0.03249198849
5.49,0.03191357393
5.5,0.03123780878
5.51,0.03046667255
5.52,0.02960243743
5.53,0.02864766266
5.54,0.02760518795
5.55,0.026478126
5.56,0.02526985413
5.57,0.02398400499
5.58,0.02262445633
5.59,0.02119531999
5.6,0.0197009299
5.61,0.01814582938
5.62,0.01653475754
5.63,0.01487263495
5.64,0.01316454858
5.65,0.01141573606
5.66,0.009631569305
5.67,0.007817537542
5.68,0.005979229862
5.69,0.00412231728
5.7,0.00225253443
5.71,0.0003756609311
5.72,0.001502497479
5.73,0.003376127981
5.74,0.005239429794
5.75,0.007086633303
5.76,0.008912019111
5.77,0.01070993694
5.78,0.01247482427
5.79,0.01420122473
5.8,0.01588380596
5.81,0.01751737716
5.82,0.01909690599
5.83,0.02061753485
5.84,0.02207459653
5.85,0.02346362909
5.86,0.02478038991
5.87,0.02602086892
5.88,0.02718130093
5.89,0.02825817695
5.9,0.02924825462
5.91,0.03014856759
5.92,0.03095643382
5.93,0.03166946295
5.94,0.03228556247
5.95,0.03280294295
5.96,0.03322012214
5.97,0.03353592803
5.98,0.03374950087
5.99,0.03386029418
6,0.03386807472
6.01,0.03377292155
6.02,0.03357522408
6.03,0.03327567924
6.04,0.03287528774
6.05,0.03237534955
6.06,0.03177745854
6.07,0.03108349633
6.08,0.03029562549
6.09,0.029416282
6.1,0.02844816708
6.11,0.02739423848
6.12,0.02625770112
6.13,0.02504199725
6.14,0.02375079609
6.15,0.02238798309
6.16,0.02095764863
6.17,0.01946407643
6.18,0.01791173152
6.19,0.01630524789
6.2,0.01464941578
6.21,0.01294916866
6.22,0.01120956993
6.23,0.009435799321
6.24,0.007633139019
6.25,0.005806959552
6.26,0.003962705416
6.27,0.002105880475
6.28,0.0002420331415
6.29,0.001623258649
6.3,0.003484402659
6.31,0.005335807738
6.32,0.007171899375
6.33,0.00898713538
6.34,0.01077602168
6.35,0.01253312821
6.36,0.01425310482
6.37,0.01593069724
6.38,0.01756076301
6.39,0.01913828733
6.4,0.02065839885
6.41,0.0221163852
6.42,0.02350770842
6.43,0.02482802004
6.44,0.02607317583
6.45,0.02723925019
6.46,0.02832255001
6.47,0.02931962806
6.48,0.03022729568
6.49,0.03104263489
6.5,0.03176300967
6.51,0.03238607646
6.52,0.03290979373
6.53,0.03333243059
6.54,0.03365257443
6.55,0.03386913737
6.56,0.03398136166
6.57,0.03398882381
6.58,0.03389143751
6.59,0.03368945529
6.6,0.0333834688
6.61,0.03297440783
6.62,0.0324635379
6.63,0.03185245655
6.64,0.03114308828
6.65,0.03033767807
6.66,0.02943878366
6.67,0.0284492665
6.68,0.02737228148
6.69,0.02621126538
6.7,0.02496992429
6.71,0.02365221986
6.72,0.0222623546
6.73,0.02080475624
6.74,0.01928406128
6.75,0.01770509776
6.76,0.01607286741
6.77,0.01439252731
6.78,0.01266937105
6.79,0.01090880958
6.8,0.009116351876
6.81,0.00729758545
6.82,0.005458156902
6.83,0.003603752529
6.84,0.001740079149
6.85,0.0001271547816
6.86,0.001992257679
6.87,0.003849572868
6.88,0.005693496091
6.89,0.007518492396
6.9,0.009319112369
6.91,0.01109000764
6.92,0.01282594566
6.93,0.0145218236
6.94,0.01617268157
6.95,0.01777371489
6.96,0.0193202856
6.97,0.0208079331
6.98,0.02223238402
6.99,0.02358956126
7,0.02487559225
7.01,0.02608681657
7.02,0.02721979275
7.03,0.02827130452
7.04,0.0292383664
7.05,0.03011822881
7.06,0.03090838259
7.07,0.03160656311
7.08,0.03221075404
7.09,0.03271919067
7.1,0.03313036298
7.11,0.03344301843
7.12,0.03365616453
7.13,0.03376907119
7.14,0.03378127289
7.15,0.03369257067
7.16,0.033503034
7.17,0.03321300242
7.18,0.03282308705
7.19,0.03233417189
7.2,0.03174741482
7.21,0.03106424847
7.22,0.03028638061
7.23,0.02941579428
7.24,0.02845474744
7.25,0.02740577208
7.26,0.02627167282
7.27,0.02505552474
7.28,0.0237606706
7.29,0.02239071708
7.3,0.02094953023
7.31,0.01944122982
7.32,0.01787018268
7.33,0.01624099484
7.34,0.01455850246
7.35,0.01282776142
7.36,0.01105403564
7.37,0.009242783923
7.38,0.0073996454
7.39,0.005530423553
7.4,0.003641068736
7.41,0.001737659276
7.42,0.0001736188594
7.43,0.002086493767
7.44,0.003994630638
7.45,0.005891656039
7.46,0.007771183334
7.47,0.009626839156
7.48,0.01145229079
7.49,0.01324127431
7.5,0.01498762336
7.51,0.01668529833
7.52,0.01832841578
7.53,0.01991127795
7.54,0.02142840206
7.55,0.02287454925
7.56,0.02424475291
7.57,0.02553434617
7.58,0.02673898835
7.59,0.02785469003
7.6,0.0288778367
7.61,0.02980521062
7.62,0.03063401063
7.63,0.03136186994
7.64,0.0319868714
7.65,0.03250756032
7.66,0.03292295452
7.67,0.0332325515
7.68,0.03343633267
7.69,0.0335347644
7.7,0.03352879593
7.71,0.03341985395
7.72,0.03320983391
7.73,0.03290108797
7.74,0.03249640962
7.75,0.03199901485
7.76,0.03141252017
7.77,0.03074091715
7.78,0.02998854378
7.79,0.02916005255
7.8,0.02826037519
7.81,0.02729468414
7.82,0.02626835041
7.83,0.02518689778
7.84,0.02405595258
7.85,0.02288118848
7.86,0.02166826492
7.87,0.02042275705
7.88,0.01915007389
7.89,0.01785535936
7.9,0.01654336704
7.91,0.01521829329
7.92,0.01388354115
7.93,0.01254136356
7.94,0.01119228315
7.95,0.009834068662
7.96,0.008459719898
7.97,0.007053313639
7.98,0.005574217075
7.99,0.0039790442
8,0.001018089891
};\addlegendentry{Current};

\end{axis}
\end{tikzpicture}

%% file: graphics/results/strip/delays.tex
\begin{tikzpicture}
\centering
\begin{axis}[width=2.5in,
height=2in,
xlabel near ticks,
ylabel near ticks,
yticklabel style = {font=\small},
xticklabel style = {font=\small},
at={(0.758in,0.481in)},
scale only axis,
grid style={line width=.1pt, draw=gray!20},major grid style={line width=.2pt,draw=gray!40},
minor tick num=4,
grid = major,
xmin=0,
xmax=10,
xlabel={\small WS mode \#},
every outer y axis line/.append style={black},
every y tick label/.append style={font=\color{black}},
every y tick/.append style={black},
ymin=-10,
ymax=2,
ylabel={\small Spatial shift [cm]},
legend style={legend cell align=left, align=left, draw=black,at={(0.18,0.7)},anchor=south}
]

\addplot[color=blue, line width=1pt]
  table[col sep=comma]
  {
1,-8.077352155
2,-8.072980972
3,-0.005738880471
4,-0.005730080038
5,-0.005715380975
6,-0.005694741786
7,-0.005668089771
8,-0.005635352382
9,-0.005596403579
10,-0.005551139569
};

\draw[red, very thick] (axis cs: 2.5, -8.5) -- (axis cs: 2.5, 1.5);
\node at (axis cs: 1, -4) {\large {\color{red} a}};
\node at (axis cs: 5, 1) {\large {\color{red} b}};
\end{axis}
\end{tikzpicture}

%% file: graphics/results/cavity/delays.tex
\begin{tikzpicture}[spy using outlines=
	{rectangle, magnification=14, connect spies}]
\centering
\begin{axis}[width=2.5in,
height=2in,
xlabel near ticks,
ylabel near ticks,
yticklabel style = {font=\small},
xticklabel style = {font=\small},
at={(0.758in,0.481in)},
scale only axis,
grid style={line width=.1pt, draw=gray!20},major grid style={line width=.2pt,draw=gray!40},
minor tick num=4,
grid = major,
xmin=0,
xmax=125,
xlabel={\small WS mode \#},
every outer y axis line/.append style={black},
every y tick label/.append style={font=\color{black}},
every y tick/.append style={black},
ymin=-10,
ymax=60,
ylabel={\small Spatial shift [cm]},
legend style={legend cell align=left, align=left, draw=black,at={(0.18,0.7)},anchor=south}
]

\addplot[color=blue, line width=1pt]
  table[col sep=comma]
  {
1,-9.280399423
2,-9.276294128
3,-9.275506318
4,-9.262884818
5,-8.466398353
6,-8.460092529
7,-8.421213792
8,-8.40865379
9,-8.405059629
10,-8.395324638
11,-8.380841598
12,-8.37496126
13,-8.343439763
14,-8.335635644
15,-8.324018841
16,-8.23869745
17,-8.22434703
18,-8.204801295
19,-8.118020962
20,-8.096113243
21,-8.078538983
22,-8.042783131
23,-8.036758335
24,-7.910529545
25,-7.891668003
26,-7.84594748
27,-7.72645602
28,-7.671643363
29,-7.645320394
30,-7.566480228
31,-7.542746235
32,-7.381110051
33,-7.343156315
34,-7.335352452
35,-7.181368464
36,-7.112283039
37,-7.047583083
38,-6.880019391
39,-6.878895812
40,-6.761392911
41,-6.69211739
42,-6.639146807
43,-6.432075954
44,-6.326267577
45,-6.237878368
46,-6.003361363
47,-5.954103819
48,-5.739793139
49,-5.688347001
50,-5.645268731
51,-5.401332715
52,-5.218532546
53,-5.094002786
54,-4.796552967
55,-4.672591939
56,-4.411646077
57,-4.238102089
58,-3.862193696
59,-3.834326468
60,-3.374596456
61,-3.343135791
62,-2.755386738
63,-2.58795047
64,-1.98595281
65,-1.761528259
66,-1.189217012
67,-1.094632902
68,-0.5770696434
69,-0.55741385
70,-0.2440069696
71,-0.2121340068
72,-0.08525772926
73,-0.06867795346
74,-0.02356684937
75,-0.02084697805
76,-0.005955233283
77,-0.00558866978
78,-0.001443721081
79,-0.001280656829
80,-0.000316619199
81,-0.0002711974172
82,-6.14811529e-05
83,-5.572648466e-05
84,-1.120919692e-05
85,-1.060232253e-05
86,-1.985795518e-06
87,-1.810072033e-06
88,-3.279707894e-07
89,-2.906436001e-07
90,-4.926929544e-08
91,-4.541436472e-08
92,-7.019864172e-09
93,-6.67755336e-09
94,-9.738684081e-10
95,-9.010451186e-10
96,-1.275341997e-10
97,-1.153404527e-10
98,-1.545804371e-11
99,-1.438116526e-11
100,-1.787145728e-12
101,-1.70296991e-12
102,-2.014324627e-13
103,-1.879961333e-13
104,-2.194366119e-14
105,-1.996924807e-14
106,-4.616576455e-15
107,-4.016108709e-15
108,-3.457520409e-15
109,-3.245850186e-15
110,-2.022980828e-15
111,2.677245559e-15
112,2.69869062e-15
113,2.833065531e-15
114,3.032300748e-15
115,3.215330089e-15
116,3.290520449e-15
117,1.888343302
118,35.37685749
119,51.72022009
120,61.59436554
};
  \coordinate (spypoint) at (axis cs:4,-8);
  \coordinate (magnifyglass) at (axis cs:40,30);
  
  \draw[red, very thick] (axis cs: 4.5, -9.4) -- (axis cs: 4.5, 15);

  \draw[red, very thick] (axis cs: 116, -4) -- (axis cs: 116, 45);

  \draw[red, very thick] (axis cs: 70, -7) -- (axis cs: 70, 10);

    
    \node at (axis cs: 30, -2) {\large {\color{red} b}};

    \node at (axis cs: 90, 3) {\large {\color{red} d}};

  \node at (axis cs: 121, 5) {\large {\color{red} c}};

\end{axis}

\node[pin={[pin distance=1.8cm]40:{%
        \begin{tikzpicture}[baseline,trim axis left,trim axis right]
            \begin{axis}[
                    no markers,
                    every axis plot post/.append style={thick},
                    footnotesize,
                    xmin=1,xmax=8,
                    ymin=-9.5,ymax=-8,
                ]
               
                \addplot  coordinates {( 1,-9.280399423) (2,-9.276294128) (3,-9.275506318) (4,-9.262884818) (5,-8.466398353) (6,-8.460092529) (7,-8.421213792) (8,-8.40865379
                )};
                
                \draw[red, very thick] (axis cs: 4.5, -9.4) -- (axis cs: 4.5, -8.2);
                \node at (axis cs: 2.5, -9) {\large {\color{red} a}};
                \node at (axis cs: 6, -8.4) {\large {\color{red} b}};
            \end{axis}
        \end{tikzpicture}%
    }},draw,circle,minimum size=0.75cm] at (spypoint) {};
    

\end{tikzpicture}

%% file: Plots/TwoDipole_delays.tex
\begin{tikzpicture}[spy using outlines=
	{rectangle, magnification=8, connect spies}, thick,every node/.style={scale=1.0}]
\centering
\begin{axis}[width=3.0in,
height=2.5in,
xlabel near ticks,
ylabel near ticks,
yticklabel style = {font=\small},
xticklabel style = {font=\small},
compat=newest,
minor tick num=4,
xlabel = {\small WS mode \#}, ylabel = {\small Spatial shift [m]}, ylabel shift = 0 pt, xmin = 00, xmax = 100, ymin = -0.1, ymax = 3]
\addplot+[ycomb, mark size = 1pt] plot table[x expr=\coordindex+1, y=Q]{Plots/Twodipoles_Q.csv};
 \coordinate (spypoint) at (axis cs:0,0);
\end{axis}

\node[pin={[pin distance=0.5cm]70:{%
     \begin{tikzpicture}[scale=0.70, every node/.style={scale=1}]
\begin{axis}[width=2.0in,
 height=1.5in,
compat=newest, xmin = 00, xmax = 4, ymin = -0.02, ymax = 0, ytick = {-0.02, -0.01, 0}]
\addplot+[ycomb, mark size = 1pt] plot table[x expr=\coordindex+1, y=Q]{Plots/Twodipoles_Q.csv};
\end{axis}
\end{tikzpicture}
        
     }},draw,circle,minimum size=0.5cm] at (spypoint) {};
\end{tikzpicture}

%% file: jnl-2020-tap-WS-1.bbl
\begin{thebibliography}{10}
\providecommand{\url}[1]{#1}
\csname url@samestyle\endcsname
\providecommand{\newblock}{\relax}
\providecommand{\bibinfo}[2]{#2}
\providecommand{\BIBentrySTDinterwordspacing}{\spaceskip=0pt\relax}
\providecommand{\BIBentryALTinterwordstretchfactor}{4}
\providecommand{\BIBentryALTinterwordspacing}{\spaceskip=\fontdimen2\font plus
\BIBentryALTinterwordstretchfactor\fontdimen3\font minus
  \fontdimen4\font\relax}
\providecommand{\BIBforeignlanguage}[2]{{%
\expandafter\ifx\csname l@#1\endcsname\relax
\typeout{** WARNING: IEEEtran.bst: No hyphenation pattern has been}%
\typeout{** loaded for the language `#1'. Using the pattern for}%
\typeout{** the default language instead.}%
\else
\language=\csname l@#1\endcsname
\fi
#2}}
\providecommand{\BIBdecl}{\relax}
\BIBdecl

\bibitem{Smith_1960}
F.~T. Smith, ``Lifetime matrix in collision theory,'' \emph{Physical Review},
  vol. 118, no.~1, p. 349–356, Jan 1960.

\bibitem{Buttiker_1982}
M.~Büttiker and R.~Landauer, ``Traversal time for tunneling,'' \emph{Physical
  Review Letters}, vol.~49, no.~23, p. 711–717, 1982.

\bibitem{Wardlaw_1988}
W.~Jaworski and D.~M. Wardlaw, ``Time delay in tunneling: Transmission and
  reflection time delays,'' \emph{Physical Review A}, vol.~37, no.~8, p.
  2843–2854, Apr 1988.

\bibitem{Gallmann_2017}
L.~Gallmann, I.~Jordan, H.~J. W{\"o}rner, L.~Castiglioni, M.~Hengsberger,
  J.~Osterwalder, C.~A. Arrell, M.~Chergui, E.~Liberatore, U.~Rothlisberger
  \emph{et~al.}, ``Photoemission and photoionization time delays and rates,''
  \emph{Structural Dynamics}, vol.~4, no.~6, p. 061502, 2017.

\bibitem{Hockett_2016}
P.~Hockett, E.~Frumker, D.~M. Villeneuve, and P.~B. Corkum, ``Time delay in
  molecular photoionization,'' \emph{Journal of Physics B: Atomic, Molecular
  and Optical Physics}, vol.~49, no.~9, p. 095602, 2016.

\bibitem{Dittes_2000}
F.~Dittes, ``The decay of quantum systems with a small number of open
  channels,'' \emph{Physics Reports}, vol. 339, no.~4, p. 215–316, Dec 2000.

\bibitem{Texier_2016}
C.~Texier, ``Wigner time delay and related concepts: Application to transport
  in coherent conductors,'' \emph{Physica E: Low-dimensional Systems and
  Nanostructures}, vol.~82, p. 16–33, Oct 2016.

\bibitem{Winful_2003}
H.~G. Winful, ``Group delay, stored energy, and the tunneling of evanescent
  electromagnetic waves,'' \emph{Physical Review E}, vol.~68, no.~1, 2003.

\bibitem{Carpenter_2015}
J.~Carpenter, B.~J. Eggleton, and J.~Schr{\"o}der, ``Observation of
  eisenbud--wigner--smith states as principal modes in multimode fibre,''
  \emph{Nature Photonics}, vol.~9, no.~11, p. 751, 2015.

\bibitem{fan2005principal}
S.~Fan and J.~M. Kahn, ``Principal modes in multimode waveguides,''
  \emph{Optics letters}, vol.~30, no.~2, pp. 135--137, 2005.

\bibitem{Durand_2019}
M.~Durand, S.~Popoff, R.~Carminati, and A.~Goetschy, ``Optimizing light storage
  in scattering media with the dwell-time operator,'' \emph{Physical Review
  Letters}, vol. 123, no.~24, p. 243901, 2019.

\bibitem{Brandstotter_2019}
A.~Brandst{\"o}tter, A.~Girschik, P.~Ambichl, and S.~Rotter, ``Shaping the
  branched flow of light through disordered media,'' \emph{Proceedings of the
  National Academy of Sciences}, vol. 116, no.~27, pp. 13\,260--13\,265, 2019.

\bibitem{Gerardin_2016}
B.~Gérardin, J.~Laurent, P.~Ambichl, C.~Prada, S.~Rotter, and A.~Aubry,
  ``Particlelike wave packets in complex scattering systems,'' \emph{Physical
  Review B}, vol.~94, no.~1, Jul 2016.

\bibitem{Bohm_2018}
J.~B{\"o}hm, A.~Brandst{\"o}tter, P.~Ambichl, S.~Rotter, and U.~Kuhl, ``In situ
  realization of particlelike scattering states in a microwave cavity,''
  \emph{Physical Review A}, vol.~97, no.~2, p. 021801, 2018.

\bibitem{Texier_2013}
C.~Texier and S.~N. Majumdar, ``Wigner time-delay distribution in chaotic
  cavities and freezing transition,'' \emph{Physical Review Letters}, vol. 110,
  no.~25, 2013.

\bibitem{Lewenkopf_1992}
C.~H. Lewenkopf, A.~Müller, and E.~Doron, ``Microwave scattering in an
  irregularly shaped cavity: Random-matrix analysis,'' \emph{Physical Review
  A}, vol.~45, no.~4, p. 2635–2636, Jan 1992.

\bibitem{Cunden_2015}
F.~D. Cunden, ``Statistical distribution of the wigner-smith time-delay matrix
  moments for chaotic cavities,'' \emph{Physical Review E}, vol.~91, no.~6, p.
  060102, 2015.

\bibitem{Orjubin_2007}
G.~Orjubin, E.~Richalot, O.~Picon, and O.~Legrand, ``Chaoticity of a
  reverberation chamber assessed from the analysis of modal distributions
  obtained by fem,'' \emph{IEEE transactions on electromagnetic compatibility},
  vol.~49, no.~4, pp. 762--771, 2007.

\bibitem{Schab_2018}
K.~Schab, L.~Jelinek, M.~Capek, C.~Ehrenborg, D.~Tayli, G.~A. Vandenbosch, and
  M.~Gustafsson, ``Energy stored by radiating systems,'' \emph{IEEE Access},
  vol.~6, pp. 10\,553--10\,568, 2018.

\bibitem{Chalas_2016}
J.~Chalas, K.~Sertel, and J.~L. Volakis, ``Computation of the {Q} limits for
  arbitrary-shaped antennas using characteristic modes,'' \emph{IEEE
  Transactions on Antennas and Propagation}, vol.~64, no.~7, p. 2637–2647,
  Jul 2016.

\bibitem{Best_2005}
A.~D. Yaghjian and S.~R. Best, ``Impedance, bandwidth, and q of antennas,''
  \emph{IEEE Transactions on Antennas and Propagation}, vol.~53, no.~4, p.
  1298–1324, Apr 2005.

\bibitem{VDB_2010}
{G. A. E. Vandenbosch}, ``Reactive energies, impedance, and {Q} factor of
  radiating structures,'' \emph{IEEE Transactions on Antennas and Propagation},
  vol.~58, no.~4, p. 1112–1127, Apr 2010.

\bibitem{Capek_2015}
M.~Capek, L.~Jelinek, and P.~Hazdra, ``On the functional relation between
  quality factor and fractional bandwidth,'' \emph{IEEE Transactions on
  Antennas and Propagation}, vol.~63, no.~6, pp. 2787--2790, 2015.

\bibitem{Gustafsson_2015}
M.~Gustafsson and L.~Jonsson, ``Stored electromagnetic energy and antenna
  {Q},'' \emph{Progress In Electromagnetics Research}, vol. 150, p. 13–27,
  2015.

\bibitem{Pozar_2005}
D.~M. Pozar, \emph{Microwave engineering}.\hskip 1em plus 0.5em minus
  0.4em\relax {John Wiley \& Sons, .}, 2005.

\bibitem{Horn2012}
{R. A. Horn and C R. Johnson}, \emph{Matrix Analysis}.\hskip 1em plus 0.5em
  minus 0.4em\relax New York: Cambridge University Press, 2012.

\bibitem{DelHougne}
P.~del Hougne, R.~Sobry, O.~Legrand, F.~Mortessagne, U.~Kuhl, and M.~Davy,
  ``Experimental realization of optimal energy storage in resonators embedded
  in scattering media,'' \emph{arXiv preprint arXiv:2001.04658}, 2020.

\bibitem{Horodynski2020}
M.~Horodynski, M.~K{\"u}hmayer, A.~Brandst{\"o}tter, K.~Pichler, Y.~V.
  Fyodorov, U.~Kuhl, and S.~Rotter, ``Optimal wave fields for micromanipulation
  in complex scattering environments,'' \emph{Nature Photonics}, vol.~14,
  no.~3, pp. 149--153, 2020.

\bibitem{Harrington_2001}
R.~F. Harrington, \emph{Time-harmonic electromagnetic fields}.\hskip 1em plus
  0.5em minus 0.4em\relax Wiley-Interscience, 2001.

\bibitem{Bucci_1989}
O.~Bucci and G.~Franceschetti, ``On the degrees of freedom of scattered
  fields,'' \emph{IEEE Transactions on Antennas and Propagation}, vol.~37,
  no.~7, p. 918–926, Jul 1989.

\bibitem{gustafsson2014q}
M.~Gustafsson, D.~Tayli, and M.~Cismasu, ``Q factors for antennas in dispersive
  media,'' \emph{arXiv preprint arXiv:1408.6834}, 2014.

\bibitem{Prony}
M.~L. Van~Blaricum and R.~Mittra, ``Problems and solutions associated with
  prony's method for processing transient data,'' \emph{IEEE Transactions on
  Electromagnetic Compatibility}, no.~1, pp. 174--182, 1978.

\bibitem{Chew1995waves}
W.~C. Chew, \emph{Waves and fields in inhomogeneous media}.\hskip 1em plus
  0.5em minus 0.4em\relax IEEE press, 1995.

\bibitem{Born2013}
M.~Born and E.~Wolf, \emph{Principles of optics: electromagnetic theory of
  propagation, interference and diffraction of light}.\hskip 1em plus 0.5em
  minus 0.4em\relax Elsevier, 2013.

\bibitem{Marcuvitz}
N.~Marcuvitz, \emph{Waveguide Handbook}.\hskip 1em plus 0.5em minus 0.4em\relax
  M. I. T. Radiation Laboratory Series, 1986.

\bibitem{Kristensson}
\BIBentryALTinterwordspacing
G.~Kristensson, ``Spherical vector waves.'' [Online]. Available:
  \url{https://www.eit.lth.se/fileadmin/eit/courses/eit080f/Literature/book.pdf}
\BIBentrySTDinterwordspacing

\bibitem{Hansen}
J.~E. Hansen, \emph{Spherical Near-field Antenna Measurements}.\hskip 1em plus
  0.5em minus 0.4em\relax The Institution of Engineering and Technology, 2008.

\bibitem{Abr64}
M.~{Abramowitz} and I.~A. {Stegun}, \emph{Handbook of Mathematical Functions
  with Formulas, Graphs, and Mathematical Tables}.\hskip 1em plus 0.5em minus
  0.4em\relax New York: Dover, 1964.

\end{thebibliography}
